\documentclass[12pt,preprint]{aastex} 
\usepackage{subfigure}

\begin{document}

\title{OH Maser Observations of Planetary Nebulae Precursors}

\shorttitle{Observations of Planetary Nebulae Precursors}

\author{R. M. Deacon \altaffilmark{1}} 
\affil{School of Physics A29,
University of Sydney, NSW Australia 2006} 
\email{rdeacon@physics.usyd.edu.au}
\author{J. M. Chapman}
\affil{CSIRO Australia Telescope National Facility \\ PO Box 76, Epping,
NSW 1710, Australia} 
\and
\author{A. J. Green} 
\affil{School of Physics, University of Sydney, 
NSW 2006, Australia}
\altaffiltext{1}{Affiliated with the Australia Telescope National
Facility, CSIRO}

\begin{abstract} 
We present OH maser observations at 1612, 1665, 1667, and 1720 MHz for
86 post-asymptotic giant branch (post-AGB) stars selected from a
survey of 1612 MHz maser sources in the Galactic Plane. The
observations were taken with the Parkes Telescope and the Australia
Telescope Compact Array between 2002 September and 2003
August. Post-AGB stars are the precursors to planetary nebulae, the
diverse morphological range of which is unexplained. The maser
observations were taken to investigate the onset and incidence of wind
asymmetries during the post-AGB phase.  We re-detected all 86 sources 
at 1612 MHz while 27 sources were
detected at 1665 and 45 at 1667 MHz. One source was re-detected at
1720 MHz.  We present a classification scheme for the maser profiles
and show that 25\% of sources in our sample are likely to have
asymmetric or bipolar outflows. From a comparison of the maser and
far-infrared properties we find that there is a likely trend in the
shape of the maser profiles with some sources evolving from
double-peaked to irregular to fully bipolar profiles.  A subset of
higher-mass sources stand out as having almost no mainline emission
and mostly double-peaked profiles.  At least 25\% of sources in the
sample are variable at one or more of the frequencies observed.  We
also confirm a previously-noted 1667 MHz overshoot phenomenon.
\end{abstract}


\section{Introduction} 

Planetary nebulae (PN) are the remnants of low-to-middle mass
stars. When stars, like our own sun, run out of core-burning hydrogen,
they go through a sequence of structural changes. The outer layers
greatly expand while the stellar cores decrease in size and become
hotter. Of particular interest in the asymptotic giant branch (AGB)
stage is that most stars shed a large fraction of their mass through
slow, dense stellar winds.  Such winds transfer more than half of a
star's original mass back into the interstellar medium. After most of
the outer layers of gas have been shed, it is thought that stars
change from losing mass in slow dense winds to losing mass in much
hotter, faster winds \citep{Kwo78}. During this post-AGB stage, the
hot winds sweep up the remaining material surrounding the stars and
the swept-up shells become visible as PN as the central stars become
hot enough to ionise the shells (Kwok et al. 1978).

The ionised shells of PN exhibit many different forms. While some
appear circular, more than half have elliptical or bipolar shapes; in
some cases with complex, filamentary structures \citep{Man00}. Hubble
Space Telescope (HST) observations indicate that approximately 50\% of
proto-planetary nebulae (the newly-optically visible precursors to PN)
also show similar structures \citep{Uet00,Su01}.  The causes for this
proliferation of geometries are not yet known and several different
theories are being debated.  One possible explanation is that magnetic
fields from the stars may constrain or collimate the stellar winds
\citep[e.g.][]{Gar99}. Hydrodynamical models show that a bipolar 
structure may form if the slow AGB wind possesses a weak asymmetry
that is amplified during the post-AGB phase by a fast-wind, slow-wind
interaction \citep[e.g.][]{Fra94}. In this case, strong bipolar
structures are not expected to be present in the early post-AGB phase
as the stars are too cool ($<$ 15 000 K) for the onset of a fast
wind. However, \citet{Sah98} have proposed that fast jets operating
in the early post-AGB stage may be the dominant shaping mechanism.
The origin of these jets is unclear but the influence of companion
stars or planets is the most plausible reason.  Bipolarity may also
arise through other mechanisms linked to the presence of companion
stars or planets \citep{Woo00,Sok01}.

OH maser observations provide a means of studying the cooler molecular
emission from the outer circumstellar envelopes of AGB and post-AGB
stars. The properties of OH 1612 MHz maser emission from OH/IR stars
at the tip of the AGB are the best understood. Almost all OH 1612
MHz spectra show a double-peaked profile which is characteristic of
maser emission from a thin spherical shell of a constant radius and
expansion velocity. In an OH 1612 MHz survey of the Galactic Plane,
\citet{Sev01} found that 86\% had double-peaked profiles, 13\% had
single-peaked profiles while less than 1\% of sources had irregular
profiles with more than two emission peaks. A high degree of spherical
symmetry in OH/IR stars has also been confirmed from aperture
synthesis observations of approximately 30 sources
\citep[e.g.][]{Cha85,Wel87}.

The post-AGB stars occupy a short-lived transition stage between AGB
and PN phases. Maser observations of post-AGB stars provide a means of
investigating the onset and development of wind asymmetries.  Most
studies of maser emission from post-AGB stars have concentrated on
sources with unusual maser properties. Aperture synthesis observations
for a sample of 10 post-AGB stars with unusually high OH expansion
velocities, for example, are consistent with bipolar outflows and
enhanced equatorial densities \citep{Zij01}.  It is apparent that for
some sources, wind asymmetries are present during the early post-AGB
stage where molecular material leftover from the AGB phase is still
present. However, \citet{Sev02I,Sev02II} has argued that only a small
fraction of both AGB and post-AGB sources have irregular OH 1612 MHz
spectra.

To investigate the onset and incidence of wind asymmetries during the
post-AGB phase, we are observing the maser properties of an
OH-selected sample of 86 post-AGB stars with well-determined
far-infrared properties. Here we present total intensity spectra
obtained for the OH ground state maser transitions at 1612, 1665,
1667, and 1720 MHz and discuss their statistical properties. Other
maser transitions and their polarisation properties will be discussed
in future papers.

\section{Sample Selection}

We have selected a well-defined sample of 86 compact sources from the
OH 1612 MHz survey of \citet{Sev97I,Sev97II,Sev01}. This survey, taken
with the Australia Telescope Compact Array (ATCA) and the Very Large
Array (VLA) (hereafter the ATCA/VLA survey) detected OH 1612 MHz maser
emission from a total of 766 compact sources. The OH 1612 MHz maser
positions were measured with a precision of typically 0.5 arcsec and
stellar velocities to 1 km s$^{-1}$. The survey is 95\% complete for
peak OH flux densities above 0.38 Jy.

\begin{figure}[b!]
	\hspace*{40pt}\rotatebox{270}{\scalebox{0.5}{\includegraphics{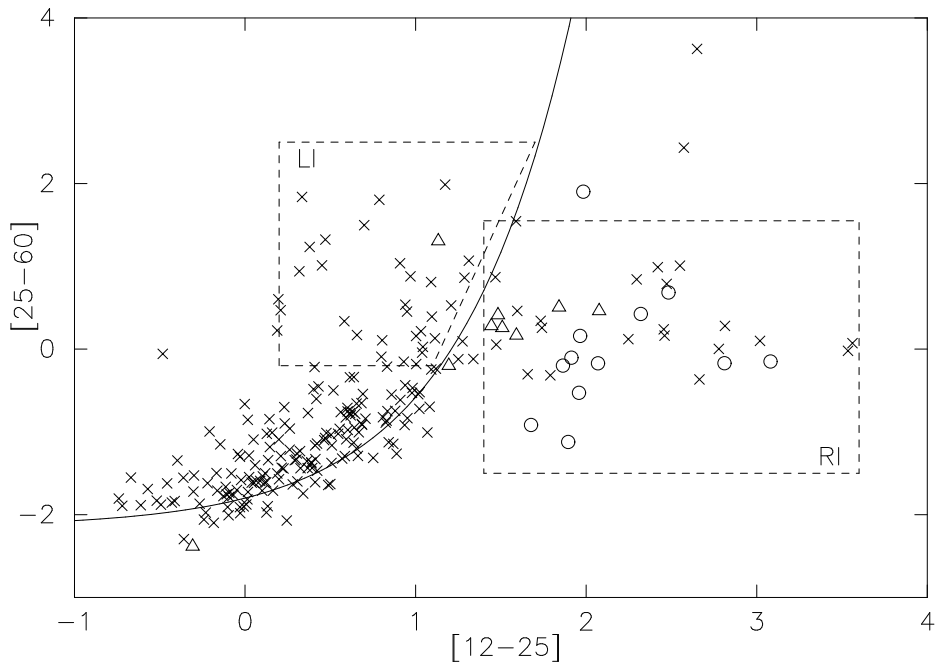}}}
	\caption{ An IRAS two-colour diagram for sources in the
	ATCA/VLA survey. The solid line is the evolutionary sequence
	for AGB stars defined by \citet{Vee88}. The regions marked
	`RI' and `'LI' are associated with post-AGB stars (Sevenster
	2002a,b). The three sources above the RI box are identified as
	star-formation regions. The crosses show sources with IRAS
	colours only. Triangles show sources with MSX colours that
	correspond to Quad IV (early post-AGB stars) of Figure \ref{Fi:MSX}. 
        Circles show sources 
        with MSX colours that correspond to Quad I (late post-AGB stars) of 
Figure \ref{Fi:MSX}. \label{Fi:IRAS}}
\end{figure}

\citet{Sev02I,Sev02II} compared the   far-infrared  IRAS (InfraRed
Astronomical Satellite) and MSX (Mid-course Space Experiment)
properties for sources in the ATCA/VLA survey and has shown that the
far-infrared colours provide a powerful tool for distinguishing
between AGB and post-AGB stars.  Figure \ref{Fi:IRAS} shows an IRAS
two-colour diagram for 239 sources in the ATCA/VLA survey that have
well-determined flux densities (quality flag 2 or 3) in the IRAS 12,
25, and 60 $\mu$m bands. The infrared colours are defined as [a-b] =
2.5 log (S$_b$/S$_a$) where S is the flux density in Jy and a, b, are
wavelengths in $\mu$m.  The solid line shows the well-known
evolutionary track for AGB stars, defined by \citet{Vee88}. The
right-hand IRAS (RI) region has traditionally been associated with
post-AGB stars that have evolved to the right of the evolutionary
track. Increasing [12-25] colour indicates thickening circumstellar
shells.  A second group of likely post-AGB stars, discussed by
\citet{Sev02II}, the left-hand IRAS (LI) group, lies to the left of
the track.  These stars have a 60 $\mu$m excess indicating large
amounts of cool dust are around these stars, probably due to very high
mass-loss rates. Sevenster has shown that this group contains stars
with higher outflow velocities ($>$ 15 km s$^{-1}$) and higher initial masses. The RI and
LI groups probably represent different evolutionary sequences for
lower and higher mass post-AGB stars respectively with an initial mass
separation between the two groups at 4 M$_\odot$.  The IRAS colours
for the RI group are [25-60] $<$ 1.5 and [12-25] $>$ 1.4. For the LI
region the IRAS colours are [25-60] $>$ -- 0.2 and [12-25] $>$ 0.2,
with [25-60] $>$ -- 2.15 + 0.35 exp(1.5 [12-25]).

\begin{figure}[h]
	\hspace*{40pt}\rotatebox{270}{\scalebox{0.5}{\includegraphics{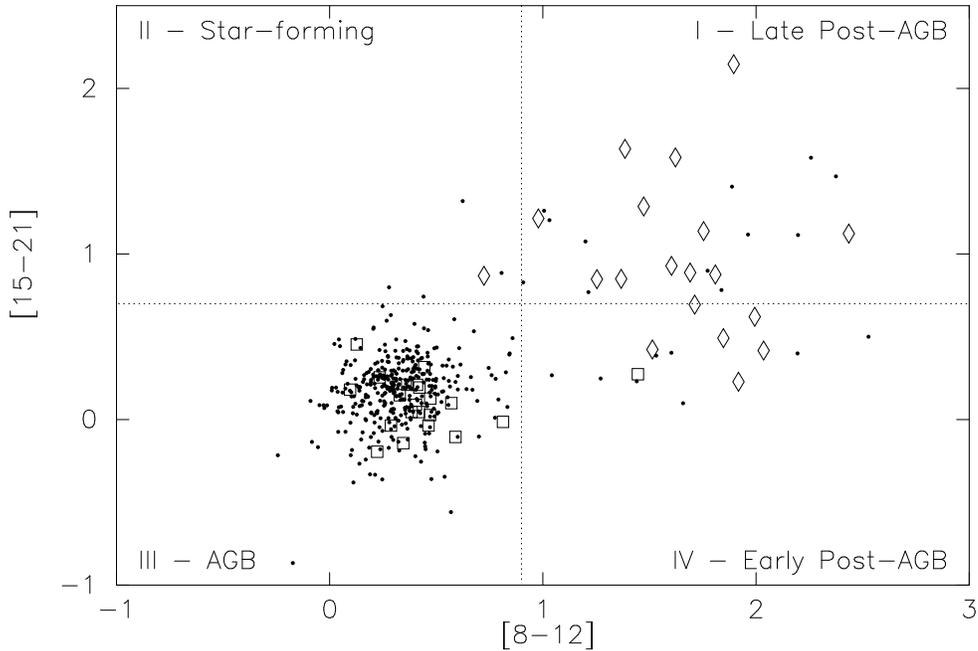}}}
	\caption {An MSX two-colour diagram for sources in the
	ATCA/VLA survey. Following \citet{Sev02I}, this is divided
	into four quadrants according to minima which are seen in
	bimodal distributions of the MSX infrared colours [8-12] and
	[15-21]. The four quadrants I-IV are thought to contain in
	turn late post-AGB stars, star-formation regions, AGB stars and 
        early post-AGB stars. The dots show sources with MSX colours only. Squares show
	sources with IRAS colours in the LI region of Figure \ref{Fi:IRAS}.
	Diamonds show sources with IRAS colours in the RI region of
	Figure \ref{Fi:IRAS}. \label{Fi:MSX}}
\end{figure}

Figure \ref{Fi:MSX} shows an MSX two-colour diagram for 424 sources in
the ATCA/VLA survey with well-determined MSX flux densities in the 8,
12, 15, and 21 $\mu$m bands. Following \citet{Sev02I}, the diagram is
split into four quadrants separated at [8-12] = 0.9 and [15-21] =
0.7.  The MSX colours provide a clear separation between AGB stars and
post-AGB stars to the right of the IRAS evolutionary track. Early
(younger) post-AGB stars are located in Quad IV and late (more evolved)
post-AGB stars in Quad I. Quad II contains star-forming
regions. Quad III includes both AGB stars and LI sources. The MSX
colours for Quad IV stars are [15-21] $<$ 0.7 and [8-12] $>$ 0.9.  For
Quad I, the MSX colours are [15-21] $>$ 0.7 and [8-12] $>$ 0.9.

Our sample is the set of 88 sources in the ATCA/VLA survey with OH
1612 MHz detections and far-infrared colours corresponding either to
the IRAS RI and LI groups or to MSX Quads I and IV. The number of
sources in each group are: 38 RI, 30 LI, 15 Quad IV, and 24 Quad I.
As shown in Figures 1 and 2 there is an overlap between the IRAS and
MSX groups; 12 stars are both Quad I and RI objects, six are Quad IV
and RI, and one source is a Quad IV and LI object.  No objects are in
both the Quad I and LI groups.

Table \ref{Ta:source} lists the source parameters for the 88 sources.
The columns give the source identifier from the ATCA/VLA survey
\citep{Sev97I,Sev97II,Sev01}, IRAS Name, Right Ascension and
Declination (J2000), radial velocity of the central star (relative to
the Local Standard of Rest), the IRAS 12, 25, and 60 $\mu$m and MSX 8,
12, 15, and 21 $\mu$m flux densities, and the classification/s of each
source as RI, LI, Quad I, or Quad IV.  All coordinates in this paper
are given for the epoch J2000 unless otherwise stated. The radial
velocities are from the ATCA/VLA survey and the infrared flux
densities are reproduced from \citet{Sev02I,Sev02II}.  Two sources in
Table \ref{Ta:source}, d29 \citep[OH323.459-0.079,][]{Cha96} and v280
\citep[OH43.165-0.028,][]{Cas04} have previously been identified as
star-formation regions and are not discussed further in this
paper. Our final sample therefore consists of 86 sources.

\section{Observations and Analysis}

\subsection{Parkes Observations}

Observations were taken for the four ground-state OH maser transitions
with rest frequencies of 1612.231, 1665.4018, 1667.359, and 1720.530
MHz, using the Parkes radio telescope in 2002 September, 2003
January/February, and 2003 August. A linear feed with a radio
frequency hybrid was used to generate two circular polarisations.  All
observations were taken with the H-OH receiver, a bandwidth of 4 MHz,
on-line Doppler tracking, and the Parkes multibeam correlator
configured for 8192 channels. Only the central 2 MHz were recorded for
1612 MHz observations.  At 18 cm the FWHM of the telescope primary
beam is 12.6 arcmin. Table \ref{Ta:obs} lists the observing dates and
frequencies, average integration times, and the typical 1$\sigma$
noise per channel of spectra in each observing run.  Severe dust
storms around Parkes on 2003 January 30 are responsible for the
anomalously high noise levels on this date.

Initial data reduction to calibrate the spectral bandpass and absolute
flux scale was carried out using SPC, a single-dish spectral line
reduction package supported by the Australia Telescope National
Facility (ATNF).  Absolute flux calibration was applied using
observations of the standard sources Hydra A and 1934-638.  Flux
densities of 38.0 Jy (1612 MHz) and 37.0 Jy (1665/1667/1720 MHz) for
Hydra A, and 15.0 Jy (1612 MHz) and 14.9 MHz (1665/1667/1720 MHz) for
1934-638 were assumed.  Further data reduction involving
Hanning-smoothing, baseline calibration, and flagging was carried out
with the Spectral Line Analysis Package \citep[SLAP,][]{Sta85}, an
interactive program for radio spectral line data reduction.  The
channel separation and velocity resolution of the spectra were 0.09
and 0.18 km s$^{-1}$ respectively. All velocities are given in the
radio convention with respect to the Local Standard of Rest.

The spectra were analysed using SLAP with peak flux densities and
velocities measured for either one (for irregular or single-peaked
sources) or two (for double-peaked sources) emission peaks. Emission
peaks were considered to be definite detections if one of the
following criteria was satisfied:

\begin{itemize}
\item Peak flux density $> 5\sigma$ and linewidth $> 10$ channels 
(0.9 km s$^{-1}$); or
\item Peak flux density $> 5\sigma$, linewidth = $6\rightarrow$ 10
channels  and velocity consistent with emission feature detected in
another OH transition; or
\item Peak flux density $> 5\sigma$, linewidth = $4\rightarrow$ 5
channels  and spectrum double-peaked with both peaks detected in
another  OH transition; or
\item Peak flux density $3\rightarrow5\sigma$, linewidth $>$ 10
channels and spectrum double-peaked with both peaks detected in
another  OH transition.
\end{itemize}

Occasionally the Parkes spectra are confused by other sources detected
within the primary beam. In some cases, confusion can be attributed to
nearby compact stellar maser sources with identifications in the
ATCA/VLA survey. Confusion also arises from OH emission and absorption
from extended gas clouds in the Galactic plane, particularly for
sources with longitudes close to the Galactic centre. This type of
confusion is characterised by broad emission and/or absorption ($>$ 10
km s$^{-1}$) with features often being present at the same velocities
at all four frequencies \citep{Cas77}.

\subsection{Australia Telescope Compact Array Observations}

Observations at 1720 MHz were taken with the EW367 array configuration
with all six antennas of the ATCA on 2003 August 8, for 18 sources
with 1720 MHz emission detected within the beam of the Parkes
telescope. The ATCA is an array of six 22-m diameter dishes that are
located on an east-west track, with a maximum baseline of 6 km
\citep{Fra92}.  Each source was observed in seven two-minute cuts
distributed over an eight-hour observing period.  A bandwidth of 4 MHz
was used with 1024 channels, giving a channel separation of 0.73 km
s$^{-1}$ and a velocity resolution of 0.88 km s$^{-1}$. The data were
reduced using the Miriad reduction package \citep{Sau95}. To search
for compact sources, images were made using natural weighting with an
angular resolution of typically 6'' x 14'', measured east-west by
north-south.

On 2003 July 23/24 and 2003 August 7, observations of the 1665 and
1667 MHz OH transitions were taken at the ATCA. Each source was
observed for typically eight two-minute cuts, distributed over a 12
hour period. The two OH mainlines were observed simultaneously using a
4 MHz bandwidth with 1024 spectral channels. The channel separation is
0.70 km s$^{-1}$ and the velocity resolution is 0.84 km s$^{-1}$. The
6D array was used, giving an angular resolution of approximately 7'' x
10''.  Table \ref{Ta:obs} gives the observing dates and frequencies,
average integration times, and the typical 1$\sigma$ noise per channel
(measured in line-free channels) for the ATCA observations.

The ATCA OH mainline spectra are not shown but have been used to
confirm maser positions and to verify whether the OH mainline
detections obtained from the Parkes data are associated with the
stellar sources, or are due to sources of confusion or interference.
As an example, Figure 3 shows the Parkes and ATCA spectra of b292 at
1720 MHz plotted over the same velocity range. The features detected
with the Parkes radio telescope at velocities below 60 km s$^{-1}$ are
completely absent in the ATCA spectrum confirming that these are due
to Galactic plane confusion. The 1720 MHz emission from b292 is
discussed further in Section 5.

\section{Results}

\subsection{OH 1612, 1665 and 1667 MHz detections} \label{Sec:Results}

All 86 sources selected from the Sevenster survey were re-detected at
1612 MHz with the Parkes telescope. Of the 86 sources, 27 (31\%) were
detected at 1665 MHz and 45 (52 \%) were detected at 1667 MHz. The
maser detections are summarized in Table \ref{Ta:1612etc}. The three
rows for each source give the 1612, 1665, and 1667 MHz results
respectively. The columns are:

\begin{itemize}
\item 1: source name, as given in the ATCA/VLA survey;
\item 2--4: velocity, peak flux density and integrated flux density for the
blue-shifted emission peak of double-peaked sources, or for the
strongest emission peak from sources with single peaks or irregular
spectra;
\item 5--7: velocity, peak flux density and integrated flux density for the
red-shifted emission peak;
\item 8: peak-to-peak flux density ratio;
\item 9: integrated flux density ratio
\item 10: total velocity range of detected emission (excluding absorption
and identified confusion)
\item 11: spectral classification (explained in Section 6.1)
\item 12: Variability, marginal detection and date flag: A `V' in this column 
indicates the peak flux of a source at 1612 MHz is more than 30 \% variable 
(see Section \ref{Sec:var}).  An `m' in this column
indicates that the detection was marginal with a peak flux density of
$3\rightarrow5$ sigma. Marginal detections are included in Figure 7
but are not included in source counts and further discussion. A `\#'
indicates that the observation was taken on 2003 August 18. All other
Parkes data for the 1612, 1665, and 1667 transitions were taken in
2003 September.
\end{itemize}

The Parkes spectra are plotted in Figure \ref{Fi:thumbnails:a} with
labels in the top-left and top-right corners giving the source name
and OH transition respectively. OH emission peaks from nearby sources
are indicated where these have been identified.

\subsection{Notes on individual sources}

The following notes refer to the OH 1612, 1665, and 1667 MHz Parkes
spectra shown in Figure 7.

{\it d3:} Confusion from d2 at 1612 MHz.

{\it d34:} Confusion from d33 at 1612 MHz.

{\it d46/d47:} These two sources suffer mutual confusion as well as
from Galactic dust clouds. Absorption at $-$20 and $-$50 km s$^{-1}$
is evident in the mainline spectra of both sources (off-scale in the
figure for d46).  The 1720 MHz line is seen at these velocities in
absorption in d46 and in emission in d47, and there is 1612 MHz
absorption at $-$20 km s$^{-1}$ in d47.  These features are attributed
to the same network of Galactic dust clouds.  The strongest 1612 MHz
peaks in d47 appear faintly in the d46 spectrum (outside range shown
in figure).

d47 is the only source with a strong, symmetrical four-peaked OH
1612-MHz profile. ATCA observations confirm that the narrow emission
peaks in the mainline spectra at $-$30 and $-$70 km s$^{-1}$ are
associated with the stellar source.  The ATCA observations detected
several weak 1667 MHz emission features between $-$30 and $-$74 km
s$^{-1}$, including one at 50 km s$^{-1}$, which was detected at
Parkes but confused with an absorption dip at that velocity.

{\it d62:} Confusion from d59 at 1612 MHz. The mainline features seen
in the Parkes spectra may include some background Galactic confusion
and variable features.

{\it d93:} The weak 1612 MHz peaks at $-$117.1 and $-$39.5 km s$^{-1}$
are likely to be radio frequency interference (RFI).

{\it b11:} The stellar velocity is revised to 13.4 km s$^{-1}$, from
the velocity of 25.3 km s$^{-1}$ given by \citet{Sev97I} as the new
observations show a double-peaked 1667 MHz spectral profile and broad
1665 MHz emission centred at this velocity.

{\it b14:} The peak at 50.4 km s$^{-1}$ at 1612 MHz is not present in
the ATCA/VLA spectrum and is attributed to an unknown confusing
source. There is no match with any other ATCA/VLA sources.  It does
not have a typical signature of RFI.

{\it d200:} The stellar velocity has been revised to $-$74.7 km
s$^{-1}$ for this source on the basis of its double-peaked 1612 MHz
profile.

{\it b25:} A small broad peak at $-$40 km s$^{-1}$ at 1612 MHz is
attributed to Galactic confusion, as it is not present in the ATCA/VLA
spectrum.

{\it b33:} This source has four peaks at 1612 MHz. The peak at $-$4 km
s$^{-1}$ is too narrow (3 channels) to be a reliable detection.  The
peaks at $-$36 and $-$12 km s$^{-1}$ were identified with this source
in the ATCA/VLA survey.  The peak at $-$2l km s$^{-1}$ may be from a
confusing source, as there is absorption near this velocity.

{\it b44:} This source shows one definite peak at $-$35.0 km s$^{-1}$
at 1612 MHz and a weak narrow peak at $-$9.5 km s$^{-1}$ that is
probably RFI.  A second peak at +30.5 km s$^{-1}$ noted in the
ATCA/VLA survey was not detected here.

{\it b62:} This is an irregular source at 1612 MHz with three peaks
apparent.  It is possible one or more emission features are due to
extended Galactic emission as there is absorption at a nearby velocity
(5 km s$^{-1}$).  The two narrow spikes at $-$9 and $-$3 km $^{-1}$ in
the 1612 MHz spectrum are likely to be RFI.

{\it b96:} This source shows four peaks at 1612 MHz.  The peaks at
$-$127 and $-$146 km s$^{-1}$ were identified in the ATCA/VLA survey.
The peaks at $-$118 and $-$136 km s$^{-1}$ have not been identified
with another source.  Higher resolution observations are needed to
determine if the latter peaks are associated with b96.

{\it b133:} The narrow peaks at $-$2 and $-$8 km s$^{-1}$ are likely
to be RFI.  The broader peaks at 18 and $-$24 km s$^{-1}$ were
identified in the ATCA/VLA survey.

{\it b155:} There is probable RFI at $-$13.5 km s$^{-1}$ at 1612 MHz.

{\it b199:} The drop-off at the red-shifted end of the baseline in
both spectra is due to a poor baseline fit to complex absorption
features in the spectrum.

{\it b246:} The stellar velocity has been revised to 138.0 km s$^{-1}$ from
the central velocity of the 1612 and 1665 MHz spectra

{\it b292:} This source is the only post-AGB object previously
detected at 1720 MHz \citep{Sev01b} and the only source detected in
the present study where the 1720 MHz emission is associated with the
stellar source (see Section 5). As shown in Figure 3, Galactic confusion
was detected at velocities of $-$20, 20, and 40 km s$^{-1}$.

{\it b301:} Confusion from b298 at 1612 MHz.

{\it b304:} This source has probable RFI at $-$5.0 km s$^{-1}$ at 1612
MHz .

{\it v50:} This source has probable RFI at $-$5 km s$^{-1}$ at 1612
MHz, and at 5, 95, and 100 km s$^{-1}$ at 1667 MHz.

{\it v53:} Galactic confusion is present in the Parkes observations.
ATCA observations at 1667 MHz confirmed only a narrow peak near 4 km
s$^{-1}$, similar in width to the 1612 MHz peaks and much narrower
than the 1667 MHz Parkes spectrum, leading to a 1667 MHz
classification as S rather than I (see Section \ref{Sec:spec}).

{\it v117:} The Parkes 1667-MHz spectrum shows confusion at velocities
above 70 km s$^{-1}$. The ATCA observations confirm that the broad
emission peak seen at all three frequencies around 40-50 km s$^{-1}$
is associated with the stellar source.

{\it v154:} Confusion from v156 at 1612 MHz.

{\it v162:} Confusion from v158 at 1612 MHz. The two narrow peaks at
86.6 and 89.1 km s$^{-1}$ are probably RFI.

{\it v268:} This source has an unidentified peak at 21 km s$^{-1}$,
likely to be RFI or confusion from interstellar clouds.  There is also
RFI at 6 km s$^{-1}$ and weakly at velocities above 90 km s$^{-1}$.

{\it v270:} There is probable RFI at 60, 115, and 135 km s$^{-1}$ at
1667 MHz.

\section {OH 1720 MHz results} \label{Sec:1720}

From the Parkes observations, a total of 18 sources showed 1720 MHz
emission features. In most cases the emission was broad and coincident
in velocity with other features detected in the OH mainlines that were
known to be from unrelated Galactic emission. These sources were not
detected in the follow-up ATCA 1720 MHz observations, confirming that
the emission was spatially extended.

For three sources, the ATCA images showed compact sources of 1720 MHz
emission within 20 arcmin of the stellar position. For d168, the ATCA
1720 MHz maser position, determined to a precision of approximately
one arcsec, is $17^h5^m11.3^s$, $-41^{\circ}28'44''$, offset by
$\sim$15 arcmin from the stellar position. The 1720 MHz detection is
probably associated with the star-formation region IRAS 17016-4124,
also known as G345.0-0.2 \citep[$17^h1^m40.268^s$,
$-41^{\circ}25'01.16''$ (B1950),][]{Gau87}.

For b33, the position of the compact 1720 MHz source is
$17^h25^m32.4^s$, $-36^{\circ}21'51''$, offset by 8 arcmin from the
stellar position. This is clearly associated with IRAS 17221-3619
($17^h25^m31.7^2$, $-36^{\circ}21'54''$), a massive young stellar
object (Chan et al. 1996).

For b292 (OH009.1-0.4) the ATCA 1720 MHz observations confirm the
presence of a compact maser source associated with the stellar source,
previously reported by \citet{Sev01b}.  The ATCA 1720 MHz position is
$18^h7^m20.9^s$, $-21^{\circ}16'10.9''$, in excellent agreement with
the OH 1612 MHz maser position (see Table 1) and with the previously
published 1720 MHz position from \citet{Sev01b}.

\begin{figure}[t]
\begin{center}
\rotatebox{270}{\scalebox{0.4}{\includegraphics{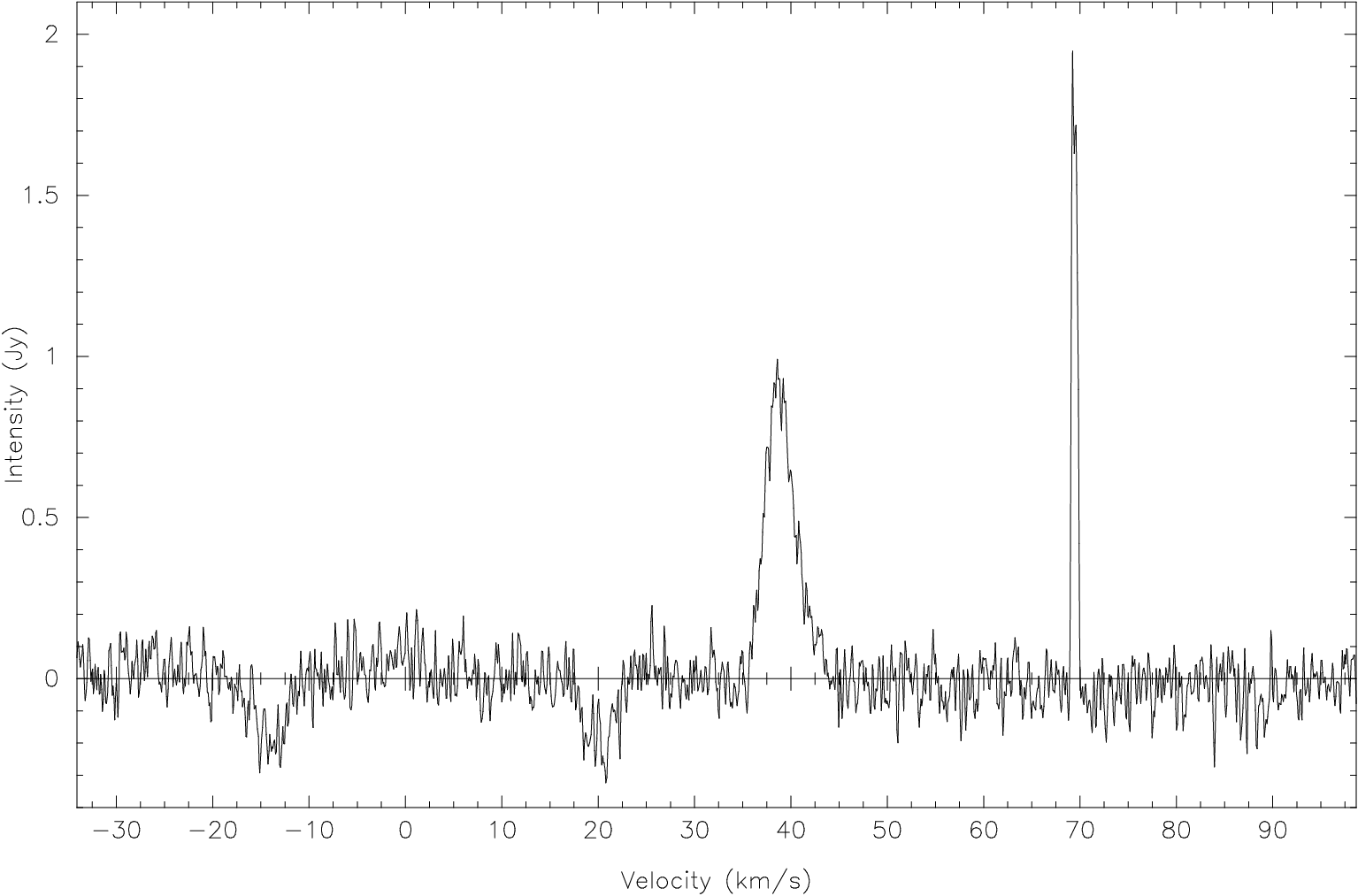}}}
\\
\vspace{12pt}
\rotatebox{270}{\scalebox{0.4}{\includegraphics{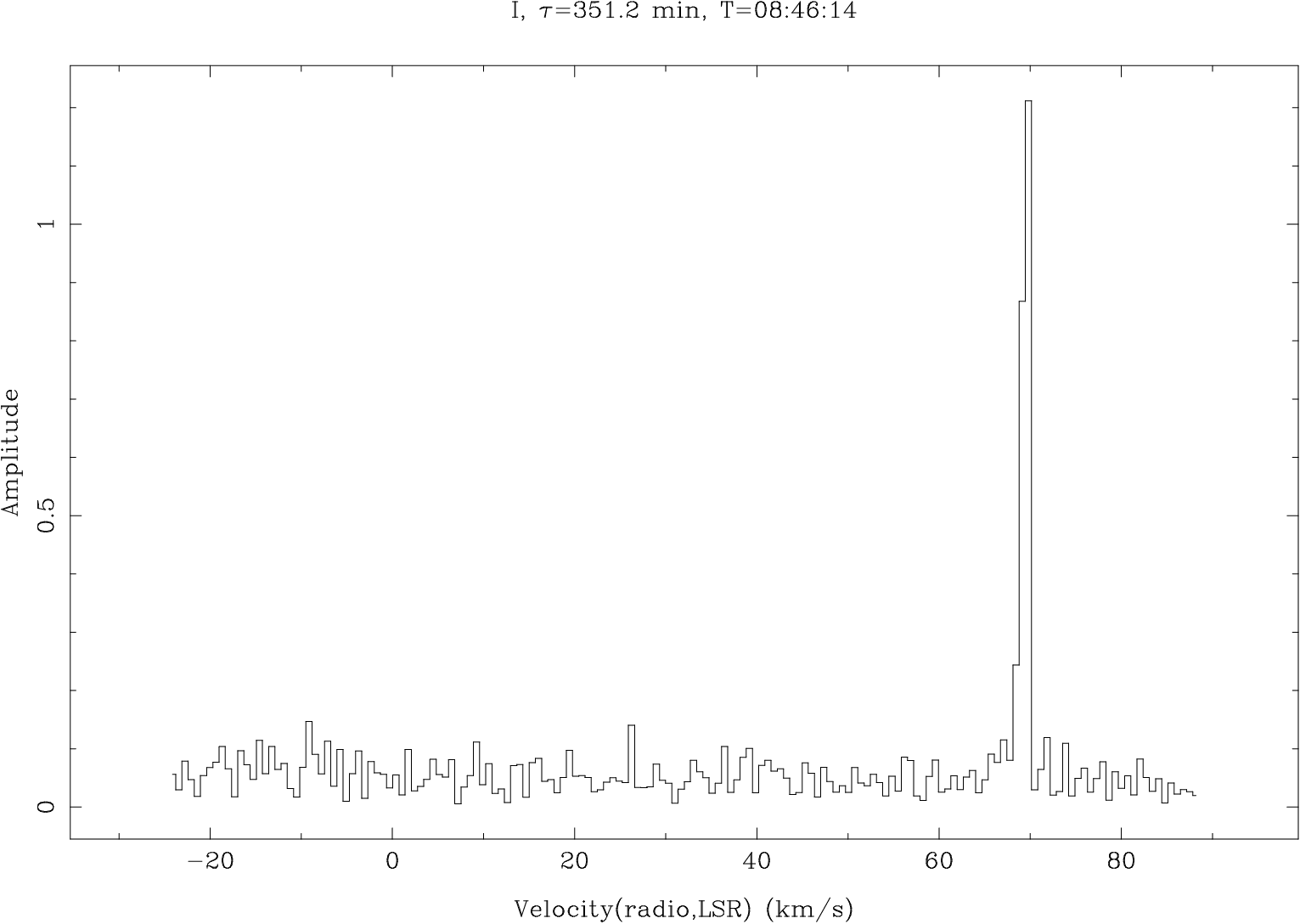}}}
\end{center}

\caption{Total intensity 1720 MHz spectra for b292. The top spectrum
is from an observation at Parkes on 2003 February 2. The
velocity resolution is 0.18 km s$^{-1}$. The bottom spectrum was
obtained with the ATCA on 2003 August 8 with a velocity resolution of
0.88 km s$^{-1}$. The broad emission and absorption features
in the Parkes spectrum are absent from the ATCA spectrum, showing that
these emission and absorption features are spatially extended and are
not associated with the compact stellar source. The smaller peak flux density
at 69.2 km s$^{-1}$ in the ATCA spectrum is due to the dilution effect caused by the broader velocity
resolution of the ATCA spectrum. \label{Fi:1720}}
\end{figure}

Figure \ref{Fi:1720} shows the 1720 MHz Parkes and ATCA spectra for
b292 obtained in February 2003 and August 2003 respectively. The
Parkes spectrum shows a narrow emission feature with a peak flux
density of 2.0 Jy at a velocity of 69.2 km s$^{-1}$, consistent with
the previous detection at ATCA of an emission peak of 1.22 Jy at a velocity of
69.4 km s$^{-1}$ reported by \citet{Sev01b}.  The lower velocity
resolution of this previous observation accounts for the lower
peak flux.

From the MSX colours, b292 is a likely young post-AGB star. It is
unusual among our sample in that it has 1665 MHz emission but no 1667
MHz emission (see Sections \ref{Sec:Results} and
\ref{Sec:colours}). Figure \ref{Fi:b292_olay} (top) shows the 2002/2003
Parkes spectra at 1612, 1665, and 1720 MHz. The strongest emission, at
1612 MHz, is one sided with a blue-shifted peak of 18.9 Jy at 70.4 km
s$^{-1}$.  A much weaker red-shifted peak was detected at velocities
near 99 km s$^{-1}$ but is not seen in Figure \ref{Fi:b292_olay} as
 scaling has reduced the feature to insignificance.  The 1665 MHz 
emission is also strongest at
blue-shifted velocities with a peak flux density of 7.4 Jy at 69.7 km
s$^{-1}$. We detected a second weaker 1665 MHz peak with a flux
density of 1.6 Jy at a velocity of 104.0 km s$^{-1}$ that was not
evident in the 1998 observations of Sevenster \& Chapman.  From the
double-peaked 1665 MHz spectrum we obtain an expansion velocity at
1665 MHz of 17.1 km s$^{-1}$ and a stellar velocity of 87.0 km
s$^{-1}$.  As previously reported, the narrow 1720 MHz emission peak
is seen at a slightly more blue-shifted velocity than the bluest 1612
and 1665 MHz emission peaks.

A comparison of our spectra with the previous spectra from
\citet{Sev01b} indicates negligible variability in the 1720 and 1612
MHz emission between 1999 and 2003. However, a comparison of the 2003
ATCA 1665 MHz spectrum with the earlier spectrum, obtained with the
same velocity resolution of 0.8 km s$^{-1}$, shows that the 1665
MHz emission has increased significantly in intensity, from a peak
flux density of 1.64 Jy in 1998/1999 to a peak flux density of 2.2 Jy
in 2003 July.

The physical conditions needed for 1720 MHz maser emission have been
tightly constrained by modelling \citep{Loc99,Eli76}.  Temperatures of
50--125 K, densities around $10^5$ cm$^{-3}$,and OH column densities
of order $10^{16}$ cm$^{-2}$ are required.  1720 MHz masers are
thought to be shock excited, rather than radiatively pumped and a
theoretical explanation of their production, based on the formation of
C-shocks in a molecular cloud is given by \citet{War99}. In addition,
the masers are strongly beamed (see below). In circumstellar shells,
the likely mechanism leading to the creation of a C-type shock is the
interaction of the fast post-AGB wind with the remnant slow AGB wind
\citep{Sev01b}, with the 1720 MHz maser being produced in the
interaction zone. This is a similar mechanism to the production of
1720 MHz masers in supernovae remnants \citep{Frail94}.

\citet{Sev01b} previously predicted that H$_2$O maser emission would
{\em not} be present in b292 since the passage of a C-type shock would
not lead to a sufficiently strong increase in the density of H$_2$O
\citep{War99}.  However for b292 we {\em have} detected H$_2$O maser
emission.  Figure \ref{Fi:b292_olay} (bottom) shows a 22 GHz spectrum
obtained from an ATCA observation on 2002 September 24 by Deacon.
Three peaks were detected.  As this was a ``snap-shot'' observation it
could not be confirmed that the H$_2$O emission was from the stellar
position, but an association is probable as the velocity range of the
H$_2$O emission is similar to that of the OH emission (between 72 and
106 km s$^{-1}$) and there are no known other maser sources within
the $2.75'$ 22 GHz ATCA beam.

The presence of H$_2$O and 1612 MHz OH maser emission in addition to
the shock excited 1720 MHz emission implies a strong variation in the
properties of the mass outflow from b292 since the 1612 and H$_2$O
masers cannot be spatially co-located within the circumstellar
envelope with the 1720 MHz maser emission. We predict that the OH 1665
and 1612 MHz and most blue- and red-shifted H$_2$O maser emission
arises from the remnant AGB wind while the OH 1720 MHz emission is in
a new bipolar outflow that is starting to disrupt the spherical AGB
shell (giving a De profile at 1612 MHz - see Section
\ref{Sec:spec}). This stream could indicate a high-speed collimated outflow
of the type proposed by \citet{Sah98}. If the outflow is not parallel
to the line of sight, the 1720 MHz velocity is seen in projection.
The central H$_2$O emission probably arises quite close to the central
star.

Our detection rate for 1720 MHz maser emission from post-AGB stars of
1/86 indicates that the conditions for 1720 MHz emission are rarely
met and/or the time period for occurrence is very short.  One factor
decreasing the detection rate is that the maser emission is beamed
parallel to the C-shock front and is tangentially amplified. This
requires that the shock must be near-perpendicular to the
line-of-sight for maser emission to be observed \citep{War99}. 1720
MHz masers are also very dependent on physical conditions.  Higher
densities first quench the 1720 MHz masers, then favour production of
1612 MHz masers while higher temperatures favour the formation of
mainline masers (Lockett et al. 1999). As discussed in Section 6,
bipolar outflows in post-AGB stars are usually traced by mainline and
1612 MHz OH emission. Given the unusual physical conditions required,
the low detection rate for 1720 MHz maser emission in post-AGB stars
is not surprising.

\section{Discussion}

\subsection{Spectral Profile Classifications} \label{Sec:spec}

In this section we discuss the maser properties of the OH 1612, 1665,
and 1667 MHz transitions. Table 4 summarises the number of OH
detections at 1612, 1665, and 1667 MHz by frequency and, using a
profile classification scheme, spectral profile. All 86 sources were
detected at 1612 MHz while OH mainline emission was detected in one or
both of the OH mainline transitions from a total of 50 sources. 23
sources were detected at 1667 MHz only, five sources were detected at
1665 MHz only, and 22 sources were detected in both mainlines. For
81/86 sources the OH 1612 MHz transition is the strongest while for
sources detected in both 1665 and 1667 MHz the 1667 MHz transition
almost always dominates. In many cases the 1667 MHz profiles are very similar
(though weaker) to the OH 1612 MHz spectra while the weaker OH 1665
MHz transition may show a different profile type. Two examples of
sources which differ from this are d46, where the strongest emission
occurs at 1665 MHz, and b11 where the OH 1667 MHz emission is
considerably stronger than the 1612 MHz emission.

We have identified six different types of spectral profiles. The
majority of our sources have double-peaked profiles at one or more of
the OH maser transitions, but distinct variations on this `classic'
profile have been recognised along with single-peaked and irregular
spectra. The six categories are listed and described below. Figure
\ref{Fi:examples} shows an example spectrum for each of the six
categories.

\begin{itemize}
\item {\bf D:} `Double' - Double-peaked profiles with peak-peak flux
density ratios $<$ 8.0. This is the ``classic'' profile type
associated with OH/IR stars on the AGB. In our sample of likely
post-AGB stars this is the most common profile type for both the 1612
and 1667 MHz transitions. Approximately 70\% of the OH 1612 MHz
spectra, and 50\% of OH 1667 MHz spectra, have D-type profiles. At
1665 MHz the incidence of D-type profiles is considerably lower
(19\%).

\item {\bf De:} `Double extreme' - Profiles with peak-peak flux
density ratios $>$ 8.0.  There is a large population of sources with
peak-to-peak ratios of 3 or less, and a small population with much larger
ratios.  A cutoff of 8.0 separates these two groups cleanly.  Four
sources (d200, b134, b292, and v223) have De profiles at 1612 MHz
while two sources (v67 and v132) have De profiles at 1667 MHz.

It appears likely that De maser profiles originate from OH shells that
are still largely spherical, rather than from bipolar outflows. The OH
spectral profiles at other transitions for the De sources are mostly
D-type indicating spherical maser shells. For v132, MERLIN images at
1612 and 1667 MHz show little evidence for an aspherical maser shell
\citep{Bai03}.

There are several possible causes for a De-type spectrum: An
asymmetric external UV field (responsible for dissociating H$_2$O to
OH) may lead to OH emission that is stronger on one side of the star.
Localised turbulence, possibly from the onset of bipolar outflows
could disrupt the velocity coherence needed for maser emission. In
sources with weak red-shifted emission, the maser emission from the
far side of the circumstellar envelope may be absorbed in an ionised
region near the star \citep{Zij01}.

\item {\bf Dw:} `Double with wings' - Double-peaked profiles with
sloping outer edges rather than the near-vertical outer edge of D-type
sources. Four sources in our sample, b70, b112, v41, and v274, have
this profile type for one or more lines. A well-known example of a Dw
source is HD101584 at 1667 MHz. Other examples are
IRAS 17253-2831 at 1612 MHz, IRAS 18491-0207 at 1667 MHz, and IRAS
22036+5306 at 1612, 1665, and 1667 MHz. Interferometric observations
of these sources reveal the masers are located in bipolar outflows,
with the blue- and red-shifted masers on opposite sides of the central
star \citep{Zij01}.  At other OH transitions these sources mostly have
irregular profiles.

\item {\bf DD:} `Double-double' - One source in our sample, d47, shows
four emission peaks at 1612 MHz, symmetrical in outflow velocity and
strength around the central stellar velocity. The OH/IR star
OH19.2-1.0, is known to have a similar OH profile \citep{Cha88} at
1612 MHz.  In her model for this star, based on interferometric
observations of the maser positions, Chapman proposed that the masers
are located on two rings that are the cross-sections of two bipolar
cones of emission expanding at terminal velocity.  This model predicts
the two outer peaks in the maser spectrum to be the strongest. However, for 
d47 the strongest emission occurs from the two inner peaks. A
plausible explanation for the stronger inner peaks, based on the
Chapman model, is that the masers are located in regions where the
outflows are accelerating.

The distinctiveness of this spectral profile and the
previously-substantiated interpretation from the literature, warrants
the DD category even though our sample contains only a single
example. It is possible that other sources in this sample,
particularly those with Dw or multiple-peaked irregular profiles, have
a DD profile too faint to detect.

\item {\bf S:} `Single' - Single-peaked profiles with one narrow emission 
peak with a steep edge on one side of the feature. Our sample includes 10
sources with an S-type profile at 1667 MHz. Of these nine sources have
a D- or De-type profile in the stronger 1612 MHz transition. This
indicates that most S-type sources are intrinsically double-peaked
with one peak below the detection level. A few sources (e.g. b11) have
strong single peaks indicating that some S profiles are intrinsically
De sources. There is no distinct grouping in infrared colour, or
preference for the red- or blue-shifted peak being absent.

\item {\bf I:} `Irregular' - These profiles exhibit a range of
features that include one wide peak, several peaks, or plateaus of
emission with peaks superimposed. The complex, multiple-peaked
spectra, such as d46, with large velocity widths almost certainly have
bipolar outflows and a remnant AGB wind still present. 
Several examples are discussed in \citet{Zij01}. The OH 1667
MHz emission from v231 is a good example of an irregular profile
showing elements of two emission peaks (indicating a remnant spherical
AGB shell), high-velocity emission (linear bipolar outflow), and
central-velocity emission plateau (inner accelerating outflow).

16/27 of the OH 1665 MHz detections show I-type profiles. For nine of
these sources, the 1665 MHz spectra have multiple peaks while the 1612
MHz spectra have an I, DD, or Dw classification at 1612 MHz,
consistent with aspherical morphologies. For four sources (d93,
d189, b258, and v45) the 1665 MHz emission shows a broad single
feature that is centred on the stellar velocity while the 1612 MHz
emission is double-peaked and wider than the 1665 MHz feature. For
these sources the circumstellar envelopes are likely to be spherically
symmetric with the 1665 MHz emission occuring in an inner accelerating
region where the logarithmic velocity gradient is $>$ 1
\citep{Cha85}. For the remaining three sources (b11, v67, and v237)
irregularities in the 1665-MHz spectra indicate some degree of
asymmetry in the 1665 MHz outflows although the 1612 MHz emission is
double-peaked.

\end{itemize}

In summary, we consider that the D-, De-, and S-type profiles indicate
spherical maser morphologies while the Dw-, DD-, and I-type profiles
indicate circumstellar shells with some asymmetries. Based
on these two groups, 18/86 sources show some degree of asymmetry at
1612 MHz. An additional three sources show irregularities at 1665 MHz
and one source (d168) appears irregular at 1667 MHz only. Taken
together, 22/86 (25 \%) sources in our sample show signs of
aspherical morphologies.

\subsection{OH 1667 MHz Overshoot} \label{Sec:vels}

All but eight sources have velocity ranges $<$ 50 km s$^{-1}$ 
(measured at a cut-off level of 3$\sigma$). The
largest velocity range of 78 km s$^{-1}$ is observed for v231 at 1667
MHz.  The average velocity ranges are 34.6 km s$^{-1}$ at 1612 MHz,
28.9 km s$^{-1}$ at 1665 MHz, and 34.8 km s$^{-1}$ at 1667 MHz.

For sources detected at both 1612 and 1667 MHz, 30 out of 35 have a
larger velocity range at 1667 MHz than at 1612 MHz (excluding
single-peaked sources).  For these sources the 1667 MHz velocity range is 
up to 18 km s$^{-1}$ larger than at 1612 MHz with an average difference of 1.5
km s$^{-1}$, corresponding to 16 spectral channels.

For {\em all} sources with double-peaked spectra at both 1612 and 1667
MHz (24) the velocity range at 1667 MHz is the same or larger than at
1612 MHz. The largest difference is 5 km s$^{-1}$ and the average is
2.3 km s$^{-1}$. A 1667 MHz overshoot is also evident in the peak
velocities. In most cases both the blue- and red-shifted peaks at 1667
MHz are outside the 1612 MHz peaks.  The largest peak shift is 7 km
s$^{-1}$ with an average of 0.78 km s$^{-1}$, corresponding to 9
spectral channels.

It is clear that 1667 MHz masers regularly overshoot 1612 MHz masers
in both spherical and irregular envelopes.  These results are very
similar to \citet{Siv99} in a study of the overshoot of mainline
spectral extent in circumstellar OH masers. Their sample included 21
AGB and post-AGB stars. Our results come from a much larger sample
size confirming that the phenomenon is real.

\citet{Siv99} suggested several possible mechanisms for this effect
including Zeeman broadening of the 1667 MHz line, clumping in the
stellar envelope, velocity fluctuations affecting 1612 MHz maser
coherence, or axisymmetric winds.  One possibility is that some
acceleration is still present at large radii where OH 1612 and 1667
MHz masers exist, and the distances at which the 1667 MHz masers are located
extends further than for 1612 MHz masers.  This is
the case for the supergiant star, IRC+10420 and HD 179821, which
may be a supergiant or a post-AGB star \citep{Bow84,Gle01}. Both
of these stars have spherical, although somewhat clumpy and
distorted, circumstellar OH shells that give De-type spectral
profiles.  In IRC+10420 the 1612 MHz masers are located between 3000
and 11000 AU but the 1667 MHz masers extend to 18000 AU.  There is a
velocity gradient of about 2 km s$^{-1}$ between the 1612 and 1667 MHz
shells which affects the velocity range of the spectral profiles: the
1667 MHz spectral width is larger than the 1612 MHz spectral width by
4-5 km s$^{-1}$.  In two other supergiants, VX Sgr and VY CMA,
velocity gradients exist to large radii \citep{Cha86,Rei78}.  However,
in these two stars the OH 1667 MHz masers are located inside the OH
1612 MHz shell. One other supergiant, NML Cyg, has inner circumstellar
regions that appear to be expanding {\em faster} than the outer
regions \citep{Dia84}.  In some Mira variables there is evidence for
velocity gradients to large radii in spherical circumstellar shells,
e.g. in U Her and R Cas \citep{Cha94}. Two OH/IR stars imaged
(OH127.8-0.0 and OH26.5+0.6) have small upper limits on possible
velocity gradients in their 1612 MHz shells \citep{Bow90}.

Imaging of v132 by \citet{Bai03} reveals that the D-type 1612 MHz
masers are contained in an ellipse with inner radius 250 mas and outer
radius 650 mas, and the De-type 1667 MHz masers trace a thinner
ellipse extending to a radius of 850 mas. However, the velocity range at 
each line is very similar and there is no evidence of a velocity
gradient at large radii.

It is apparent that there are a range of characteristics of the 1667
and 1612 MHz OH masers in late-type stars.  In explaining the 1667 MHz
overshoot phenomenon for sources with double-peaked spectra at both
1612 and 1667 MHz, a mechanism that simply broadens 1667 MHz emission
in the standard maser envelope model may be possible but it is likely
that the 1667 MHz masers are often located at radii beyond the 1612
MHz masers.

\subsection{IRAS and MSX Colours} \label{Sec:colours}

\begin{figure}
	\centering 
	\rotatebox{270}{\scalebox{0.3}{\includegraphics{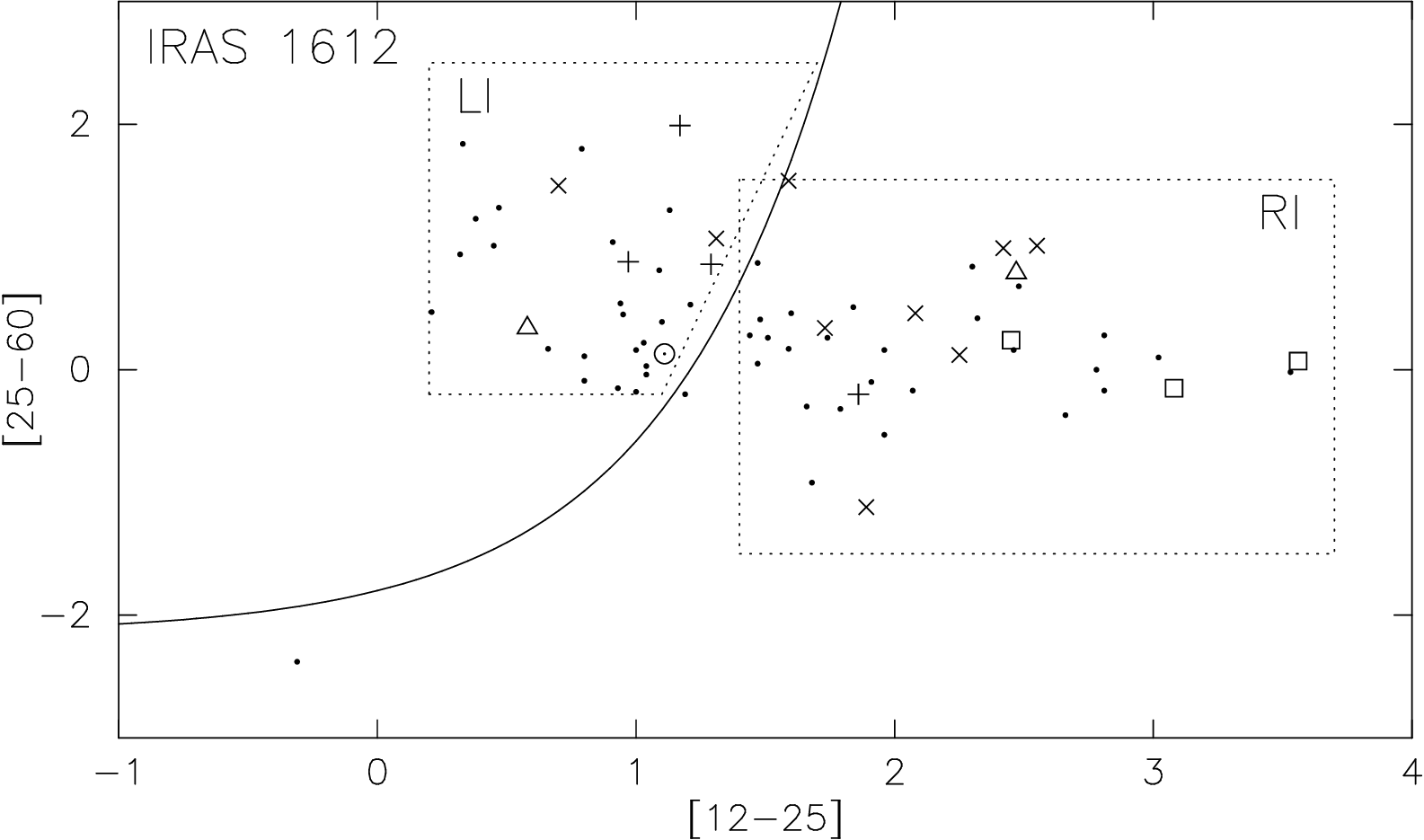}}}
	\rotatebox{270}{\scalebox{0.3}{\includegraphics{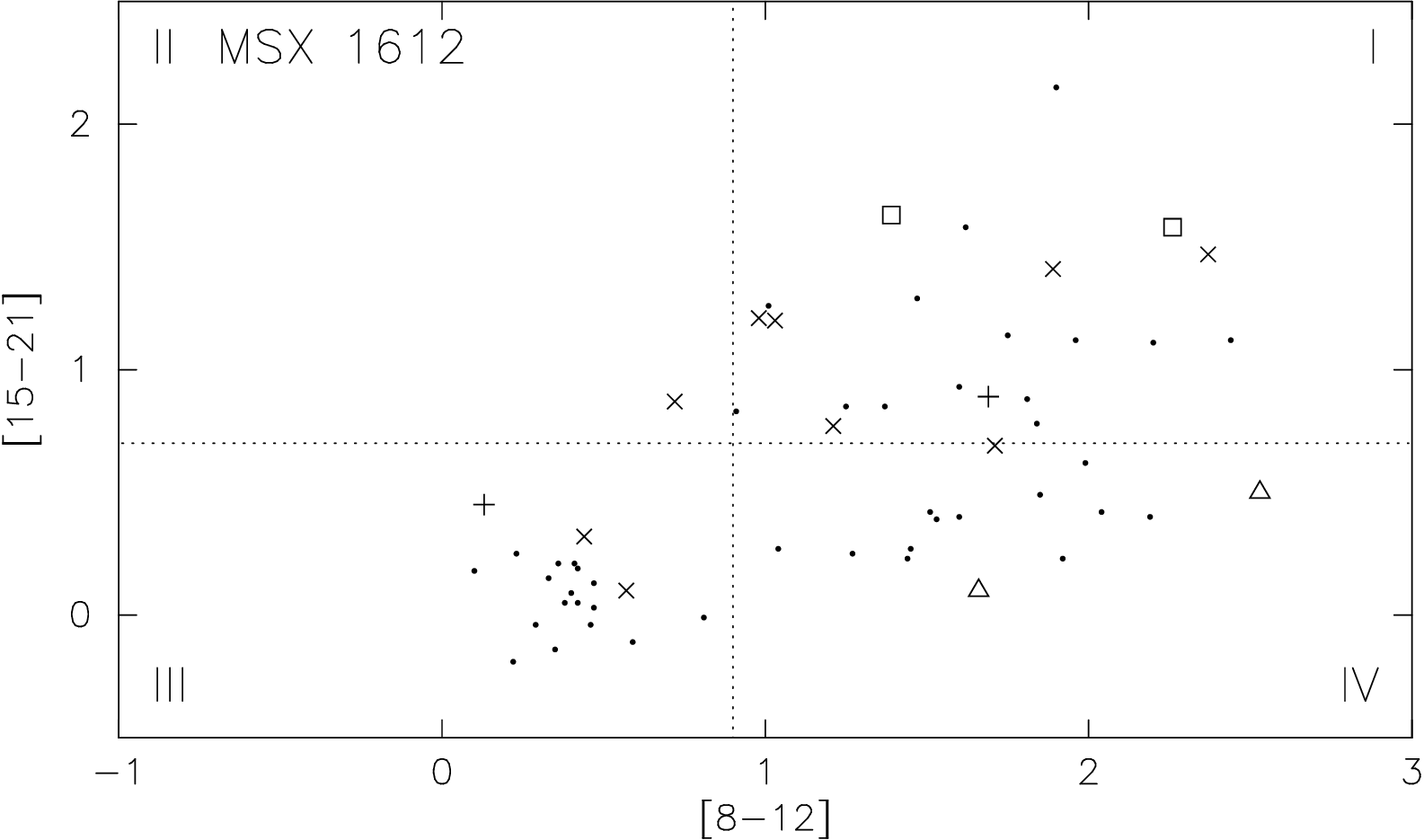}}} \\
	\vspace{25pt}
	\rotatebox{270}{\scalebox{0.3}{\includegraphics{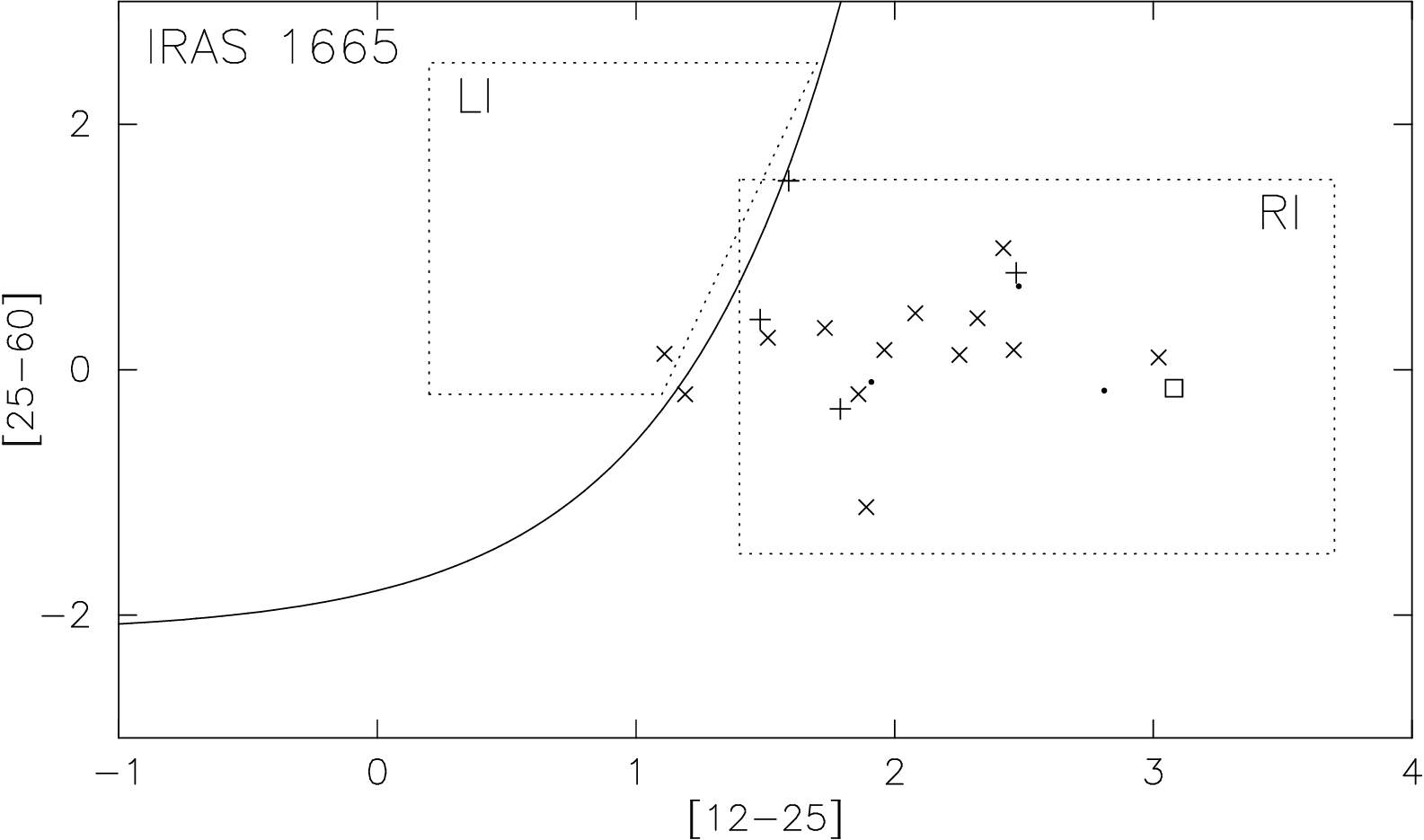}}} 
	\rotatebox{270}{\scalebox{0.3}{\includegraphics{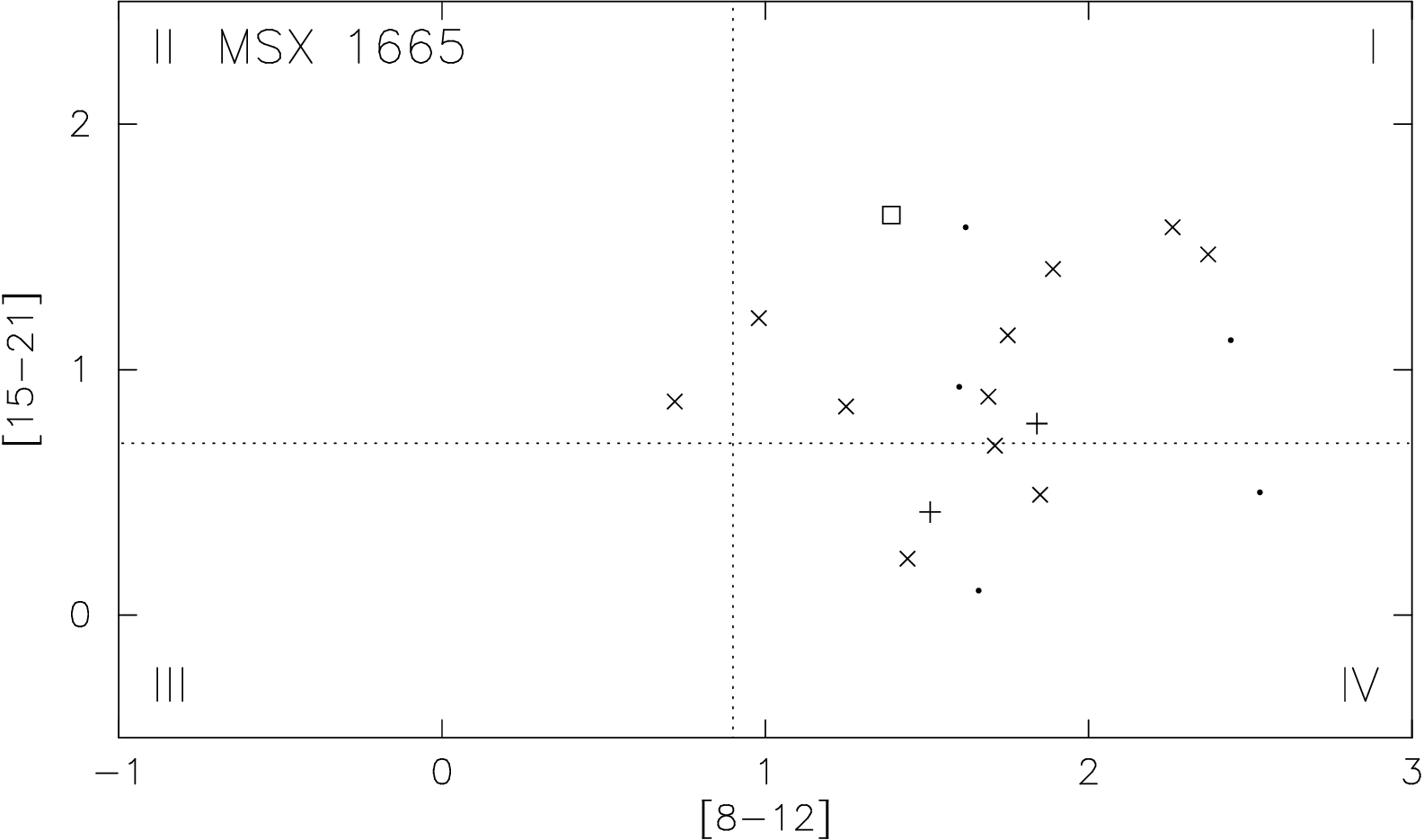}}} \\
	\vspace{25pt}
	\rotatebox{270}{\scalebox{0.3}{\includegraphics{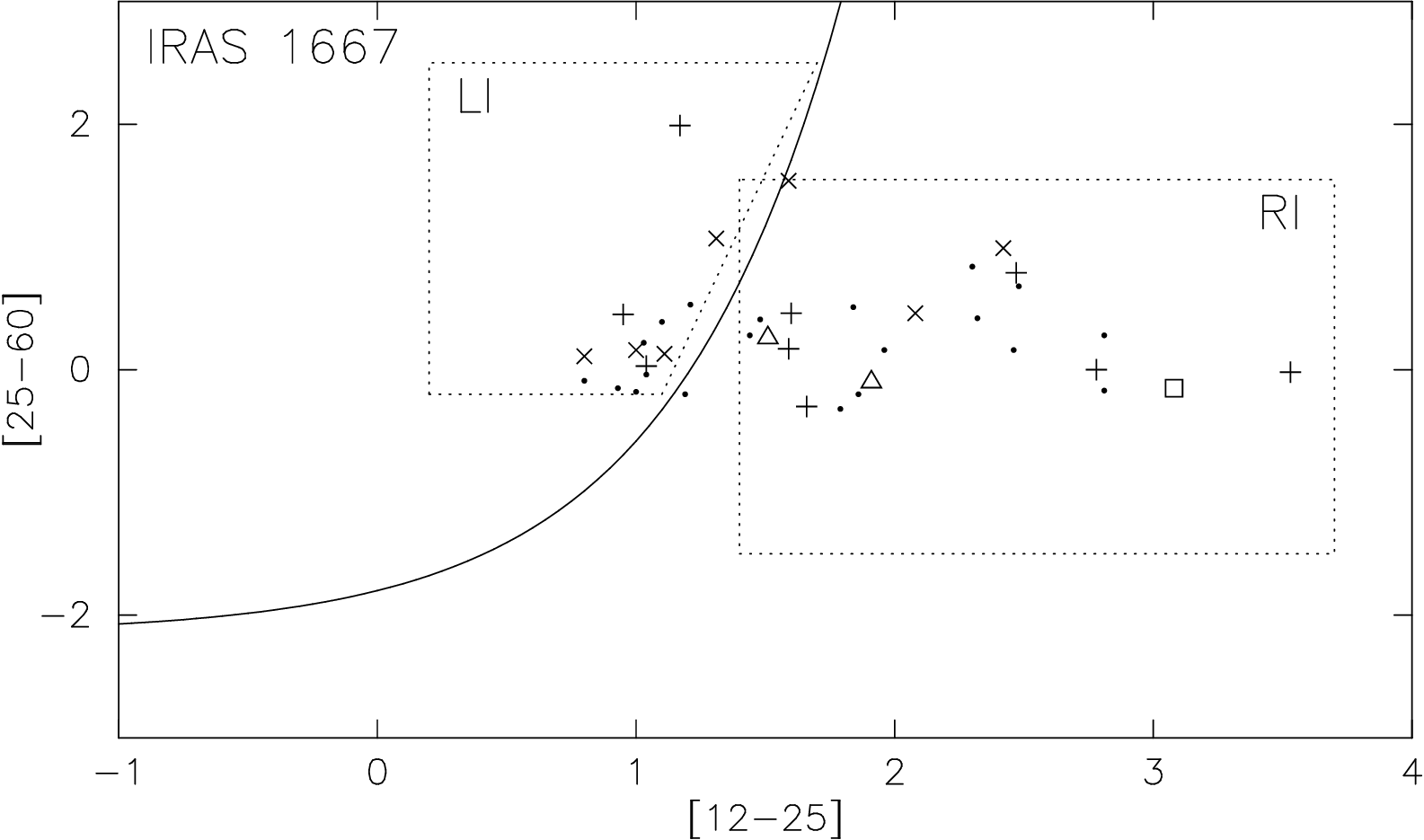}}}
	\rotatebox{270}{\scalebox{0.3}{\includegraphics{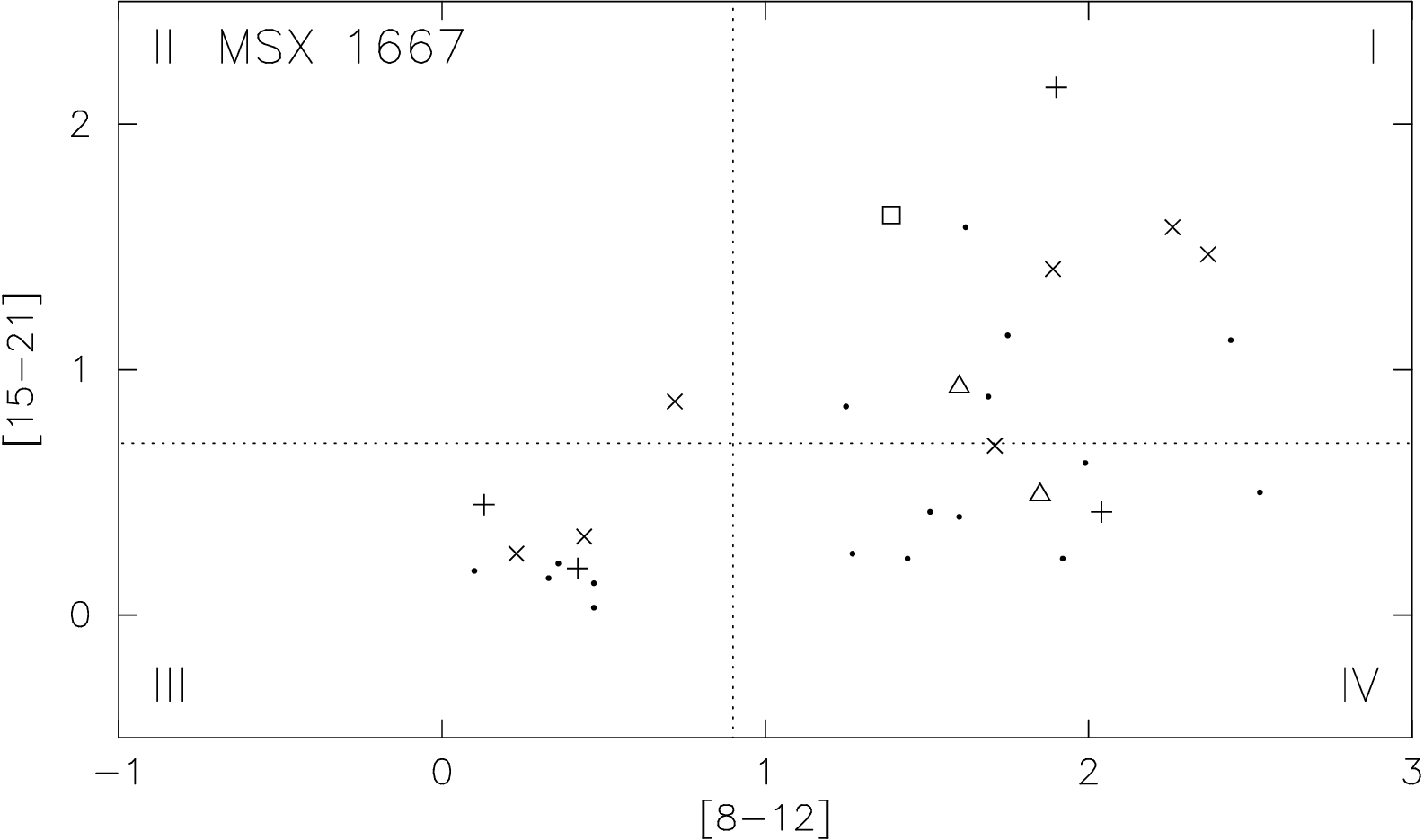}}}
	\caption{IRAS (left) and MSX (right) two-colour diagrams with
	sources with 1612, 1665 and 1667 MHz spectra 
	plotted.  Symbols are according to spectral profile classification: 
        D $\cdot$  Dw $\square $ De $\triangle$ DD $\odot $ I $\times$ and S + . \label{Fi:2colourdets}}
\end{figure}

\begin{figure}
\begin{center}
\rotatebox{270}{\scalebox{0.33}{\includegraphics{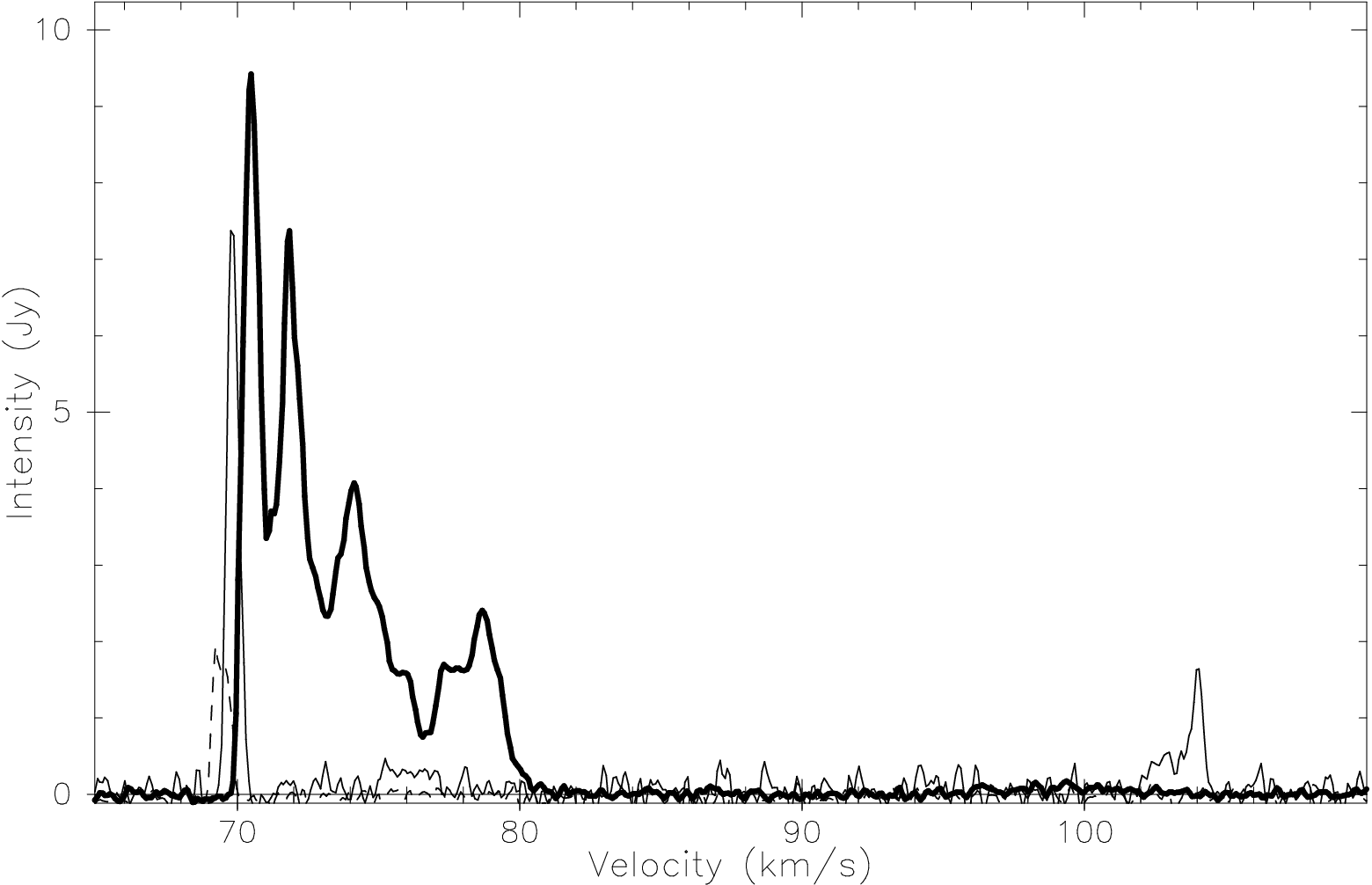}}}
\\
\vspace{12pt}
\rotatebox{270}{\scalebox{0.7}{\includegraphics{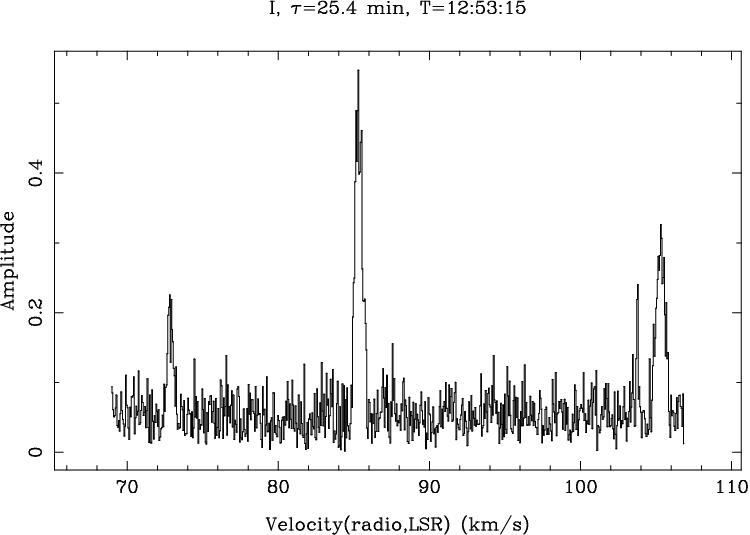}}}
\end{center}
\caption{(top) Overlay of Parkes 1612 (bold), 1665 (thin) and 1720 MHz
(dashed) spectra for b292.  The 1612 MHz spectrum is scaled by 0.5.
(bottom) ATCA water maser spectrum of b292. These spectra show the
complex velocity structure of the different maser types around b292.
\label{Fi:b292_olay}}
\end{figure}

\begin{figure}
\begin{center}
\scalebox{0.7}{\includegraphics{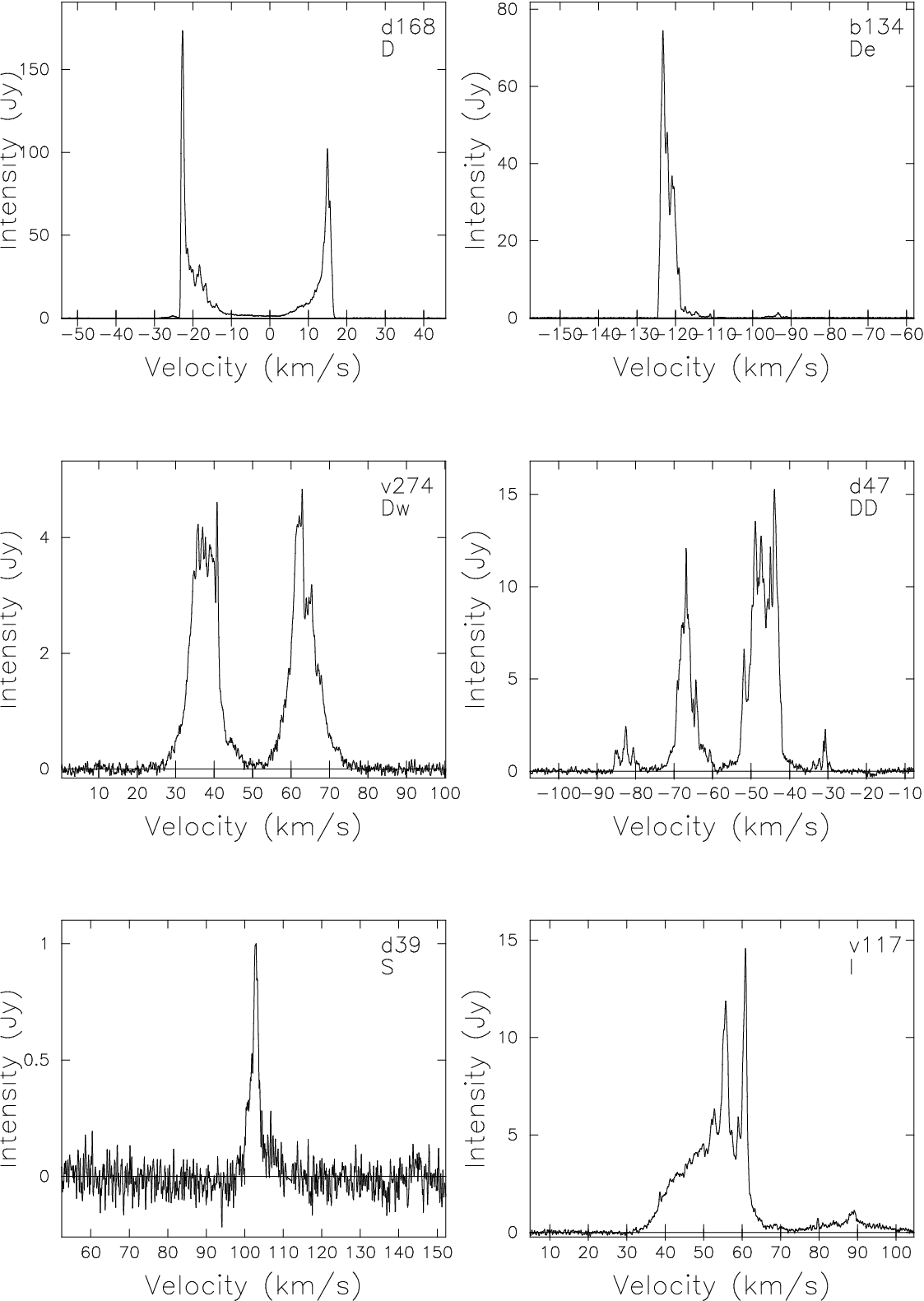}}
\end{center}
\caption{Examples of spectra from each classification category.  The
catalogue name of each star is given in the top right-hand corner with
the category type given below the name. \label{Fi:examples}}
\end{figure}

\begin{figure}
	\centering \subfigure[]{
	\scalebox{0.8}{\includegraphics{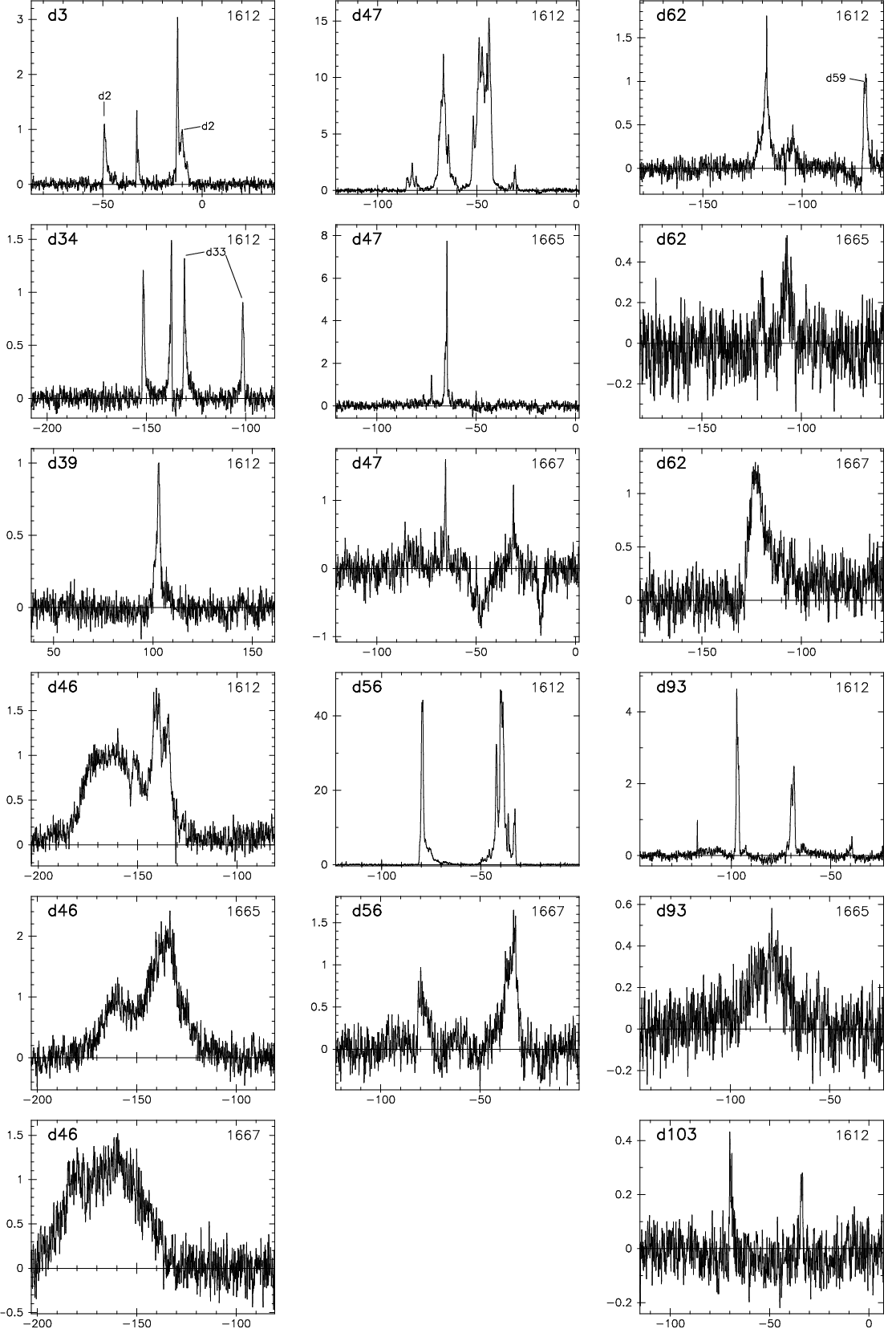}}}
	\caption{1612, 1665 and 1667 MHz spectra with detections 
	from Parkes telescope observations.
	Source names and frequencies are noted on individual
	spectra. \label{Fi:thumbnails:a}}
\end{figure}

Table 5 summarises the number of detections for each OH maser
transition for the four groups used in the infrared selection criteria
discussed in Section 2, namely the IRAS LI and RI groups, the MSX Quad IV
(early post-AGB) and Quad I (late post-AGB) groups. As previously
described there is an overlap between these groups. The IRAS LI
sources correspond to MSX Quad III while the RI sources correspond to
MSX Quads I and IV. IRAS and MSX colour-colour plots of sources
detected at 1612, 1665, and 1667 MHz are plotted in Figure
\ref{Fi:2colourdets}, combining graphically the results in Tables
\ref{Ta:dets_by_profile} and \ref{Ta:all_line-stats}.

Figure \ref{Fi:2colourdets} shows that for both LI and RI regions and both 
MSX Quads 1 \&
IV, the majority of sources have double-peaked (D-type) profiles, at
1612 MHz.  In detail, 13 of the 15 early post-AGB stars (Quad IV) have D-type
spectra while the other two sources have De-type spectra. We therefore
find no evidence for strongly disturbed outflows in the early post-AGB
region. In contrast, for the late post-AGB stars (Quad I), eight out
of 22 sources have I- or Dw-type profiles at 1612 MHz. Based on this
sample we therefore estimate that approximately one-third of post-AGB
stars develop aspherical morphologies during the late post-AGB stage.

The location on the colour-colour plots of the I- and Dw-type sources
indicates a progression with the maser profiles of some sources
evolving from a spherical outflow with a D-type spectrum, to a
disturbed outflow with an I-type spectrum, and finally in some cases
to a fully bipolar outflow and Dw-type spectrum. Both the
IRAS and MSX diagrams show that the Dw sources have the most extreme
far-infrared colours with [12-25] $>$ 2.3 and [15-21] $>$ 1.5. This
progression can be seen at both 1612 and 1667 MHz as the 1667 MHz
spectral characteristics largely mirror those at 1612 MHz.  The
small number of sources with De-type profiles may come from an
intermediate stage in some cases between the D and I profile types that occurs when 
a spherical outflow is first disturbed by a jet outflow as proposed by \citet{Sah98}. 

The OH 1665 MHz emission is the weakest of the three transitions and
is detected most frequently from sources with irregular outflows. Of
the 60 sources with double-peaked profiles at 1612 MHz, only 12 were
detected at 1665 MHz whereas for 13 sources with irregular 1612 MHz
spectra, eight were detected at 1665 MHz.  In several late-type 
stars it has been found that 1665 MHz masers are located closer to the 
central star than 1612 and/or 1667 MHz masers.  For example, in images of VX Sgr 
\citep{Cha86}, Roberts 22 and IRAS 18491-0207 \citep{Zij01}, and U Ori 
\citep{Cha91}.  The irregular profiles could indicate the beginnings of 
bipolar outflows disrupting the inner masers, and/or it could reflect the 
clumpy nature of the maser distribution.  Possibly for the 1665 
MHz profiles showing a broad hump centred on the stellar velocity, 
they are from an inner accelerating outflow (see Section \ref{Sec:spec}).
All this implies that the 1665 MHz transition is likely to be a good 
tracer of the inner envelope, where early departures from
spherical symmetry are seen. 1665 or 1667 MHz emission
can dominate in sources with highly irregular outflows. Four out of
the five objects with strongest emission at 1665 or 1667 MHz (d46,
b11, b262, and b263) are late post-AGB objects with irregular profiles
at one or more OH maser lines. (v189, an LI object with S-type
spectra, is the exception.)

Our sample includes a total of 30 LI sources. As discussed by
\citet{Sev02II} this group of sources are likely to have higher
initial stellar masses and AGB mass-loss rates than the RI
sources. The average outflow velocity (half the peak-to-peak velocity width) of
the LI sources is 18.25 km s$^{-1}$, compared to 13.00 km s$^{-1}$
for RI sources.  This is close to the outflow velocities from stars with 
initial masses of 1.6 and $\sim$4 M$_{\odot}$ as calculated
by \citet{Vee89} and quoted by \citet{Sev02II}, and supports the theory
that the LI sources are more massive stars than those in the RI category.  
The post-AGB evolutionary track \citep{Hoo97} highlighted by
\citet{Sev02II} for LI stars passes
through this section of the IRAS [12-25]/[25-60] diagram before
looping up to higher [25-60] colour. The LI region overlaps with the
AGB and may include some stars that are still in the AGB evolutionary
phase. \citet{Sev02II} has proposed that the LI stars are likely
precursors to bipolar planetary nebulae although this is not
confirmed.  Of the 30 sources at 1612 MHz, 26 have double- or weak
single-peaked profiles indicating that their circumstellar envelopes
are still fairly homogeneous. Two sources (d62 and b165) have
irregular profiles, one (d200) is a De source and one (d47) is a DD
source with four emission peaks. 14 LI sources were detected at 1667
MHz. These are located close to the AGB evolutionary track. A
remarkable observational result is that only one LI source was
detected at 1665 MHz while the `extreme' LI region away from the
evolutionary track is devoid of OH mainline maser emission. For the
small region defined by the IRAS colours [12-25] $>$ 0.75 and [25-60]
$<$ 0.7 there is a 100\% detection rate for 1667 MHz maser
emission. This group of sources are likely to be the youngest LI
objects and may indeed still be AGB stars as they are located near the AGB
evolutionary track at the point where the RI and LI sources diverge.

\subsection{OH 1612 MHz Variability} \label{Sec:var}

As an indicator of OH 1612 MHz variability we have compared the peak
flux densities from our Parkes data, taken in 2002--2003 with a
velocity resolution of 0.2 km s$^{-1}$, with the ATCA/VLA survey peak
flux densities, obtained in 1993--1995 with much lower velocity
resolution of 1.5 and 2.3 kms $^{-1}$ \citep{Sev97I,Sev97II,Sev01}. To 
compare the two sets of data obtained with different
velocity resolution, the Parkes data were smoothed using a Gaussian
smoothing function to approximately the same velocity resolution as
the ATCA/VLA data. After smoothing, the average peak flux densities of
the Parkes spectra were a factor of 1.3 times greater than for the
ATCA/VLA spectra. The reason for this discrepancy is unclear but we
assume that this is a systematic bias and not an effect intrinsic to
the sources.

Overall, for 26 of the 86 sources in our sample (30\%), the OH 1612
MHz peak flux densities measured in 1993--1995 differ by at least 30\%
from those measured in 2002--2003 (after smoothing and removing the
bias). These likely variable sources are indicated in Table 3 by a `V'
in the last column. For 16 of the 26 sources the flux densities differ
by at least 50\%. In all cases the profile shapes are unchanged. We
find no obvious correlations between variability and the profile type
or infrared colours.

OH 1612 MHz variability is well understood for AGB stars but is poorly
understood for post-AGB stars. For AGB stars, the OH 1612 MHz maser
emission is strongly saturated and radiatively coupled to the stellar
pulsation cycle. The peak OH 1612 MHz maser emission typically varies
in intensity, over the stellar light cycle, by a factor of
two. Monitoring studies have shown OH light curves with periods from
several hundred to several thousand days \citep[e.g.][]{Her85,Lan90,Cha95}.

It is often assumed that post-AGB stars are non-variable at 1612 MHz
since for post-AGB stars the stellar pulsations are expected to have
ceased. The comparison of the Parkes data with the earlier ATCA/VLA
data shows that while the 1612 MHz variability {\em is} much weaker
during the post-AGB phase, some variability is present for at least
one-third of the sources in our sample.  Monitoring data are needed to
determine variability timescales and whether the 1612 MHz variability
during the post-AGB phase is associated with (weak) stellar pulsations
and/or with other mechanisms such as structural changes in the inner
envelopes.

OH mainline variability will be discussed further in a later paper.
Here we note that in at least some sources, strong 1665 MHz
variability is present. As an example, for d47 the Parkes data taken (by
us) in 2002 September and 2003 August show that the peak 1665 MHz flux
density increased from 0.94 Jy to 7.74 Jy.

\section{Conclusions} 

The results of an extensive multi-transition OH maser survey of 86 post-AGB stars have
been presented.  Selected from an OH 1612 MHz survey of the Galactic
Plane, most sources are also strongest at 1612 MHz and have
double-peaked profiles.  All 86 sources were re-detected at 1612 MHz,
27 were detected at 1665 MHz and 45 at 1667 MHz.  One source
previously detected at 1720 MHz (b292) was re-detected.

A classification scheme is used to distinguish six profile types
found.  22 sources in the sample have spectral profiles that indicate
aspherical and/or bipolar outflows.  There are signs of an evolutionary
trend in the OH maser profiles of the lower-mass stars (56) in the
sample.  A fraction of sources with double-peaked profiles will evolve
to irregular profiles (associated with bipolar outflows and remnant
spherical shells), and then to winged double-peaked profiles
indicative of fully-bipolar outflows.

A significantly higher percentage of aspherical post-AGB sources have
1665 MHz emission than spherical sources.  1665 MHz emission may be a
good tracer of initial departures from spherical morphology in inner
circumstellar envelopes. Highly irregular, late post-AGB sources also tend
to be dominant at one of the mainlines rather than at 1612 MHz.

The more massive and bluer post-AGB stars, known as `LI' sources, are
thought to evolve into bipolar PN.  This distinctive group are
noteworthy in our sample in having mostly double-peaked profiles. They
also show an almost complete lack of mainline emission, except in the
youngest sources.  The outflow velocities of the LI sources are on
average 10 km s$^{-1}$ larger than those of the redder `RI' objects,
supporting their probable higher initial mass status.

At least 30\% of sources show variability at 1612 MHz, showing that
some maser variability is still present in many post-AGB stars.  The
degree of variability is significantly smaller than in AGB stars.

Most sources with 1612 and 1667 MHz emission have broader emission at
1667 MHz.  This overshoot phenomenon has been previously noted by
\citet{Siv99}.  The origin of this effect is unclear but it appears likely
that in many cases the 1667 MHz masers are found out to distances beyond the
1612 MHz masers in regions where small velocity gradients are still
present in the outer radii of circumstellar envelopes.

\begin{center}
{\bf ACKNOWLEDGEMENTS} 
\end{center}

We wish to thank Jim Caswell for many helpful discussions, and Aidan
Hotan for observing assistance at Parkes in August 2003.

\begin{deluxetable}{c c c c r r r r r r r r l l}
\tabletypesize{\scriptsize} \rotate \tablewidth{0pt}
\tablecaption{Source data.  \label{Ta:source}} 
\tablehead{ \colhead{} & \colhead{} & \colhead{} & \colhead{} & \colhead{} &
\multicolumn{3}{l}{IRAS Fluxes}  & \multicolumn{4}{l}{MSX Fluxes} & \colhead{} \\ 
\colhead{Identifier} & \colhead{Iras Name} & \colhead{RA} & \colhead{Dec} & \colhead{$V_{lsr}$} &
\colhead{$S_{12}$} & \colhead{$S_{25}$} & \colhead{$S_{60}$} &
\colhead{$S_8$} & \colhead{$S_{12}$} & \colhead{$S_{15}$} &
\colhead{$S_{21}$} & \multicolumn{2}{c}{Classification} \\ 
\colhead{} & \colhead{} & \colhead{(J2000)} & \colhead{(J2000)} & \colhead{(km/s)}
& \colhead{(Jy)} & \colhead{(Jy)} & \colhead{(Jy)} & \colhead{(Jy)} &
\colhead{(Jy)} & \colhead{(Jy)} & \colhead{(Jy)} & \colhead{(MSX)} & \colhead{(IRAS)}  }
\startdata
d3   & $14341-6211$ & 14:38:04.961 & $-$62:24:46.83 & $ -22.7$ &    2.82 &   13.21 &    5.68 &   0.84 &   2.96 &   5.22 &  11.42 & Quad I  & RI      \\
d29  & $15254-5621$ & 15:29:19.347 & $-$56:31:22.33 & $ -70.8$ &   84.08 &  522.40 & 3011.00 &  35.01 & 105.70 & 157.74 & 424.89 & Quad I  & RI      \\
d34  & $15338-5202$ & 15:37:31.772 & $-$52:12:13.15 & $-144.6$ &  \nodata  &  \nodata  &  \nodata  &   0.80 &   2.02 &   3.41 &  10.90 & Quad I  & \nodata \\
d39  & $15367-5420$ & 15:40:38.278 & $-$54:30:18.59 & $-102.3$ &    1.85 &    6.05 &   13.39 &  \nodata  &  \nodata  &  \nodata  &  \nodata  & \nodata & LI      \\
d46  & $15445-5449$ & 15:48:19.498 & $-$54:58:21.52 & $-140.2$ &  \nodata  &  \nodata  &  \nodata  &   0.57 &   5.08 &  13.65 &  52.81 & Quad I  & \nodata \\
d47  & $15452-5459$ & 15:49:11.398 & $-$55:08:51.77 & $ -57.8$ &   87.05 &  242.70 &  273.60 &  \nodata  &  \nodata  &  \nodata  &  \nodata  & \nodata & LI      \\
d56  & $15514-5323$ & 15:55:20.690 & $-$53:32:45.07 & $ -59.8$ &   32.06 &   87.93 &  126.20 &  26.61 &  41.05 &  72.38 &  74.20 & \nodata & LI      \\
d62  & $15544-5332$ & 15:58:18.830 & $-$53:40:40.20 & $-118.3$ &    4.64 &   15.54 &   41.55 &   2.85 &   4.29 &   6.42 &   8.59 & \nodata & LI      \\
d93  & $16209-4714$ & 16:24:33.912 & $-$47:21:29.56 & $ -82.6$ &    1.17 &   18.87 &   20.68 &  \nodata  &  \nodata  &  \nodata  &  \nodata  & \nodata & RI      \\
d103 & $16314-5018$ & 16:35:14.944 & $-$50:24:15.19 & $ -51.9$ &   27.74 &   20.91 &    2.33 &   4.91 &  12.82 &  10.10 &  12.92 & Quad IV & \nodata \\
d117 & $16372-4808$ & 16:40:55.829 & $-$48:13:58.34 & $ -90.6$ &  \nodata  &  \nodata  &  \nodata  &   0.77 &   1.99 &   3.66 &  11.10 & Quad I  & \nodata \\
d150 & $16507-4810$ & 16:54:30.953 & $-$48:15:21.24 & $ -96.4$ &    1.08 &   12.56 &    8.96 &  \nodata  &  \nodata  &  \nodata  &  \nodata  & \nodata & RI      \\
d168 & $17004-4119$ & 17:03:56.206 & $-$41:23:59.25 & $  -4.5$ &  128.10 &  322.60 &  373.10 & 118.24 & 146.74 & 254.77 & 322.18 & \nodata & LI      \\
d189 & $17088-4221$ & 17:12:22.806 & $-$42:25:09.82 & $   2.8$ &   42.70 &  128.30 &  106.80 &  11.91 &  44.88 &  89.24 & 110.41 & Quad IV & \nodata \\
b5   & $17097-3624$ & 17:13:04.994 & $-$36:27:54.34 & $ -33.0$ &    1.52 &    6.61 &   10.11 &  \nodata  &  \nodata  &  \nodata  &  \nodata  & \nodata & RI      \\
d190 & $17103-3702$ & 17:13:44.339 & $-$37:06:10.95 & $ -41.0$ &   32.08 &  335.90 &  849.70 &  \nodata  &  \nodata  &  \nodata  &  \nodata  & \nodata & RI      \\
b11  & $17150-3224$ & 17:18:19.889 & $-$32:27:21.96 & $  25.3$ &   57.92 &  322.30 &  268.30 &  13.07 &  62.04 & 130.90 & 296.48 & Quad I  & RI      \\
d197 & $17151-3845$ & 17:18:34.776 & $-$38:48:57.37 & $-178.2$ &  \nodata  &  \nodata  &  \nodata  &   0.23 &   1.25 &   3.05 &   6.27 & Quad I  & \nodata \\
b14  & $17164-3226$ & 17:19:40.776 & $-$32:29:51.69 & $  87.4$ &    0.96 &    9.45 &   17.74 &   0.19 &   1.79 &   2.25 &   6.33 & Quad I  & RI      \\
b15  & $17162-3751$ & 17:19:42.067 & $-$37:54:55.16 & $ -79.0$ &  \nodata  &  \nodata  &  \nodata  &   0.51 &   1.56 &   1.88 &   3.82 & Quad I  & \nodata \\
b17  & $17168-3736$ & 17:20:15.047 & $-$37:39:34.31 & $  -6.0$ &    9.78 &   36.98 &   47.89 &   1.72 &  10.06 &  23.32 &  28.77 & Quad IV & RI      \\
d200 & $17188-3838$ & 17:22:18.780 & $-$38:41:40.13 & $ -92.1$ &    9.31 &   15.92 &   21.70 &  \nodata  &  \nodata  &  \nodata  &  \nodata  & \nodata & LI      \\
b25  & $17193-3546$ & 17:22:42.634 & $-$35:49:31.60 & $  17.4$ &    3.09 &    6.38 &   33.58 &   0.87 &   1.84 &   3.14 &   3.10 & \nodata & LI      \\
b30  & $17205-3556$ & 17:23:57.403 & $-$35:58:50.31 & $  70.7$ &    2.36 &    3.21 &   17.44 &   3.36 &   5.15 &   5.87 &   5.68 & \nodata & LI      \\
b31  & $17207-3632$ & 17:24:07.277 & $-$36:35:40.45 & $-143.9$ &    6.23 &   14.94 &   22.68 &  \nodata  &  \nodata  &  \nodata  &  \nodata  & \nodata & LI      \\
b33  & $17227-3623$ & 17:26:07.344 & $-$36:26:15.08 & $ -24.2$ &  \nodata  &  \nodata  &  \nodata  &   0.20 &   1.51 &   2.00 &   2.89 & Quad IV & \nodata \\
b34  & $17230-3348$ & 17:26:18.783 & $-$33:51:20.81 & $-208.2$ &    2.60 &    7.38 &   24.55 &   0.61 &   2.31 &   4.36 &   5.61 & Quad IV & LI      \\
d202 & $17245-3951$ & 17:28:04.637 & $-$39:53:44.20 & $ -97.9$ &    3.36 &   44.73 &   38.23 &   0.66 &   2.94 &   8.52 &  36.61 & Quad I  & RI      \\
b44  & $17256-3258$ & 17:28:55.751 & $-$33:00:41.86 & $  -2.3$ &    1.76 &    4.30 &    9.66 &   0.58 &  -1.35 &   1.89 &  -2.60 & \nodata & LI      \\
b62  & $17293-3302$ & 17:32:39.109 & $-$33:04:15.89 & $  18.9$ &    4.12 &   17.80 &   73.83 &  \nodata  &  \nodata  &  \nodata  &  \nodata  & \nodata & RI      \\
b68  & $17310-3432$ & 17:34:20.788 & $-$34:34:55.41 & $-286.3$ &    0.84 &   10.85 &   10.87 &  \nodata  &  \nodata  &  \nodata  &  \nodata  & \nodata & RI      \\
b70  & $17317-2743$ & 17:34:53.287 & $-$27:45:11.18 & $  47.4$ &    1.04 &   27.65 &   29.54 &  \nodata  &  \nodata  &  \nodata  &  \nodata  & \nodata & RI      \\
b96  & $17359-2902$ & 17:39:07.703 & $-$29:04:02.60 & $-136.6$ &    2.04 &   12.38 &    7.61 &   0.43 &   1.67 &   3.13 &  10.24 & Quad I  & RI      \\
b106 & $17367-3134$ & 17:39:57.498 & $-$31:35:57.28 & $  70.7$ &  \nodata  &  \nodata  &  \nodata  &   0.23 &   1.74 &   1.29 &   3.60 & Quad I  & \nodata \\
b112 & $17370-3357$ & 17:40:20.199 & $-$33:59:13.43 & $  25.0$ &    0.78 &    7.48 &    9.32 &  \nodata  &  \nodata  &  \nodata  &  \nodata  & \nodata & RI      \\
b114 & $17371-2747$ & 17:40:23.069 & $-$27:49:11.42 & $ 115.2$ &    2.14 &   10.56 &   14.47 &  \nodata  &  \nodata  &  \nodata  &  \nodata  & \nodata & RI      \\
b128 & $17385-3332$ & 17:41:52.278 & $-$33:33:40.30 & $-234.9$ &    2.88 &   13.25 &   10.01 &  \nodata  &  \nodata  &  \nodata  &  \nodata  & \nodata & RI      \\
b130 & $17390-2809$ & 17:42:15.569 & $-$28:10:36.37 & $   0.6$ &  \nodata  &  \nodata  &  \nodata  &   0.50 &   2.05 &   3.42 &   4.88 & Quad IV & \nodata \\
b133 & $17392-3020$ & 17:42:30.517 & $-$30:22:07.96 & $  -3.8$ &    6.32 &   17.26 &   36.37 &  \nodata  &  \nodata  &  \nodata  &  \nodata  & \nodata & LI      \\
b134 & $17393-2727$ & 17:42:33.161 & $-$27:28:24.62 & $-108.2$ &    1.83 &   17.83 &   36.85 &  \nodata  &  \nodata  &  \nodata  &  \nodata  & \nodata & RI      \\
b143 & $17404-2713$ & 17:43:38.042 & $-$27:14:44.30 & $  43.7$ &    3.99 &   20.74 &   15.49 &  \nodata  &  \nodata  &  \nodata  &  \nodata  & \nodata & RI      \\
b155 & $17414-3108$ & 17:44:39.771 & $-$31:10:05.40 & $  46.6$ &    2.74 &    6.51 &   10.67 &   1.20 &   1.75 &   3.47 &   4.21 & \nodata & LI      \\
b165 & $17426-2804$ & 17:45:46.658 & $-$28:05:28.46 & $ -26.4$ &    5.30 &   10.09 &   40.07 &   3.04 &   3.62 &   4.87 &   3.59 & \nodata & LI      \\
b199 & $17461-2741$ & 17:49:20.912 & $-$27:41:54.11 & $-127.9$ &  \nodata  &  \nodata  &  \nodata  &   0.45 &   1.97 &   3.91 &   5.67 & Quad IV & \nodata \\
b209 & $17479-3032$ & 17:51:12.148 & $-$30:33:39.79 & $ -15.9$ &    2.62 &   13.02 &   16.51 &  \nodata  &  \nodata  &  \nodata  &  \nodata  & \nodata & RI      \\
b210 & $17482-2501$ & 17:51:22.846 & $-$25:01:52.72 & $  72.9$ &    1.34 &    5.18 &   11.53 &  \nodata  &  \nodata  &  \nodata  &  \nodata  & \nodata & RI      \\
b228 & $17506-2955$ & 17:53:50.715 & $-$29:55:28.29 & $ 182.0$ &    1.62 &    6.29 &    6.61 &  \nodata  &  \nodata  &  \nodata  &  \nodata  & \nodata & RI      \\
b246 & $17543-3102$ & 17:57:33.587 & $-$31:03:02.08 & $ 121.4$ &    2.76 &   21.87 &   24.44 &  \nodata  &  \nodata  &  \nodata  &  \nodata  & \nodata & RI      \\
b250 & $17548-2753$ & 17:57:58.251 & $-$27:53:19.89 & $-185.2$ &    1.27 &   16.95 &   21.95 &  \nodata  &  \nodata  &  \nodata  &  \nodata  & \nodata & RI      \\
b251 & $17550-2120$ & 17:58:04.953 & $-$21:21:07.07 & $ -21.5$ &    5.38 &   21.08 &   30.81 &   1.20 &   4.84 &  11.41 &  16.85 & Quad IV & RI      \\
b258 & $17560-2027$ & 17:59:05.007 & $-$20:27:23.95 & $ 204.3$ &    1.59 &   15.32 &   17.79 &  \nodata  &  \nodata  &  \nodata  &  \nodata  & \nodata & RI      \\
b262 & $17574-2403$ & 18:00:30.391 & $-$24:04:01.29 & $  -7.9$ &  \nodata  &  \nodata  &  \nodata  &  26.99 & 153.50 & 318.37 & 162.80 & Quad I  & \nodata \\
b263 & $17576-2653$ & 18:00:49.500 & $-$26:53:12.52 & $ 138.9$ &    2.82 &   16.15 &    5.75 &   0.97 &   2.39 &   4.53 &  13.87 & Quad I  & RI      \\
b266 & $17582-2619$ & 18:01:21.546 & $-$26:19:36.75 & $ 174.7$ &    1.38 &    9.28 &    7.92 &   0.34 &   1.80 &   3.31 &   7.42 & Quad I  & RI      \\
b292 & $18043-2116$ & 18:07:20.861 & $-$21:16:10.86 & $  87.3$ &  \nodata  &  \nodata  &  \nodata  &   0.43 &   1.98 &   4.63 &   5.07 & Quad IV & \nodata \\
b300 & $18051-2415$ & 18:08:12.819 & $-$24:14:36.73 & $ 120.3$ &    1.88 &    8.14 &    9.49 &   0.27 &   1.76 &   3.93 &   5.77 & Quad IV & RI      \\
b301 & $18052-2016$ & 18:08:16.376 & $-$20:16:11.51 & $  51.7$ &   21.42 &   32.45 &   82.36 &  16.49 &  28.39 &  29.92 &  27.16 & \nodata & LI      \\
b304 & $18070-2332$ & 18:10:04.804 & $-$23:32:09.83 & $ -63.5$ &  \nodata  &  \nodata  &  \nodata  &   1.25 &   4.03 &   8.38 &  10.53 & Quad IV & \nodata \\
v41  & $18076-1853$ & 18:10:38.665 & $-$18:52:58.08 & $  42.0$ &  \nodata  &  \nodata  &  \nodata  &   1.35 &  10.80 &  33.05 & 141.82 & Quad I  & \nodata \\
v45  & $18087-1440$ & 18:11:34.086 & $-$14:39:54.25 & $  12.4$ &    2.58 &   21.86 &   32.31 &   0.32 &   1.61 &   5.55 &  15.83 & Quad I  & RI      \\
v50  & $18092-2347$ & 18:12:20.411 & $-$23:46:56.95 & $  48.8$ &   10.95 &   25.81 &   22.44 &  \nodata  &  \nodata  &  \nodata  &  \nodata  & \nodata & LI      \\
v53  & $18100-1915$ & 18:13:03.099 & $-$19:14:18.66 & $  18.1$ &    8.70 &   18.25 &   20.12 &  \nodata  &  \nodata  &  \nodata  &  \nodata  & \nodata & LI      \\
v56  & $18103-1738$ & 18:13:20.240 & $-$17:37:17.35 & $  17.0$ &    5.34 &    7.16 &   17.01 &   3.10 &   4.04 &   5.19 &   5.02 & \nodata & LI      \\
v67  & $18135-1456$ & 18:16:25.688 & $-$14:55:17.16 & $  -1.1$ &   31.02 &  124.40 &  157.60 &   4.95 &  27.13 &  61.13 &  96.13 & Quad IV & RI      \\
v87  & $18182-1504$ & 18:21:06.939 & $-$15:03:22.11 & $  21.5$ &   75.13 &  194.30 &  237.00 &  \nodata  &  \nodata  &  \nodata  &  \nodata  & \nodata & LI      \\
v117 & $18246-1032$ & 18:27:23.764 & $-$10:30:21.71 & $  54.5$ &    2.18 &   20.25 &   50.37 &   0.96 &   1.87 &   6.17 &  13.73 & \nodata & RI      \\
v120 & $18257-1052$ & 18:28:30.818 & $-$10:50:52.99 & $ 136.2$ &   12.38 &   37.68 &   61.23 &  11.54 &  15.62 &  33.54 &  38.40 & \nodata & LI      \\
v121 & $18257-1000$ & 18:28:30.937 & $-$09:58:14.33 & $ 115.8$ &   46.32 &  120.50 &  115.90 &   9.75 &  10.65 &  21.33 &  25.19 & \nodata & LI      \\
v132 & $18276-1431$ & 18:30:30.676 & $-$14:28:57.78 & $  61.2$ &   22.65 &  132.00 &  120.00 &   4.76 &  20.85 &  45.39 & 106.73 & Quad I  & RI      \\
v146 & $18310-0806$ & 18:33:49.578 & $-$08:04:01.38 & $ 106.7$ &  \nodata  &  \nodata  &  \nodata  &   0.31 &   1.89 &   4.17 &  11.67 & Quad I  & \nodata \\
v149 & $18314-0900$ & 18:34:11.303 & $-$08:58:02.55 & $  35.0$ &   22.92 &   35.36 &  119.60 &  10.32 &  14.70 &  18.39 &  19.18 & \nodata & LI      \\
v154 & $18327-0715$ & 18:35:29.202 & $-$07:13:11.01 & $  42.0$ &   44.41 &   81.27 &   95.03 &  22.60 &  33.24 &  43.10 &  44.95 & \nodata & LI      \\
v162 & $18342-0655$ & 18:36:57.999 & $-$06:53:25.07 & $  87.3$ &  \nodata  &  \nodata  &  \nodata  &   2.14 &   4.94 &   5.55 &  11.90 & Quad I  & \nodata \\
v169 & $18355-0712$ & 18:38:15.423 & $-$07:09:54.16 & $ 142.9$ &    1.73 &   14.34 &   31.14 &  \nodata  &  \nodata  &  \nodata  &  \nodata  & \nodata & RI      \\
v172 & $18361-0647$ & 18:38:50.529 & $-$06:44:49.85 & $  36.3$ &   14.79 &   34.16 &   88.73 &   4.12 &   5.98 &   8.30 &   9.01 & \nodata & LI      \\
v189 & $18389-0601$ & 18:41:35.902 & $-$05:58:51.02 & $  90.8$ &    3.52 &   10.37 &   64.61 &  11.11 &  12.49 &  16.75 &  25.44 & \nodata & LI      \\
v204 & $18420-0512$ & 18:44:41.660 & $-$05:09:17.00 & $ 105.6$ &    1.03 &   26.72 &   26.22 &   0.26 &   1.49 &   3.01 &  21.74 & Quad I  & RI      \\
v211 & $18432-0149$ & 18:45:52.377 & $-$01:46:42.78 & $  66.9$ &   25.06 &   52.32 &   48.09 &  12.83 &  19.81 &  30.09 &  33.82 & \nodata & LI      \\
v212 & $18434-0202$ & 18:46:05.779 & $-$01:59:17.62 & $  37.5$ &   13.93 &   19.75 &   61.53 &   7.21 &   9.91 &  12.21 &  10.71 & \nodata & LI      \\
v223 & $18450-0148$ & 18:47:41.130 & $-$01:45:11.76 & $  33.9$ &  \nodata  &  \nodata  &  \nodata  &   2.33 &  23.90 &  55.87 &  88.56 & Quad IV & \nodata \\
v228 & $18460-0254$ & 18:48:41.947 & $-$02:50:29.23 & $  98.6$ &  111.10 &  279.90 &  237.00 & 100.32 & 140.24 & 227.94 & 276.72 & \nodata & LI      \\
v231 & $18467-0238$ & 18:49:19.446 & $-$02:34:49.90 & $  68.1$ &    5.55 &   37.55 &   57.46 &   1.20 &   5.81 &  12.63 &  23.93 & Quad IV & RI      \\
v237 & $18485+0642$ & 18:50:59.501 & $+$06:45:57.49 & $  95.3$ &    3.58 &   21.86 &   25.32 &   0.98 &   3.11 &   7.73 &  16.88 & Quad I  & RI      \\
v239 & $18488-0107$ & 18:51:26.252 & $-$01:03:52.23 & $  75.9$ &   16.45 &   42.98 &   44.03 &  21.72 &  32.01 &  49.24 &  58.86 & \nodata & LI      \\
v268 & $18588+0428$ & 19:01:20.032 & $+$04:32:31.25 & $  53.2$ &   10.77 &   13.05 &   20.11 &  15.84 &  19.45 &  19.25 &  16.09 & \nodata & LI      \\
v270 & $18596+0315$ & 19:02:06.259 & $+$03:20:15.47 & $  88.3$ &    2.60 &   14.17 &   22.57 &   0.40 &   2.51 &   5.95 &  10.54 & Quad IV & RI      \\
v274 & $19024+0044$ & 19:05:02.155 & $+$00:48:51.19 & $  50.0$ &    2.86 &   48.83 &   42.53 &   0.67 &   2.40 &   8.44 &  38.04 & Quad I  & RI      \\
v280 & $19070+0859$ & 19:10:21.766 & $+$09:05:01.43 & $  18.2$ &  \nodata  &  \nodata  &  \nodata  &  24.12 & 123.46 & 226.82 & 518.96 & Quad I  & \nodata \\
\\ \enddata
\end{deluxetable}

\begin{deluxetable}{c c c c}
\tablecolumns{4} \tabletypesize{\small} \tablewidth{0pt}
\tablecaption{Schedule of Observations \label{Ta:obs}} \tablehead{
\colhead{Frequency} & \colhead{UT date} &  \colhead{Typical
integration time} & \colhead{Typical rms noise} \\ \colhead{(MHz)} &
\colhead{} & \colhead{(min)} & \colhead{(Jy)}  } 
\startdata
\cutinhead{Parkes Telescope}   
1612 & 2002 Sep 5-7 & 10 & 0.07 \\ 
1665/67 & 2002 Sep 7 & 10 & 0.07 \\
1665/67 & 2002 Sep 7-8 & 10 & 0.12 \\ 
1720 & 2003 Jan 30 & 20 & 0.23 \\ 
1720 & 2003 Jan 31 - Feb 1 & 15 & 0.06 \\ 
1720 & 2003 Feb 1-2 &  10 & 0.07 \\ 
1720 & 2003 Feb 3 &  8 & 0.09 \\ 
1720 & 2003 Aug 17 & 20 & 0.10 \\ 
1612/65/67 & 2003 Aug 18 & 10 & 0.10 \\ 
\cutinhead{ATCA}  
1665/67 & 2003 Jul 23/24 + Aug 7 & 14 & 0.03 \\ 
1720 & 2003 Aug 08 & 14 & 0.04 \\ 
\enddata
\end{deluxetable}

\begin{deluxetable}{c r r r r r r r r c c c}
\tabletypesize{\scriptsize} 
\tablewidth{485pt} 
\tablecaption{Results from 1612, 1665 and 1667 MHz observations. \label{Ta:1612etc}}
\tablehead{ \colhead{Name} & \colhead{V$_{b}$} & \colhead{S$_b$} &
\colhead{I$_{b}$} & \colhead{V$_{r}$} & \colhead{S$_r$} &
\colhead{I$_{r}$} & \colhead{$\frac{S_b}{S_r}$} &
\colhead{$\frac{I_b}{I_r}$} & \colhead{Vrange} & \colhead{Type} & \colhead{Flag}\\ \colhead{} & \colhead{{\scriptsize
(km s$^{-1}$)}} & \colhead{{\scriptsize(Jy)}} &
\colhead{{\scriptsize(Jy km s$^{-1}$)}} & \colhead{{\scriptsize(km
s$^{-1}$)}} & \colhead{{\scriptsize(Jy)}} & \colhead{{\scriptsize(Jy
km s$^{-1}$)}} & \colhead{} & \colhead{} & \colhead{{\scriptsize(km
s$^{-1}$)}} & \colhead{} & \colhead{}  }  
\tablecolumns{12} 
\startdata
d3    & $ -32.95$ & $  1.34$ & $     1.3$ & $ -12.44$ & $  3.04$ & $     3.3$ & $ 0.44$ & $  0.4$ & 23 & D  & \tablenotemark{V}   \\ *
 & \nodata & \nodata & \nodata & \nodata & \nodata & \nodata & \nodata & \nodata & \nodata & \nodata & \\* 
 & \nodata & \nodata & \nodata & \nodata & \nodata & \nodata & \nodata & \nodata & \nodata & \nodata & \\ 
d34   & $-151.71$ & $  1.21$ & $     1.7$ & $-137.47$ & $  1.49$ & $     1.7$ & $ 0.81$ & $  1.0$ & 17 & D  &     \\ *
 & \nodata & \nodata & \nodata & \nodata & \nodata & \nodata & \nodata & \nodata & \nodata & \nodata & \\* 
 & \nodata & \nodata & \nodata & \nodata & \nodata & \nodata & \nodata & \nodata & \nodata & \nodata & \\ 
d39   & $ -101.50$ & $  1.00$ & $     2.7$ & \nodata & \nodata & \nodata & \nodata & \nodata & 13 & S  &     \\*
 & \nodata & \nodata & \nodata & \nodata & \nodata & \nodata & \nodata & \nodata & \nodata & \nodata & \\* 
 & \nodata & \nodata & \nodata & \nodata & \nodata & \nodata & \nodata & \nodata & \nodata & \nodata & \\ 
d46   & $ -140.67$ & $  1.75$ & $    49.7$ & \nodata & \nodata & \nodata & \nodata & \nodata & 77 & I  & \tablenotemark{V}   \\*
& $-133.36$ & $  2.41$ & $    55.4$ & \nodata & \nodata & \nodata & \nodata & \nodata & 62 & I  &     \\*
& $-159.61$ & $  1.52$ & $    50.6$ & \nodata & \nodata & \nodata & \nodata & \nodata & 72 & I  &     \\
d47   & $ -66.97$ & $ 12.07$ & $    42.2$ & $ -44.01$ & $ 15.27$ & $    97.6$ & $ 0.79$ & $  0.4$ & 57 & DD &     \\ *
 & $ -82.68$ & $  2.42$ & $     5.8$ & $ -30.76$ & $  2.26$ & $     3.0$ & $ 1.07$ & $  2.0$ & 57 & DD & 2nd \\*
& $ -64.75$ & $  7.74$ & $     9.9$ & \nodata & \nodata & \nodata & \nodata & \nodata & 28 & I  & \tablenotemark{\#}   \\*
& $ -65.46$ & $  1.60$ & $     2.6$ & \nodata & \nodata & \nodata & \nodata & \nodata & 42 & I  &     \\
d56   & $ -79.32$ & $ 44.25$ & $    84.0$ & $ -40.30$ & $ 46.99$ & $   198.5$ & $ 0.94$ & $  0.4$ & 50 & D  & \tablenotemark{V}   \\ *
 & \nodata & \nodata & \nodata & \nodata & \nodata & \nodata & \nodata & \nodata & \nodata & \nodata & \\* 
 & $ -79.91$ & $  0.97$ & $     4.0$ & $ -33.19$ & $  1.65$ & $     9.4$ & $ 0.59$ & $  0.4$ & 55 & D  &     \\
d62   & $-117.95$ & $  1.75$ & $     4.7$ & $-104.79$ & $  0.50$ & $     1.8$ & $ 3.51$ & $  2.7$ & 23 & I  &     \\ *
& $-107.22$ & $  0.53$ & $     1.5$ & \nodata & \nodata & \nodata & \nodata & \nodata & 21 & -  & \tablenotemark{m}  \\*
& $-123.22$ & $  1.29$ & $    16.5$ & \nodata & \nodata & \nodata & \nodata & \nodata & 31 & I  &     \\
d93   & $ -97.35$ & $  4.63$ & $     5.7$ & $ -68.58$ & $  2.48$ & $     6.0$ & $ 1.86$ & $  0.9$ & 32 & D  & \tablenotemark{V}   \\ *
& $ -79.26$ & $  0.58$ & $     7.6$ & \nodata & \nodata & \nodata & \nodata & \nodata & 37 & I  & \tablenotemark{\#}   \\*
 & \nodata & \nodata & \nodata & \nodata & \nodata & \nodata & \nodata & \nodata & \nodata & \nodata & \\ 
d103  & $ -70.17$ & $  0.43$ & $     0.4$ & $ -33.69$ & $  0.28$ & $     0.3$ & $ 1.54$ & $  1.7$ & 40 & D  & \tablenotemark{V}   \\ *
 & \nodata & \nodata & \nodata & \nodata & \nodata & \nodata & \nodata & \nodata & \nodata & \nodata & \\* 
 & \nodata & \nodata & \nodata & \nodata & \nodata & \nodata & \nodata & \nodata & \nodata & \nodata & \\ 
d117  & $  -89.51$ & $  2.36$ & $    19.8$ & \nodata & \nodata & \nodata & \nodata & \nodata & 23 & I  &     \\*
 & \nodata & \nodata & \nodata & \nodata & \nodata & \nodata & \nodata & \nodata & \nodata & \nodata & \\* 
 & \nodata & \nodata & \nodata & \nodata & \nodata & \nodata & \nodata & \nodata & \nodata & \nodata & \\ 
d150  & $-108.40$ & $  0.81$ & $     2.2$ & $ -83.81$ & $  3.22$ & $     3.2$ & $ 0.25$ & $  0.7$ & 27 & D  & \tablenotemark{V}   \\ *
 & \nodata & \nodata & \nodata & \nodata & \nodata & \nodata & \nodata & \nodata & \nodata & \nodata & \\* 
 & \nodata & \nodata & \nodata & \nodata & \nodata & \nodata & \nodata & \nodata & \nodata & \nodata & \\ 
d168  & $ -22.82$ & $173.40$ & $   356.2$ & $  14.85$ & $102.20$ & $   267.1$ & $ 1.70$ & $  1.3$ & 41 & D  & \tablenotemark{V}   \\ *
 & \nodata & \nodata & \nodata & \nodata & \nodata & \nodata & \nodata & \nodata & \nodata & \nodata & \\* 
& $ -23.29$ & $ 14.15$ & $   116.7$ & \nodata & \nodata & \nodata & \nodata & \nodata & 44 & I  &     \\
d189  & $ -11.25$ & $ 45.39$ & $    60.0$ & $  18.07$ & $ 48.36$ & $    83.4$ & $ 0.94$ & $  0.7$ & 33 & D  &     \\ *
& $ -12.22$ & $  0.86$ & $     5.7$ & \nodata & \nodata & \nodata & \nodata & \nodata & 31 & I  &     \\*
 & $ -13.19$ & $  2.78$ & $     6.6$ & $  17.64$ & $  0.76$ & $     5.1$ & $ 3.66$ & $  1.3$ & 34 & D  &     \\
b5    & $ -45.69$ & $  1.28$ & $     1.2$ & $ -19.55$ & $  1.21$ & $     1.8$ & $ 1.06$ & $  0.7$ & 28 & D  &     \\ *
 & \nodata & \nodata & \nodata & \nodata & \nodata & \nodata & \nodata & \nodata & \nodata & \nodata & \\* 
& $ -46.34$ & $  0.95$ & $     1.0$ & \nodata & \nodata & \nodata & \nodata & \nodata &  7 & S  &     \\
d190  & $  -41.45$ & $  0.73$ & $     4.3$ & \nodata & \nodata & \nodata & \nodata & \nodata & 20 & I  &     \\*
 & \nodata & \nodata & \nodata & \nodata & \nodata & \nodata & \nodata & \nodata & \nodata & \nodata & \\* 
 & \nodata & \nodata & \nodata & \nodata & \nodata & \nodata & \nodata & \nodata & \nodata & \nodata & \\ 
b11   & \nodata & \nodata & \nodata & $  24.96$ & $  3.01$ & $     4.0$ & \nodata & \nodata &  3 & S  &     \\*
& $  14.23$ & $  1.72$ & $    16.0$ & \nodata & \nodata & \nodata & \nodata & \nodata & 24 & I  &     \\*
 & $   1.41$ & $ 10.19$ & $    25.5$ & $  25.83$ & $ 13.11$ & $    34.8$ & $ 0.78$ & $  0.7$ & 31 & D  &     \\
d197  & $-193.29$ & $  7.16$ & $    10.7$ & $-163.17$ & $  3.22$ & $     6.4$ & $ 2.22$ & $  1.7$ & 33 & D  &     \\ *
& $-194.08$ & $  0.60$ & $     0.4$ & \nodata & \nodata & \nodata & \nodata & \nodata &  4 & S  & \tablenotemark{\#}   \\*
 & \nodata & \nodata & \nodata & $-162.32$ & $  0.60$ & $     0.6$ & \nodata & \nodata &  3 & -  & \tablenotemark{m}   \\
b14   & $  72.67$ & $  5.79$ & $     5.9$ & $ 101.54$ & $  3.93$ & $     4.7$ & $ 1.47$ & $  1.2$ & 33 & D  &     \\ *
 & $  72.13$ & $  0.43$ & $     0.2$ & $ 101.91$ & $  0.37$ & $     0.4$ & $ 1.16$ & $  0.4$ & 33 & D  &     \\*
 & $  71.86$ & $  1.46$ & $     0.6$ & $ 101.56$ & $  2.13$ & $     2.6$ & $ 0.69$ & $  0.2$ & 33 & D  &     \\
b15   & $  -75.10$ & $  0.63$ & $     1.5$ & \nodata & \nodata & \nodata & \nodata & \nodata & 39 & I  &     \\*
 & \nodata & \nodata & \nodata & \nodata & \nodata & \nodata & \nodata & \nodata & \nodata & \nodata & \\* 
 & \nodata & \nodata & \nodata & \nodata & \nodata & \nodata & \nodata & \nodata & \nodata & \nodata & \\ 
b17   & $ -18.33$ & $ 37.18$ & $    42.1$ & $   7.27$ & $ 18.57$ & $    31.1$ & $ 2.00$ & $  1.4$ & 29 & D  &     \\ *
 & \nodata & \nodata & \nodata & \nodata & \nodata & \nodata & \nodata & \nodata & \nodata & \nodata & \\* 
 & $ -19.11$ & $  1.00$ & $     0.2$ & $   7.15$ & $  5.54$ & $     8.0$ & $ 0.18$ & $  0.0$ & 30 & D  &     \\
d200  & $ -92.13$ & $  5.37$ & $    16.8$ & $ -57.20$ & $  0.62$ & $     1.9$ & $ 8.60$ & $  8.9$ & 42 & De &     \\ *
 & \nodata & \nodata & \nodata & \nodata & \nodata & \nodata & \nodata & \nodata & \nodata & \nodata & \\* 
 & \nodata & \nodata & \nodata & \nodata & \nodata & \nodata & \nodata & \nodata & \nodata & \nodata & \\ 
b25   & $   7.89$ & $  2.85$ & $     2.2$ & $  28.04$ & $  1.60$ & $     1.5$ & $ 1.78$ & $  1.5$ & 23 & D  &     \\ *
 & \nodata & \nodata & \nodata & \nodata & \nodata & \nodata & \nodata & \nodata & \nodata & \nodata & \\* 
 & \nodata & \nodata & \nodata & \nodata & \nodata & \nodata & \nodata & \nodata & \nodata & \nodata & \\ 
b30   & $  52.02$ & $  0.31$ & $     0.4$ & $  86.80$ & $  0.26$ & $     0.7$ & $ 1.19$ & $  0.7$ & 40 & D  & \tablenotemark{V}   \\ *
 & \nodata & \nodata & \nodata & \nodata & \nodata & \nodata & \nodata & \nodata & \nodata & \nodata & \\* 
 & \nodata & \nodata & \nodata & \nodata & \nodata & \nodata & \nodata & \nodata & \nodata & \nodata & \\ 
b31   & $-161.07$ & $  3.80$ & $     3.7$ & $-127.59$ & $  1.91$ & $     1.4$ & $ 1.99$ & $  2.7$ & 36 & D  &     \\ *
 & \nodata & \nodata & \nodata & \nodata & \nodata & \nodata & \nodata & \nodata & \nodata & \nodata & \\* 
 & \nodata & \nodata & \nodata & $-127.42$ & $  1.24$ & $     1.7$ & \nodata & \nodata &  9 & S  &     \\
b33   & $ -36.27$ & $  1.92$ & $     3.1$ & $ -11.76$ & $  0.86$ & $     2.0$ & $ 2.23$ & $  1.5$ & 36 & D  &     \\ *
 & \nodata & \nodata & \nodata & \nodata & \nodata & \nodata & \nodata & \nodata & \nodata & \nodata & \\* 
 & \nodata & \nodata & \nodata & \nodata & \nodata & \nodata & \nodata & \nodata & \nodata & \nodata & \\ 
b34   & $-218.28$ & $  7.46$ & $     9.4$ & $-198.59$ & $  2.10$ & $     2.8$ & $ 3.56$ & $  3.4$ & 23 & D  &     \\ *
& $-219.08$ & $  0.67$ & $     0.3$ & \nodata & \nodata & \nodata & \nodata & \nodata &  3 & -  & \tablenotemark{m}   \\*
 & \nodata & \nodata & \nodata & $-198.45$ & $  0.39$ & $     0.5$ & \nodata & \nodata &  4 & -  & \tablenotemark{m\#}  \\
d202  & $-109.88$ & $ 29.49$ & $    38.3$ & $ -84.93$ & $ 36.99$ & $    55.7$ & $ 0.80$ & $  0.7$ & 27 & D  &     \\ *
 & $-105.20$ & $  1.77$ & $    10.3$ & $ -86.85$ & $  2.28$ & $     9.9$ & $ 0.77$ & $  1.0$ & 25 & D  &     \\*
 & $-110.90$ & $  2.10$ & $     2.1$ & $ -84.48$ & $  5.34$ & $     5.3$ & $ 0.39$ & $  0.4$ & 28 & D  &     \\
b44   & $  -34.98$ & $  0.78$ & $     0.7$ & \nodata & \nodata & \nodata & \nodata & \nodata &  7 & S  & \tablenotemark{V}   \\*
 & \nodata & \nodata & \nodata & \nodata & \nodata & \nodata & \nodata & \nodata & \nodata & \nodata & \\* 
 & \nodata & \nodata & \nodata & \nodata & \nodata & \nodata & \nodata & \nodata & \nodata & \nodata & \\ 
b62   & $   20.01$ & $  1.80$ & $     5.4$ & \nodata & \nodata & \nodata & \nodata & \nodata & 14 & I  &     \\*
& $  18.62$ & $  0.80$ & $     0.4$ & \nodata & \nodata & \nodata & \nodata & \nodata &  3 & S  &     \\*
& $  18.97$ & $  0.87$ & $     8.0$ & \nodata & \nodata & \nodata & \nodata & \nodata & 24 & I  &     \\
b68   & $-298.54$ & $  1.19$ & $     2.7$ & $-274.42$ & $  4.69$ & $     4.8$ & $ 0.25$ & $  0.6$ & 27 & D  &     \\ *
 & \nodata & \nodata & \nodata & \nodata & \nodata & \nodata & \nodata & \nodata & \nodata & \nodata & \\* 
& $-299.32$ & $  0.70$ & $     0.7$ & \nodata & \nodata & \nodata & \nodata & \nodata &  7 & S  &     \\
b70   & $  36.51$ & $  4.72$ & $    16.8$ & $  59.76$ & $  8.54$ & $    22.5$ & $ 0.55$ & $  0.8$ & 32 & Dw &     \\ *
 & \nodata & \nodata & \nodata & \nodata & \nodata & \nodata & \nodata & \nodata & \nodata & \nodata & \\* 
 & \nodata & \nodata & \nodata & \nodata & \nodata & \nodata & \nodata & \nodata & \nodata & \nodata & \\ 
b96   & $-145.68$ & $  5.75$ & $     5.1$ & $-126.99$ & $  5.40$ & $     5.9$ & $ 1.06$ & $  0.9$ & 21 & D  & \tablenotemark{V}   \\ *
 & \nodata & \nodata & \nodata & \nodata & \nodata & \nodata & \nodata & \nodata & \nodata & \nodata & \\* 
 & \nodata & \nodata & \nodata & \nodata & \nodata & \nodata & \nodata & \nodata & \nodata & \nodata & \\ 
b106  & $  60.43$ & $  4.03$ & $     7.2$ & $  81.50$ & $  3.13$ & $     6.0$ & $ 1.29$ & $  1.2$ & 25 & D  &     \\ *
 & \nodata & \nodata & \nodata & \nodata & \nodata & \nodata & \nodata & \nodata & \nodata & \nodata & \\* 
 & \nodata & \nodata & \nodata & \nodata & \nodata & \nodata & \nodata & \nodata & \nodata & \nodata & \\ 
b112  & $  12.40$ & $  4.11$ & $    12.9$ & $  37.00$ & $  2.80$ & $    12.9$ & $ 1.47$ & $  1.0$ & 32 & Dw &     \\ *
 & \nodata & \nodata & \nodata & \nodata & \nodata & \nodata & \nodata & \nodata & \nodata & \nodata & \\* 
 & \nodata & \nodata & \nodata & \nodata & \nodata & \nodata & \nodata & \nodata & \nodata & \nodata & \\ 
b114  & $  115.42$ & $  1.48$ & $    14.2$ & \nodata & \nodata & \nodata & \nodata & \nodata & 21 & I  &     \\*
& $ 114.94$ & $  0.69$ & $     4.0$ & \nodata & \nodata & \nodata & \nodata & \nodata & 15 & I  &     \\*
 & \nodata & \nodata & \nodata & \nodata & \nodata & \nodata & \nodata & \nodata & \nodata & \nodata & \\ 
b128  & $-245.34$ & $  5.38$ & $     8.6$ & $-224.11$ & $  8.59$ & $    10.9$ & $ 0.63$ & $  0.8$ & 26 & D  &     \\ *
 & \nodata & \nodata & \nodata & \nodata & \nodata & \nodata & \nodata & \nodata & \nodata & \nodata & \\* 
 & \nodata & \nodata & \nodata & $-223.68$ & $  0.53$ & $     0.8$ & \nodata & \nodata &  3 & S  &     \\
b130  & $ -11.74$ & $  2.08$ & $     1.9$ & $  13.50$ & $  0.91$ & $     1.2$ & $ 2.27$ & $  1.6$ & 28 & D  &     \\ *
 & \nodata & \nodata & \nodata & \nodata & \nodata & \nodata & \nodata & \nodata & \nodata & \nodata & \\* 
 & \nodata & \nodata & \nodata & \nodata & \nodata & \nodata & \nodata & \nodata & \nodata & \nodata & \\ 
b133  & $ -24.59$ & $  1.42$ & $     7.4$ & $  18.52$ & $  0.66$ & $     1.0$ & $ 2.16$ & $  7.4$ & 49 & D  &     \\ *
 & \nodata & \nodata & \nodata & \nodata & \nodata & \nodata & \nodata & \nodata & \nodata & \nodata & \\* 
 & \nodata & \nodata & \nodata & \nodata & \nodata & \nodata & \nodata & \nodata & \nodata & \nodata & \\ 
b134  & $-123.37$ & $ 74.44$ & $   207.8$ & $ -93.42$ & $  1.44$ & $     3.1$ & $51.80$ & $ 67.4$ & 35 & De &     \\ *
& $-115.40$ & $  0.92$ & $     3.3$ & \nodata & \nodata & \nodata & \nodata & \nodata & 11 & S  &     \\*
& $-124.53$ & $  1.44$ & $     2.8$ & \nodata & \nodata & \nodata & \nodata & \nodata & 12 & S  &     \\
b143  & $  30.44$ & $  7.78$ & $    12.8$ & $  57.67$ & $  8.15$ & $    10.8$ & $ 0.96$ & $  1.2$ & 30 & D  &     \\ *
 & \nodata & \nodata & \nodata & $  58.12$ & $  0.52$ & $     0.2$ & \nodata & \nodata &  2 & S  & \tablenotemark{\#}   \\
 & $  29.57$ & $  0.57$ & $     0.4$ & $  54.78$ & $  1.81$ & $     2.4$ & $ 0.31$ & $  0.2$ & 32 & D  & \tablenotemark{\#}   \\
b155  & $  30.70$ & $  0.54$ & $     1.4$ & $  62.29$ & $  1.11$ & $     2.8$ & $ 0.49$ & $  0.5$ & 35 & D  &     \\ *
 & \nodata & \nodata & \nodata & \nodata & \nodata & \nodata & \nodata & \nodata & \nodata & \nodata & \\* 
 & \nodata & \nodata & \nodata & \nodata & \nodata & \nodata & \nodata & \nodata & \nodata & \nodata & \\ 
b165  & $  -26.12$ & $  4.51$ & $    35.5$ & \nodata & \nodata & \nodata & \nodata & \nodata & 59 & I  & \tablenotemark{V}   \\*
 & \nodata & \nodata & \nodata & \nodata & \nodata & \nodata & \nodata & \nodata & \nodata & \nodata & \\* 
 & \nodata & \nodata & \nodata & \nodata & \nodata & \nodata & \nodata & \nodata & \nodata & \nodata & \\ 
b199  & $-141.10$ & $  6.85$ & $     7.6$ & $-114.33$ & $ 10.31$ & $    10.2$ & $ 0.66$ & $  0.7$ & 30 & D  &     \\ *
 & \nodata & \nodata & \nodata & \nodata & \nodata & \nodata & \nodata & \nodata & \nodata & \nodata & \\* 
 & $-141.71$ & $  0.87$ & $     2.6$ & $-114.49$ & $  0.81$ & $     2.0$ & $ 1.08$ & $  1.3$ & 34 & D  &     \\
b209  & $ -29.68$ & $  5.98$ & $     5.2$ & $  -2.45$ & $  7.02$ & $     7.2$ & $ 0.85$ & $  0.7$ & 29 & D  &     \\ *
 & \nodata & \nodata & \nodata & \nodata & \nodata & \nodata & \nodata & \nodata & \nodata & \nodata & \\* 
 & \nodata & \nodata & \nodata & \nodata & \nodata & \nodata & \nodata & \nodata & \nodata & \nodata & \\ 
b210  & $  58.49$ & $  2.71$ & $     3.0$ & $  87.82$ & $  1.32$ & $     0.9$ & $ 2.05$ & $  3.5$ & 32 & D  &     \\ *
 & \nodata & \nodata & \nodata & \nodata & \nodata & \nodata & \nodata & \nodata & \nodata & \nodata & \\* 
 & \nodata & \nodata & \nodata & \nodata & \nodata & \nodata & \nodata & \nodata & \nodata & \nodata & \\ 
b228  & $ 170.21$ & $  0.45$ & $     1.1$ & $ 192.64$ & $  2.11$ & $     3.4$ & $ 0.21$ & $  0.3$ & 27 & D  &     \\ *
 & \nodata & \nodata & \nodata & \nodata & \nodata & \nodata & \nodata & \nodata & \nodata & \nodata & \\* 
 & \nodata & \nodata & \nodata & \nodata & \nodata & \nodata & \nodata & \nodata & \nodata & \nodata & \\ 
b246  & $  122.04$ & $  2.78$ & $    15.1$ & \nodata & \nodata & \nodata & \nodata & \nodata & 42 & I  & \tablenotemark{V}   \\*
& $ 126.07$ & $  0.51$ & $     6.0$ & \nodata & \nodata & \nodata & \nodata & \nodata & 39 & I  &     \\*
 & \nodata & \nodata & \nodata & \nodata & \nodata & \nodata & \nodata & \nodata & \nodata & \nodata & \\ 
b250  & $-202.45$ & $  1.31$ & $     2.0$ & $-168.88$ & $  3.32$ & $     3.4$ & $ 0.39$ & $  0.6$ & 35 & D  &     \\ *
 & \nodata & \nodata & \nodata & \nodata & \nodata & \nodata & \nodata & \nodata & \nodata & \nodata & \\* 
 & $-202.66$ & $  0.74$ & $     0.9$ & $-168.61$ & $  1.16$ & $     2.8$ & $ 0.64$ & $  0.3$ & 39 & D  &     \\
b251  & $ -34.55$ & $ 12.91$ & $    18.7$ & $  -8.13$ & $ 12.36$ & $    18.4$ & $ 1.04$ & $  1.0$ & 29 & D  &     \\ *
& $ -35.20$ & $  1.16$ & $     1.5$ & \nodata & \nodata & \nodata & \nodata & \nodata &  5 & S  &     \\*
 & $ -35.28$ & $  1.69$ & $     2.3$ & $  -7.62$ & $  3.25$ & $     5.1$ & $ 0.52$ & $  0.4$ & 30 & D  &     \\
b258  & $ 191.06$ & $  7.22$ & $    12.4$ & $ 217.58$ & $  3.83$ & $     8.0$ & $ 1.88$ & $  1.5$ & 29 & D  &     \\ *
& $ 192.17$ & $  0.59$ & $     6.3$ & \nodata & \nodata & \nodata & \nodata & \nodata & 40 & I  &     \\*
 & $ 190.06$ & $  0.67$ & $     0.9$ & $ 219.77$ & $  0.41$ & $     0.9$ & $ 1.62$ & $  1.0$ & 33 & D  &     \\
b262  & $  -21.31$ & $ 11.58$ & $    32.4$ & \nodata & \nodata & \nodata & \nodata & \nodata & 50 & I  &     \\*
& $  10.90$ & $  6.24$ & $    47.0$ & \nodata & \nodata & \nodata & \nodata & \nodata & 50 & I  &     \\*
& $ -29.33$ & $ 18.37$ & $   218.2$ & \nodata & \nodata & \nodata & \nodata & \nodata & 62 & I  &     \\
b263  & $  138.30$ & $  0.69$ & $     9.1$ & \nodata & \nodata & \nodata & \nodata & \nodata & 48 & I  &     \\*
& $ 138.65$ & $  1.09$ & $     9.3$ & \nodata & \nodata & \nodata & \nodata & \nodata & 26 & I  &     \\*
 & \nodata & \nodata & \nodata & \nodata & \nodata & \nodata & \nodata & \nodata & \nodata & \nodata & \\ 
b266  & $ 164.10$ & $  3.58$ & $     3.1$ & $ 186.81$ & $  3.71$ & $     3.1$ & $ 0.97$ & $  1.0$ & 25 & D  &     \\ *
 & \nodata & \nodata & \nodata & \nodata & \nodata & \nodata & \nodata & \nodata & \nodata & \nodata & \\* 
 & \nodata & \nodata & \nodata & \nodata & \nodata & \nodata & \nodata & \nodata & \nodata & \nodata & \\ 
b292  & $  70.40$ & $ 18.85$ & $    58.8$ & $  99.28$ & $  0.35$ & $     0.9$ & $53.84$ & $ 62.0$ & 36 & De &     \\ *
 & $  69.70$ & $  7.37$ & $     5.2$ & $ 103.97$ & $  1.64$ & $     1.9$ & $ 4.50$ & $  2.7$ & 36 & D  &     \\*
 & \nodata & \nodata & \nodata & \nodata & \nodata & \nodata & \nodata & \nodata & \nodata & \nodata & \\ 
b300  & $ 105.60$ & $  2.48$ & $     2.2$ & $ 135.48$ & $  0.80$ & $     1.4$ & $ 3.12$ & $  1.6$ & 32 & D  &     \\ *
 & \nodata & \nodata & \nodata & \nodata & \nodata & \nodata & \nodata & \nodata & \nodata & \nodata & \\* 
& $ 106.87$ & $  0.53$ & $     0.5$ & \nodata & \nodata & \nodata & \nodata & \nodata &  8 & S  & \tablenotemark{\#}   \\
b301  & $  23.30$ & $  1.15$ & $     6.4$ & $  79.59$ & $  3.03$ & $    13.2$ & $ 0.38$ & $  0.5$ & 64 & D  &     \\ *
 & \nodata & \nodata & \nodata & \nodata & \nodata & \nodata & \nodata & \nodata & \nodata & \nodata & \\* 
 & \nodata & \nodata & \nodata & \nodata & \nodata & \nodata & \nodata & \nodata & \nodata & \nodata & \\ 
b304  & $ -74.39$ & $  4.74$ & $     6.7$ & $ -52.25$ & $  3.91$ & $     6.0$ & $ 1.21$ & $  1.1$ & 25 & D  &     \\ *
& $ -74.76$ & $  0.36$ & $     0.2$ & \nodata & \nodata & \nodata & \nodata & \nodata &  2 & -  & \tablenotemark{m\#}  \\*
 & $ -75.37$ & $  0.49$ & $     0.5$ & $ -51.93$ & $  0.77$ & $     0.5$ & $ 0.63$ & $  0.9$ & 25 & D  & \tablenotemark{\#}   \\
v41   & $  20.51$ & $  7.06$ & $    46.2$ & $  64.09$ & $ 33.71$ & $   274.1$ & $ 0.21$ & $  0.2$ & 65 & Dw &     \\ *
 & \nodata & \nodata & \nodata & $  62.46$ & $  1.68$ & $    11.1$ & \nodata & \nodata & 16 & I  &     \\
 & $  13.01$ & $  0.98$ & $     9.0$ & $  65.97$ & $  2.27$ & $    13.9$ & $ 0.43$ & $  0.6$ & 75 & I  &     \\
v45   & $  -2.37$ & $ 11.28$ & $     9.1$ & $  26.77$ & $ 11.54$ & $     7.4$ & $ 0.98$ & $  1.2$ & 31 & D  &     \\ *
 & \nodata & \nodata & \nodata & $  26.27$ & $  0.45$ & $     0.1$ & \nodata & \nodata & 18 & I  & \tablenotemark{\#}   \\
 & $  -2.81$ & $  0.78$ & $     1.4$ & $  27.14$ & $  0.71$ & $     0.8$ & $ 1.11$ & $  1.7$ & 32 & D  &     \\
v50   & $  30.81$ & $  0.68$ & $     0.9$ & $  66.85$ & $  0.39$ & $     1.1$ & $ 1.72$ & $  0.7$ & 38 & D  &     \\ *
 & \nodata & \nodata & \nodata & \nodata & \nodata & \nodata & \nodata & \nodata & \nodata & \nodata & \\* 
 & $  29.03$ & $  0.38$ & $     1.3$ & $  66.99$ & $  0.38$ & $     1.3$ & $ 0.99$ & $  0.9$ & 42 & D  &     \\
v53   & $   3.49$ & $  1.72$ & $     1.6$ & $  31.45$ & $  3.21$ & $     1.9$ & $ 0.54$ & $  0.8$ & 32 & D  &     \\ *
& $   4.05$ & $  0.26$ & $     1.1$ & \nodata & \nodata & \nodata & \nodata & \nodata & 10 & -  & \tablenotemark{m}   \\*
& $   3.17$ & $  0.44$ & $     3.3$ & \nodata & \nodata & \nodata & \nodata & \nodata & 36 & I  &     \\
v56   & $  -3.77$ & $  1.04$ & $     1.9$ & $  39.71$ & $  0.53$ & $     1.3$ & $ 1.96$ & $  1.4$ & 48 & D  & \tablenotemark{V}   \\ *
 & \nodata & \nodata & \nodata & \nodata & \nodata & \nodata & \nodata & \nodata & \nodata & \nodata & \\* 
 & \nodata & \nodata & \nodata & \nodata & \nodata & \nodata & \nodata & \nodata & \nodata & \nodata & \\ 
v67   & $ -15.26$ & $ 61.79$ & $    84.7$ & $  13.06$ & $ 47.87$ & $    66.5$ & $ 1.29$ & $  1.3$ & 33 & D  & \tablenotemark{V}   \\ *
& $  -9.43$ & $  0.98$ & $     5.9$ & \nodata & \nodata & \nodata & \nodata & \nodata & 32 & I  &     \\*
 & $ -14.52$ & $  1.33$ & $     6.4$ & $  15.25$ & $ 11.44$ & $    12.9$ & $ 0.12$ & $  0.5$ & 33 & De &     \\
v87   & $  -1.45$ & $  9.55$ & $    53.0$ & $  43.03$ & $ 18.62$ & $    57.8$ & $ 0.51$ & $  0.9$ & 48 & D  &     \\ *
 & \nodata & \nodata & \nodata & \nodata & \nodata & \nodata & \nodata & \nodata & \nodata & \nodata & \\* 
 & $  -1.78$ & $  2.10$ & $     9.2$ & $  44.69$ & $  3.41$ & $    24.0$ & $ 0.62$ & $  0.4$ & 51 & D  &     \\
v117  & $   60.75$ & $ 14.56$ & $   131.6$ & \nodata & \nodata & \nodata & \nodata & \nodata & 74 & I  & \tablenotemark{V}   \\*
& $  58.37$ & $  4.31$ & $    42.0$ & \nodata & \nodata & \nodata & \nodata & \nodata & 50 & I  &     \\*
& $  40.54$ & $  0.99$ & $    19.6$ & \nodata & \nodata & \nodata & \nodata & \nodata & 89 & I  &     \\
v120  & $ 118.08$ & $ 13.69$ & $    24.5$ & $ 154.58$ & $  6.40$ & $    22.7$ & $ 2.14$ & $  1.1$ & 39 & D  &     \\ *
 & \nodata & \nodata & \nodata & \nodata & \nodata & \nodata & \nodata & \nodata & \nodata & \nodata & \\* 
 & $ 117.31$ & $  3.28$ & $     8.2$ & $ 155.45$ & $  2.96$ & $    11.7$ & $ 1.11$ & $  0.7$ & 41 & D  &     \\
v121  & $  97.82$ & $ 22.60$ & $    67.7$ & $ 134.14$ & $ 24.89$ & $    40.7$ & $ 0.91$ & $  1.7$ & 41 & D  & \tablenotemark{V}   \\ *
 & \nodata & \nodata & \nodata & \nodata & \nodata & \nodata & \nodata & \nodata & \nodata & \nodata & \\* 
 & $  96.04$ & $  3.40$ & $     7.3$ & $ 132.51$ & $  0.83$ & $     7.2$ & $ 4.12$ & $  1.0$ & 46 & D  &     \\
v132  & $  48.71$ & $ 60.29$ & $   118.6$ & $  72.95$ & $ 89.96$ & $   170.7$ & $ 0.67$ & $  0.7$ & 28 & D  & \tablenotemark{\#}   \\ *
 & $  47.75$ & $  5.47$ & $     3.3$ & $  72.43$ & $  1.09$ & $     3.3$ & $ 5.02$ & $  1.0$ & 29 & D  &     \\*
 & $  47.66$ & $  0.86$ & $     1.6$ & $  74.36$ & $ 15.69$ & $    14.5$ & $ 0.05$ & $  0.1$ & 29 & De &     \\
v146  & $  98.00$ & $  1.84$ & $     7.8$ & $ 117.16$ & $ 10.57$ & $    19.8$ & $ 0.17$ & $  0.4$ & 25 & D  & \tablenotemark{V}  \\ *
 & \nodata & \nodata & \nodata & \nodata & \nodata & \nodata & \nodata & \nodata & \nodata & \nodata & \\* 
& $  94.33$ & $  0.34$ & $     1.1$ & \nodata & \nodata & \nodata & \nodata & \nodata & 27 & -  & \tablenotemark{m}   \\
v149  & $  20.38$ & $ 12.02$ & $    26.3$ & $  49.25$ & $ 13.71$ & $    25.8$ & $ 0.88$ & $  1.0$ & 32 & D  &     \\ *
 & \nodata & \nodata & \nodata & \nodata & \nodata & \nodata & \nodata & \nodata & \nodata & \nodata & \\* 
 & \nodata & \nodata & \nodata & \nodata & \nodata & \nodata & \nodata & \nodata & \nodata & \nodata & \\ 
v154  & $  21.41$ & $  5.93$ & $    20.8$ & $  59.54$ & $  3.24$ & $    18.1$ & $ 1.83$ & $  1.1$ & 46 & D  & \tablenotemark{V}   \\ *
 & \nodata & \nodata & \nodata & \nodata & \nodata & \nodata & \nodata & \nodata & \nodata & \nodata & \\* 
 & \nodata & \nodata & \nodata & \nodata & \nodata & \nodata & \nodata & \nodata & \nodata & \nodata & \\ 
v162  & $  73.31$ & $  3.67$ & $     5.9$ & $ 102.91$ & $  1.69$ & $     4.9$ & $ 2.17$ & $  1.2$ & 35 & D  & \tablenotemark{V}   \\ *
 & \nodata & \nodata & \nodata & \nodata & \nodata & \nodata & \nodata & \nodata & \nodata & \nodata & \\* 
 & \nodata & \nodata & \nodata & \nodata & \nodata & \nodata & \nodata & \nodata & \nodata & \nodata & \\ 
v169  & $ 130.30$ & $  8.67$ & $    11.2$ & $ 155.18$ & $  6.90$ & $    10.0$ & $ 1.26$ & $  1.1$ & 27 & D  &     \\ *
 & \nodata & \nodata & \nodata & \nodata & \nodata & \nodata & \nodata & \nodata & \nodata & \nodata & \\* 
 & $ 128.85$ & $  0.42$ & $     0.2$ & $ 155.39$ & $  0.86$ & $     1.1$ & $ 0.49$ & $  0.2$ & 29 & D  & \tablenotemark{\#}   \\
v172  & $  19.98$ & $  6.41$ & $    16.1$ & $  53.75$ & $  4.63$ & $    15.5$ & $ 1.39$ & $  1.0$ & 37 & D  & \tablenotemark{V}   \\ *
 & \nodata & \nodata & \nodata & \nodata & \nodata & \nodata & \nodata & \nodata & \nodata & \nodata & \\* 
 & \nodata & \nodata & \nodata & \nodata & \nodata & \nodata & \nodata & \nodata & \nodata & \nodata & \\ 
v189  & $   89.42$ & $  0.59$ & $     1.4$ & \nodata & \nodata & \nodata & \nodata & \nodata &  8 & S  & \tablenotemark{V}   \\*
 & \nodata & \nodata & \nodata & \nodata & \nodata & \nodata & \nodata & \nodata & \nodata & \nodata & \\* 
& $  89.24$ & $  1.13$ & $     1.4$ & \nodata & \nodata & \nodata & \nodata & \nodata &  8 & S  &     \\
v204  & $  93.17$ & $  7.64$ & $    17.2$ & $ 118.25$ & $  5.67$ & $    19.1$ & $ 1.35$ & $  0.9$ & 29 & D  &     \\ *
 & \nodata & \nodata & \nodata & \nodata & \nodata & \nodata & \nodata & \nodata & \nodata & \nodata & \\* 
& $  92.61$ & $  2.58$ & $     3.4$ & \nodata & \nodata & \nodata & \nodata & \nodata &  4 & S  &     \\
v211  & $  48.93$ & $  8.44$ & $    15.3$ & $  84.07$ & $  5.30$ & $    11.4$ & $ 1.59$ & $  1.3$ & 37 & D  &     \\ *
 & \nodata & \nodata & \nodata & \nodata & \nodata & \nodata & \nodata & \nodata & \nodata & \nodata & \\* 
 & $  48.34$ & $  0.58$ & $     1.2$ & $  84.45$ & $  2.15$ & $     0.6$ & $ 0.27$ & $  2.0$ & 41 & D  &     \\
v212  & $  18.38$ & $  1.05$ & $     4.0$ & $  56.60$ & $  5.30$ & $     7.0$ & $ 0.20$ & $  0.6$ & 44 & D  & \tablenotemark{V}  \\ *
 & \nodata & \nodata & \nodata & \nodata & \nodata & \nodata & \nodata & \nodata & \nodata & \nodata & \\* 
 & \nodata & \nodata & \nodata & \nodata & \nodata & \nodata & \nodata & \nodata & \nodata & \nodata & \\ 
v223  & $  27.73$ & $  2.71$ & $     2.1$ & $  40.35$ & $ 38.59$ & $    33.5$ & $ 0.07$ & $  0.1$ & 17 & De & \tablenotemark{V}   \\ *
 & $  27.69$ & $  3.39$ & $     4.2$ & $  40.51$ & $  2.91$ & $     5.4$ & $ 1.16$ & $  0.8$ & 16 & D  &     \\*
 & $  27.95$ & $  3.00$ & $     7.4$ & $  40.69$ & $  3.99$ & $     2.9$ & $ 0.75$ & $  2.5$ & 21 & D  &     \\
v228  & $  78.52$ & $ 74.78$ & $   216.0$ & $ 119.29$ & $ 90.35$ & $   279.1$ & $ 0.83$ & $  0.8$ & 44 & D  &  \tablenotemark{V}   \\ *
 & \nodata & \nodata & \nodata & \nodata & \nodata & \nodata & \nodata & \nodata & \nodata & \nodata & \\* 
 & $  77.35$ & $  3.50$ & $     8.7$ & $ 120.49$ & $ 10.04$ & $    25.1$ & $ 0.35$ & $  0.3$ & 46 & D  &     \\
v231  & $   60.19$ & $  9.79$ & $    56.2$ & \nodata & \nodata & \nodata & \nodata & \nodata & 71 & I  &     \\*
& $  59.87$ & $  5.55$ & $    75.9$ & \nodata & \nodata & \nodata & \nodata & \nodata & 64 & I  &     \\*
& $  55.39$ & $  1.47$ & $    39.1$ & \nodata & \nodata & \nodata & \nodata & \nodata & 78 & I  &     \\
v237  & $  82.18$ & $ 16.50$ & $    11.8$ & $ 108.42$ & $ 36.46$ & $    28.1$ & $ 0.45$ & $  0.4$ & 28 & D  &     \\ *
 & \nodata & \nodata & \nodata & $ 109.07$ & $  1.08$ & $     4.5$ & \nodata & \nodata & 50 & I  &     \\
 & $  81.39$ & $  1.21$ & $     0.7$ & $ 109.24$ & $  1.76$ & $     1.3$ & $ 0.69$ & $  0.6$ & 30 & D  &     \\
v239  & $  55.20$ & $  7.62$ & $    26.0$ & $  95.87$ & $  4.52$ & $    13.3$ & $ 1.69$ & $  2.0$ & 44 & D  & \tablenotemark{V}   \\ *
 & \nodata & \nodata & \nodata & \nodata & \nodata & \nodata & \nodata & \nodata & \nodata & \nodata & \\* 
 & \nodata & \nodata & \nodata & $  96.62$ & $  1.67$ & $     4.5$ & \nodata & \nodata & 12 & S  & \tablenotemark{\#}   \\
v268  & $  29.44$ & $  0.79$ & $     2.7$ & $  74.11$ & $  0.57$ & $     1.8$ & $ 1.39$ & $  1.5$ & 51 & D  &     \\ *
 & \nodata & \nodata & \nodata & \nodata & \nodata & \nodata & \nodata & \nodata & \nodata & \nodata & \\* 
 & \nodata & \nodata & \nodata & \nodata & \nodata & \nodata & \nodata & \nodata & \nodata & \nodata & \\ 
v270  & $  76.77$ & $  8.98$ & $    24.2$ & $ 101.92$ & $  8.47$ & $    20.7$ & $ 1.06$ & $  1.2$ & 31 & D  &     \\ *
 & \nodata & \nodata & \nodata & $ 101.48$ & $  0.33$ & $     0.6$ & \nodata & \nodata & 34 & -  & \tablenotemark{m}   \\
 & $  74.77$ & $  0.45$ & $     0.7$ & $ 103.59$ & $  0.71$ & $     0.8$ & $ 0.63$ & $  0.9$ & 32 & D  &     \\
v274  & $  40.65$ & $  4.61$ & $    35.4$ & $  62.81$ & $  4.84$ & $    31.6$ & $ 0.95$ & $  1.1$ & 50 & Dw &     \\ *
 & $  40.42$ & $  1.95$ & $     4.7$ & $  61.32$ & $  1.54$ & $     8.3$ & $ 1.27$ & $  0.6$ & 33 & Dw &     \\*
 & $  32.33$ & $  0.42$ & $     4.8$ & $  76.00$ & $  0.28$ & $     2.4$ & $ 1.51$ & $  2.0$ & 68 & Dw &     \\
\enddata 
\tablenotetext{V}{Sources which are more than 30\% variable at 1612 MHz}
\tablenotetext{2nd}{d47 has two lines for 1612 MHz data} 
\tablenotetext{m}{Marginal detection, Sept 2002 data} 
\tablenotetext{\#}{August 2003 data}
\tablenotetext{m\#}{Marginal detection, August 2003 data}
\end{deluxetable}

\clearpage

\begin{deluxetable}{c c c c}
\tablewidth{0pt} 
\tablecaption{Number of detections by frequency and
spectral profile. \label{Ta:dets_by_profile}} 
\tablehead{
\colhead{Profile class} & \colhead{1612 MHz} & \colhead{1665 MHz} & \colhead{1667 MHz}} 
\startdata 
{\bf D} & 60 & 5 & 23 \\ 
{\bf DD} & 1 & 0 & 0 \\ 
{\bf De} & 4 & 0 & 2 \\ 
{\bf Dw} & 4 & 1 & 1 \\ 
{\bf I} & 13 & 16 & 10 \\ 
{\bf S} & 4 & 5 & 9 \\ 
\tableline 
& {\bf 86} & {\bf 27} &{\bf 45} \\ \enddata
\end{deluxetable}

\begin{deluxetable}{c c c c c} 
\tablewidth{0pt} \tablecaption{The number of objects from each source
group with emission at each of 1612, 1665 and 1667 MHz.
\label{Ta:all_line-stats}} 
\tablehead{ \colhead{} & \colhead{RI} & \colhead{LI} & \colhead{Quad I} & 
\colhead{Quad IV} } 
\startdata 
1612 MHz & 38 & 30 & 22 & 15 \\ 
1665 MHz & 19 & 1  & 12 & 6 \\ 
1667 MHz & 24 & 14 & 11 & 10 \\ 
\enddata
\end{deluxetable}

\clearpage

\appendix

\newpage
\section{Appendix}

\addtocounter{figure}{-1} \addtocounter{subfigure}{1}
\begin{figure}
	\centering \subfigure[]{
	\scalebox{0.8}{\includegraphics{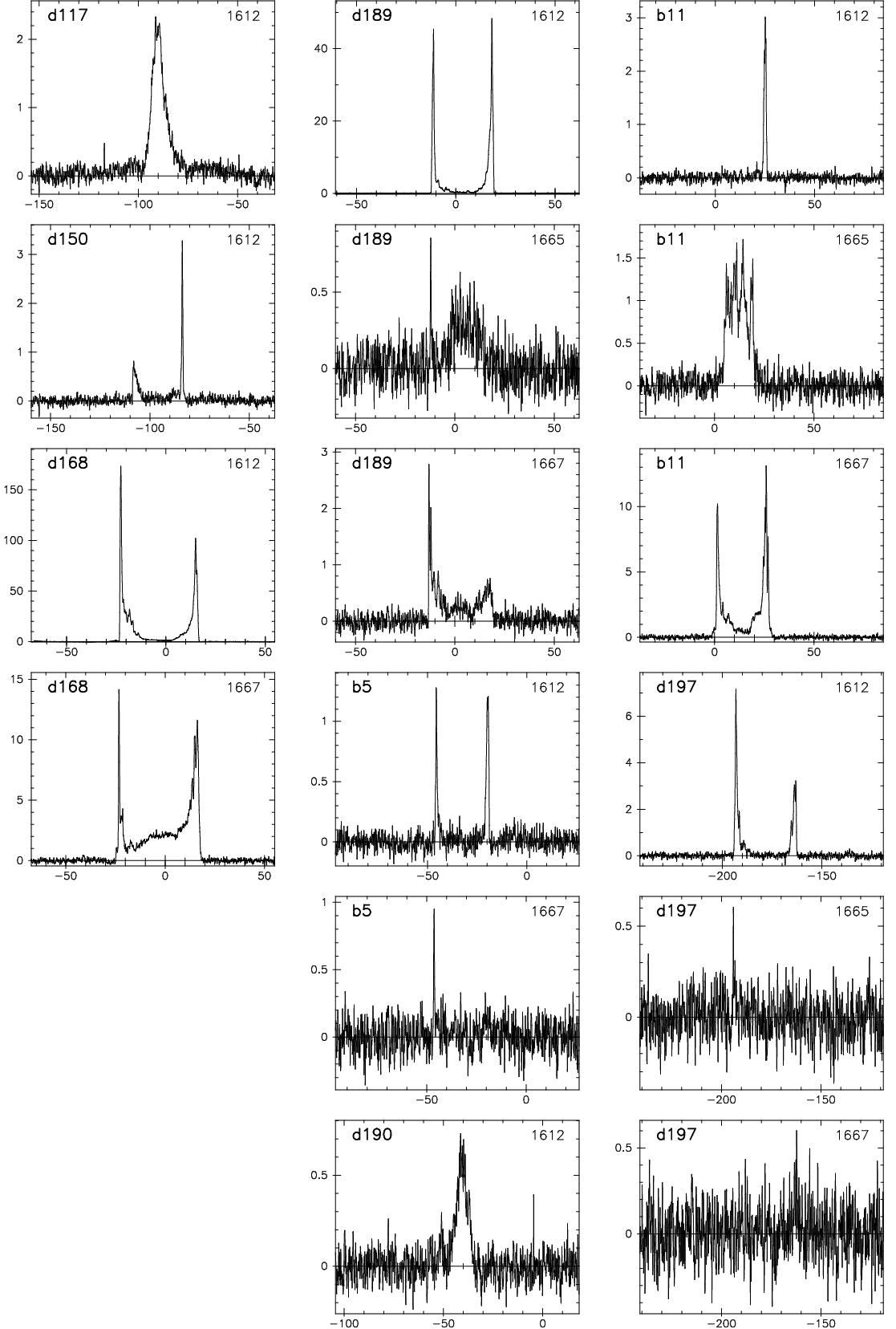}}}
	\caption{1612, 1665 and 1667 MHz spectra
	continued. \label{Fi:thumbnails:b}}
\end{figure}
\addtocounter{figure}{-1} \addtocounter{subfigure}{2}
\begin{figure}
	\centering \subfigure[]{
	\scalebox{0.8}{\includegraphics{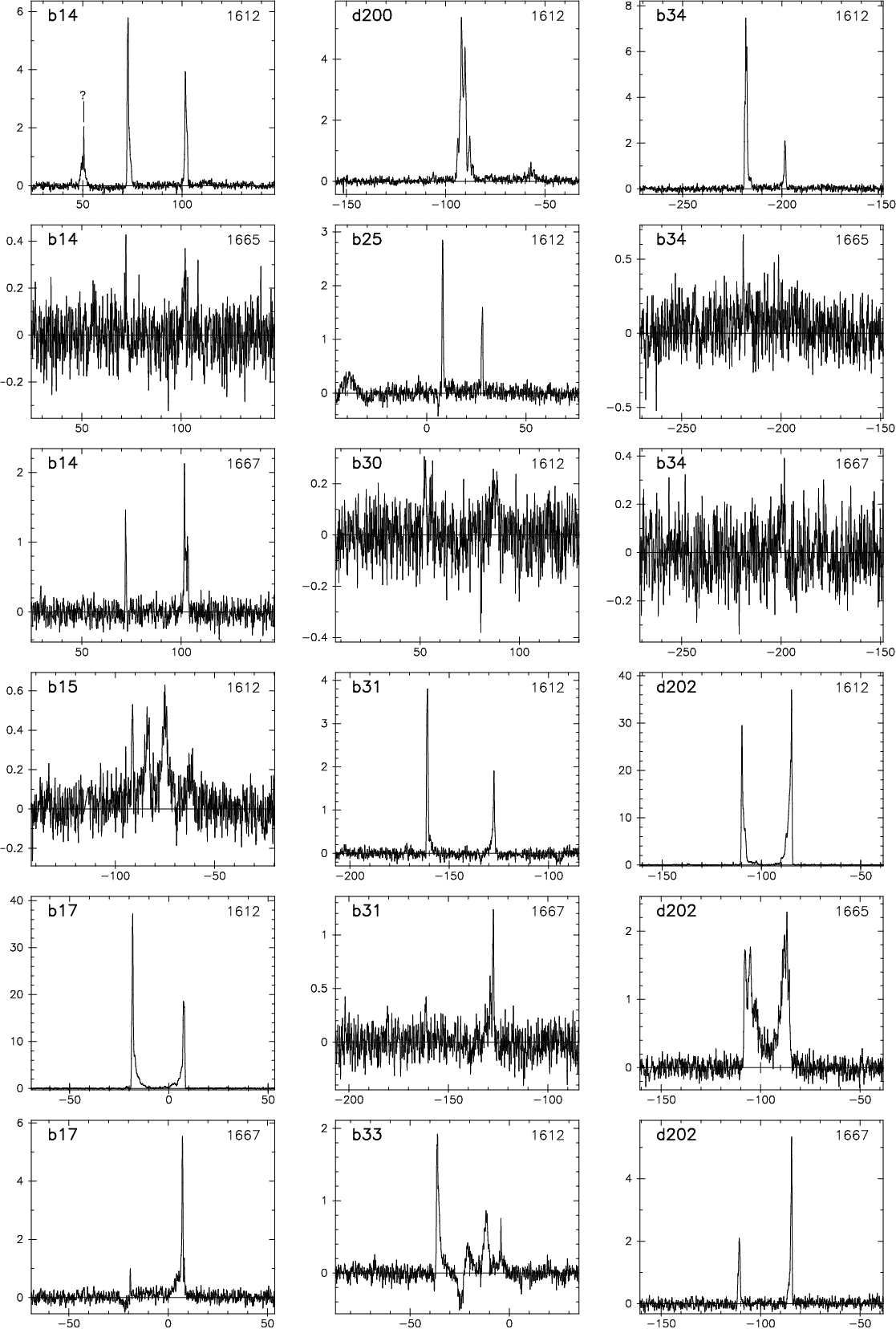}}}
	\caption{1612, 1665 and 1667 MHz spectra
	continued.\label{Fi:thumbnails:c}}
\end{figure}
\addtocounter{figure}{-1} \addtocounter{subfigure}{3}
\begin{figure}
	\centering \subfigure[]{
	\scalebox{0.8}{\includegraphics{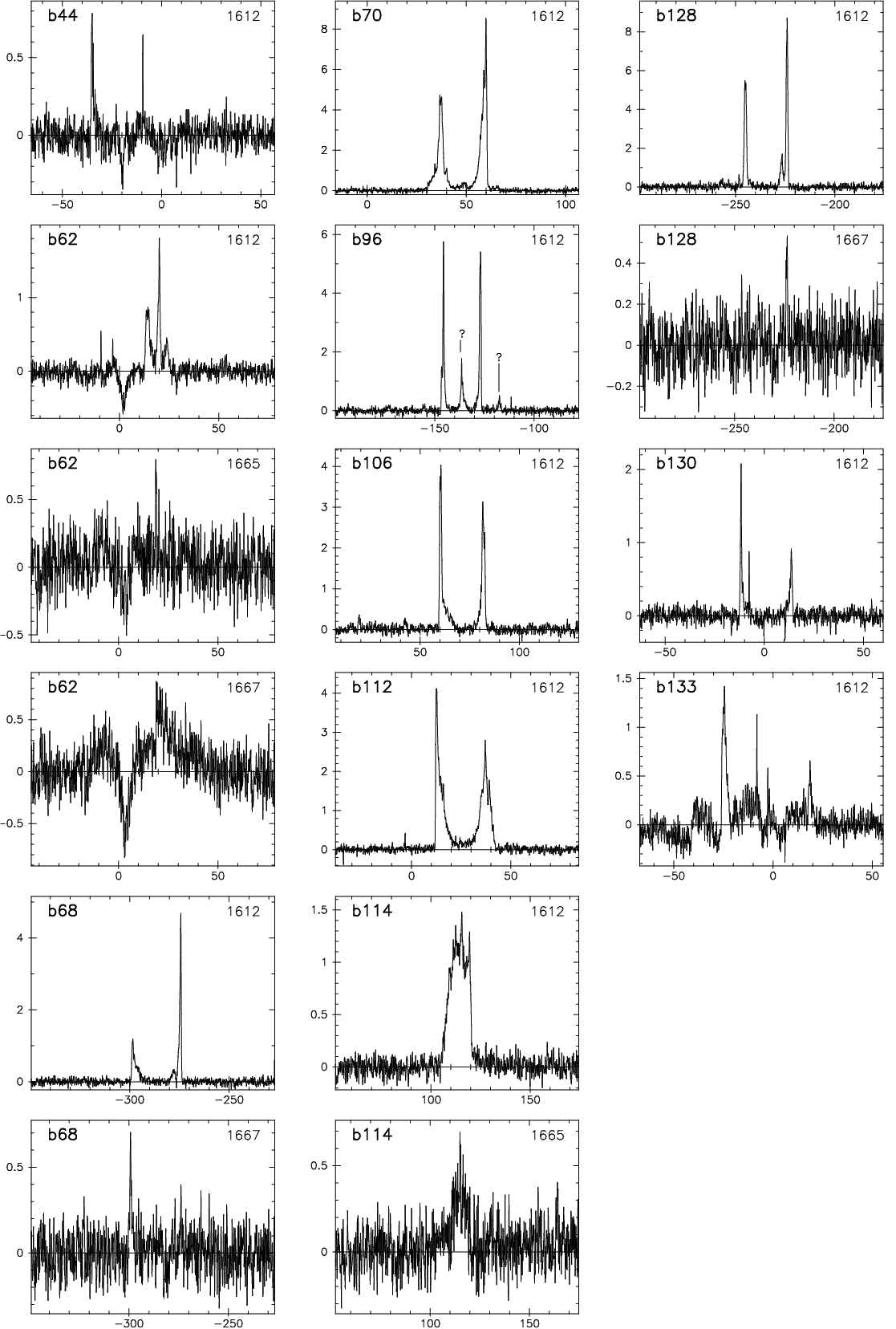}}}
	\caption{1612, 1665 and 1667 MHz spectra
	continued.\label{Fi:thumbnails:d}}
\end{figure}
\addtocounter{figure}{-1} \addtocounter{subfigure}{4}
\begin{figure}
	\centering \subfigure[]{
	\scalebox{0.8}{\includegraphics{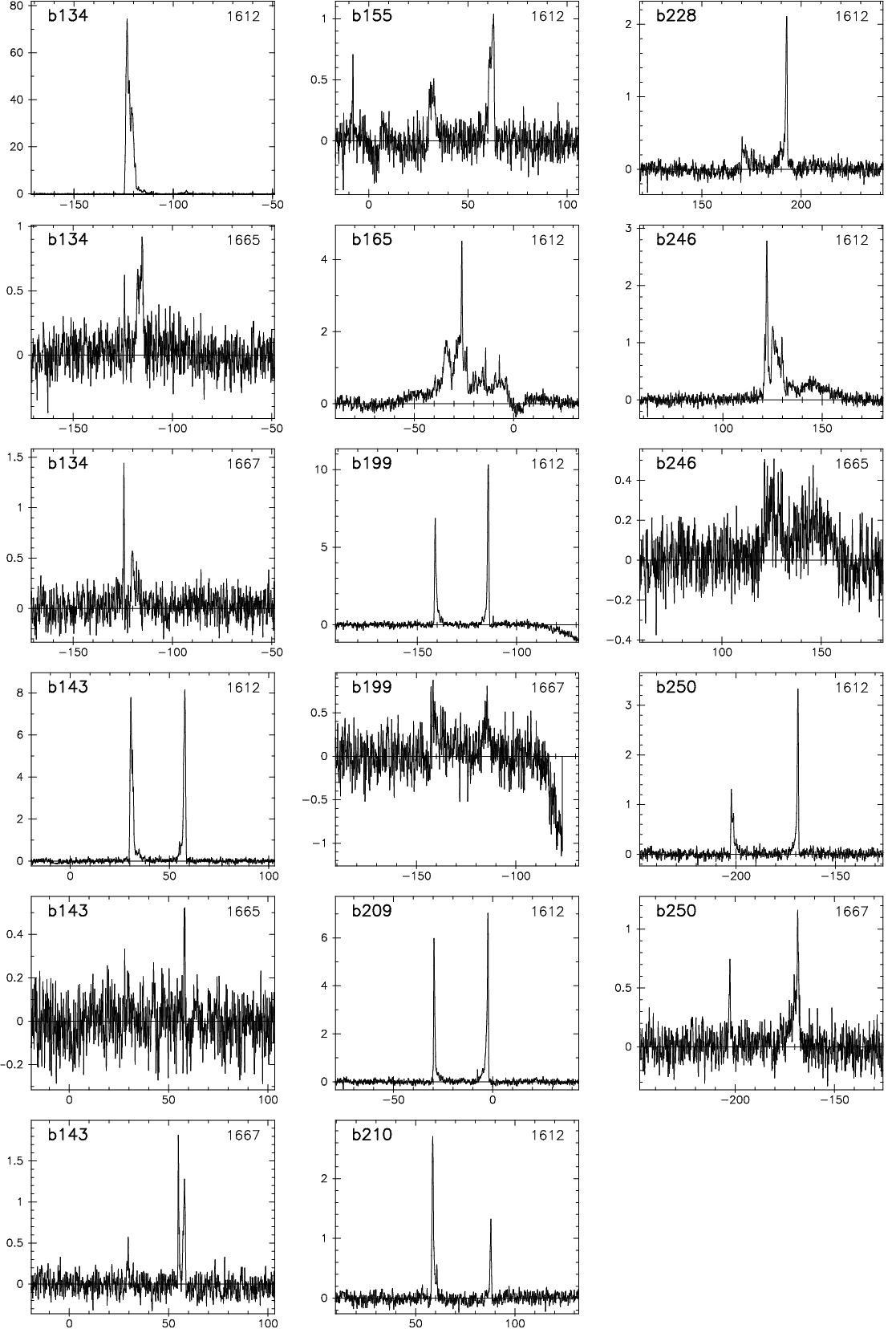}}}
	\caption{1612, 1665 and 1667 MHz spectra
	continued.\label{Fi:thumbnails:e}}
\end{figure}
\addtocounter{figure}{-1} \addtocounter{subfigure}{5}
\begin{figure}
	\centering \subfigure[]{
	\scalebox{0.8}{\includegraphics{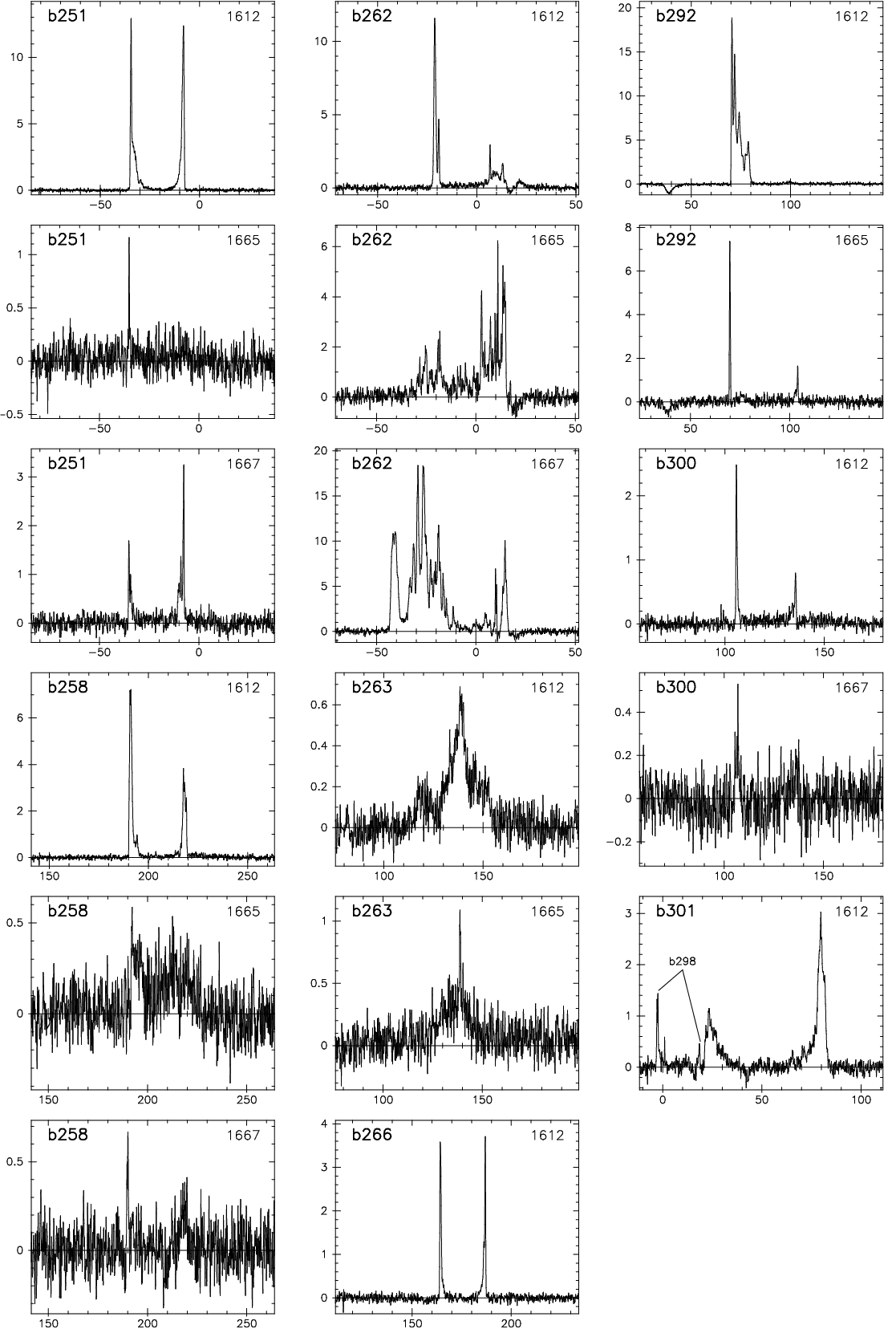}}}
	\caption{1612, 1665 and 1667 MHz spectra
	continued.\label{Fi:thumbnails:f}}
\end{figure}
\addtocounter{figure}{-1} \addtocounter{subfigure}{6}
\begin{figure}
	\centering \subfigure[]{
	\scalebox{0.8}{\includegraphics{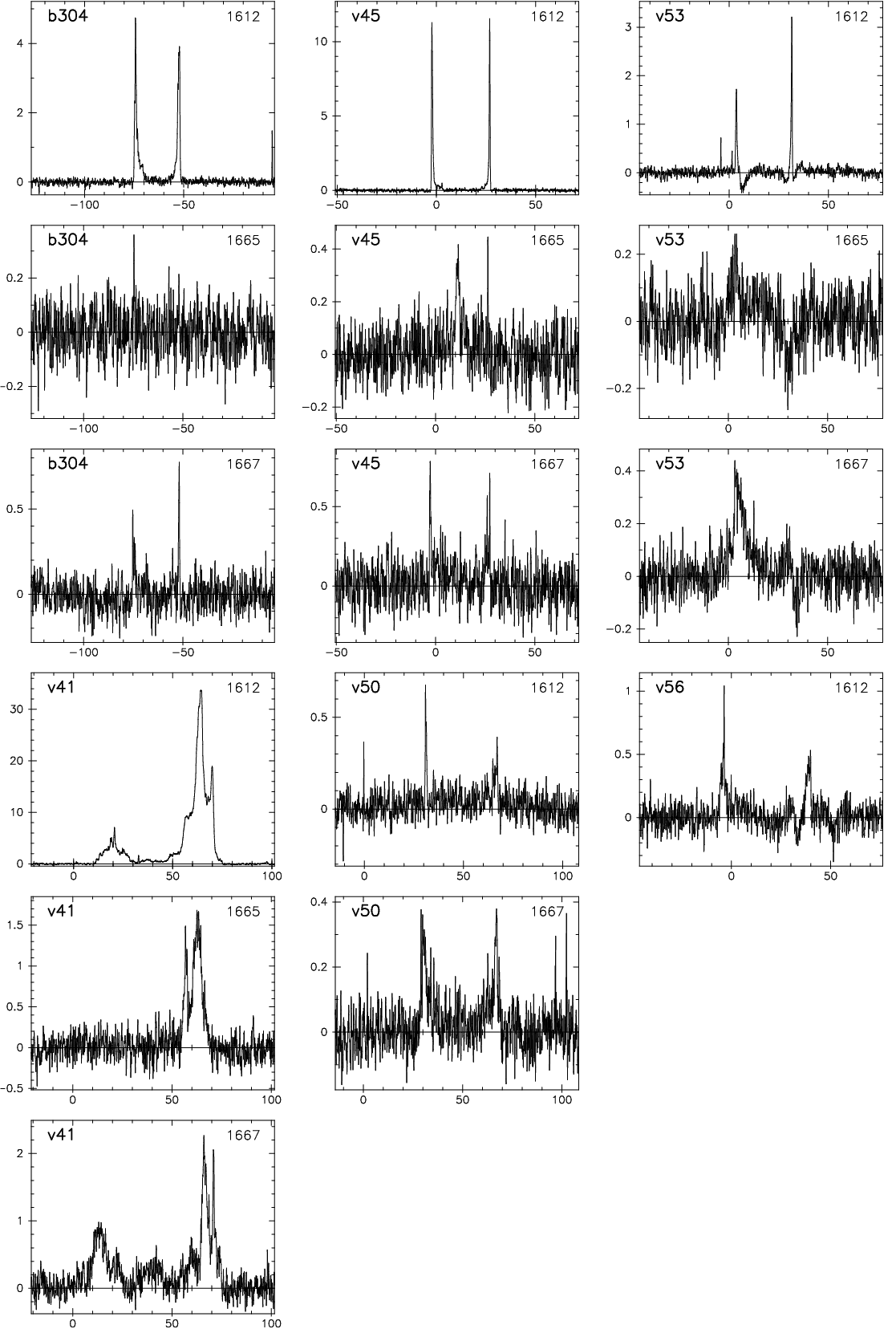}}}
	\caption{1612, 1665 and 1667 MHz spectra
	continued.\label{Fi:thumbnails:g}}
\end{figure}
\addtocounter{figure}{-1} \addtocounter{subfigure}{7}
\begin{figure}
	\centering \subfigure[]{
	\scalebox{0.8}{\includegraphics{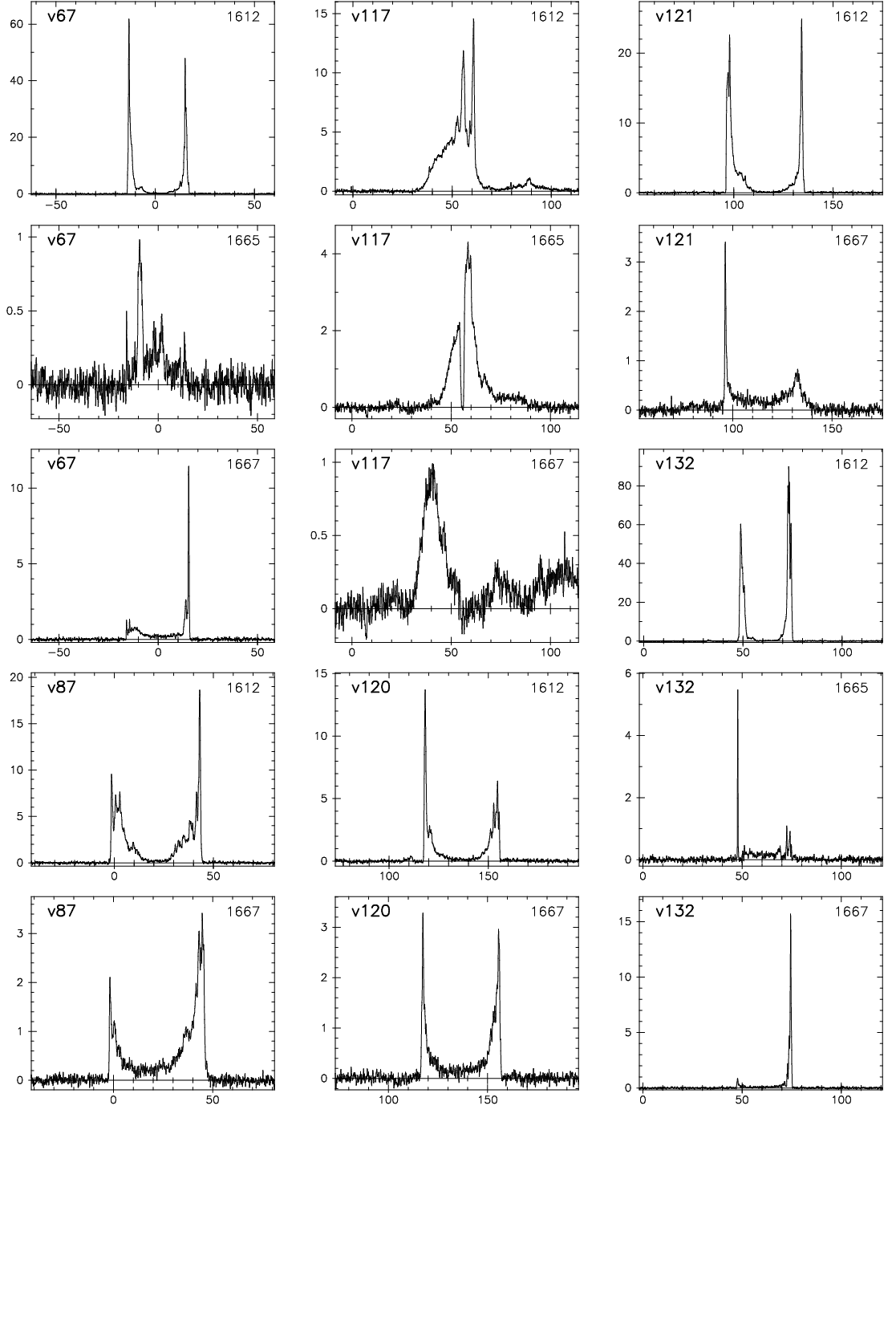}}}
	\caption{1612, 1665 and 1667 MHz spectra
	continued.\label{Fi:thumbnails:h}}
\end{figure}
\addtocounter{figure}{-1} \addtocounter{subfigure}{8}
\begin{figure}
	\centering \subfigure[]{
	\scalebox{0.8}{\includegraphics{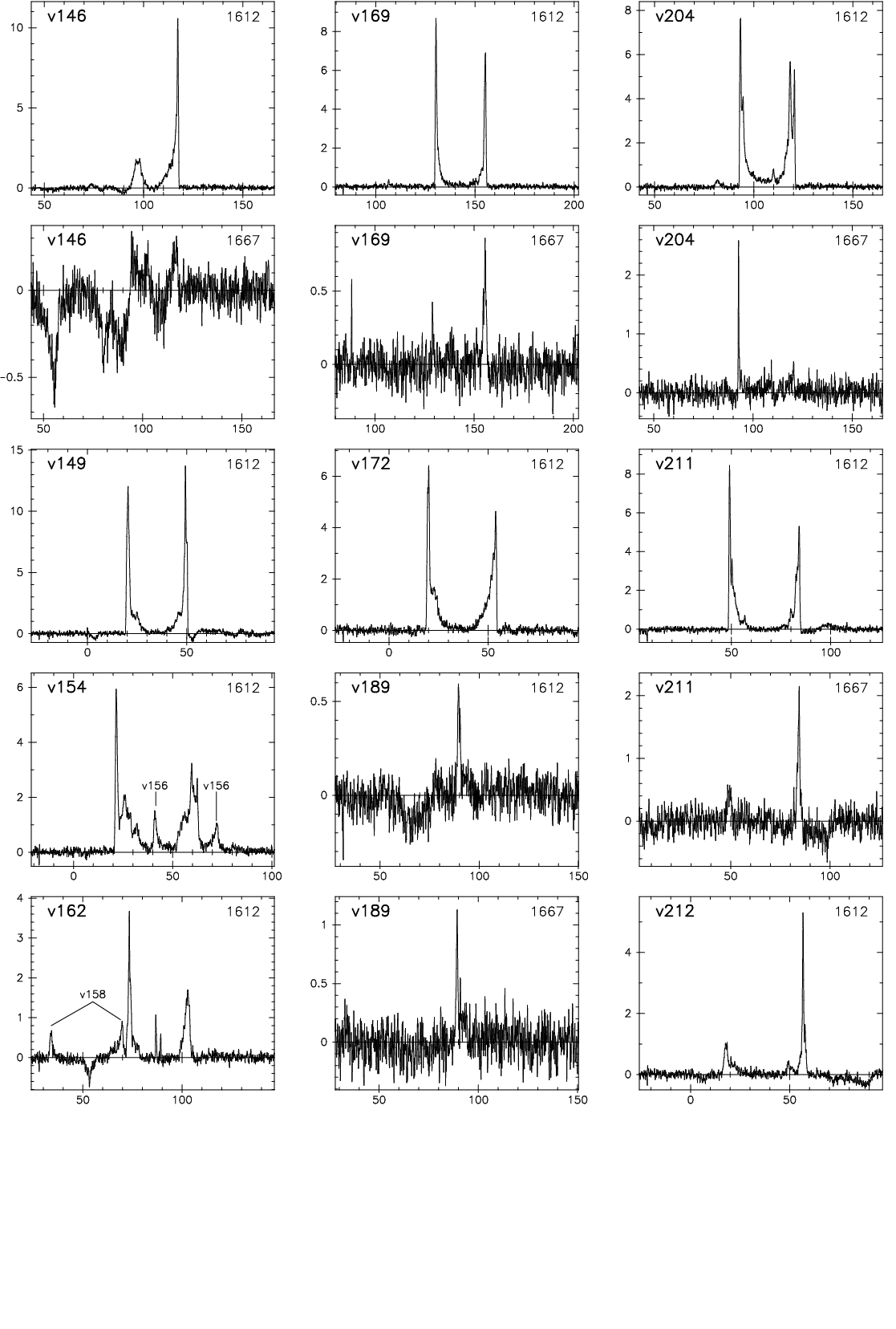}}}
	\caption{1612, 1665 and 1667 MHz spectra
	continued.\label{Fi:thumbnails:i}}
\end{figure}
\addtocounter{figure}{-1} \addtocounter{subfigure}{9}
\begin{figure}
	\centering \subfigure[]{
	\scalebox{0.8}{\includegraphics{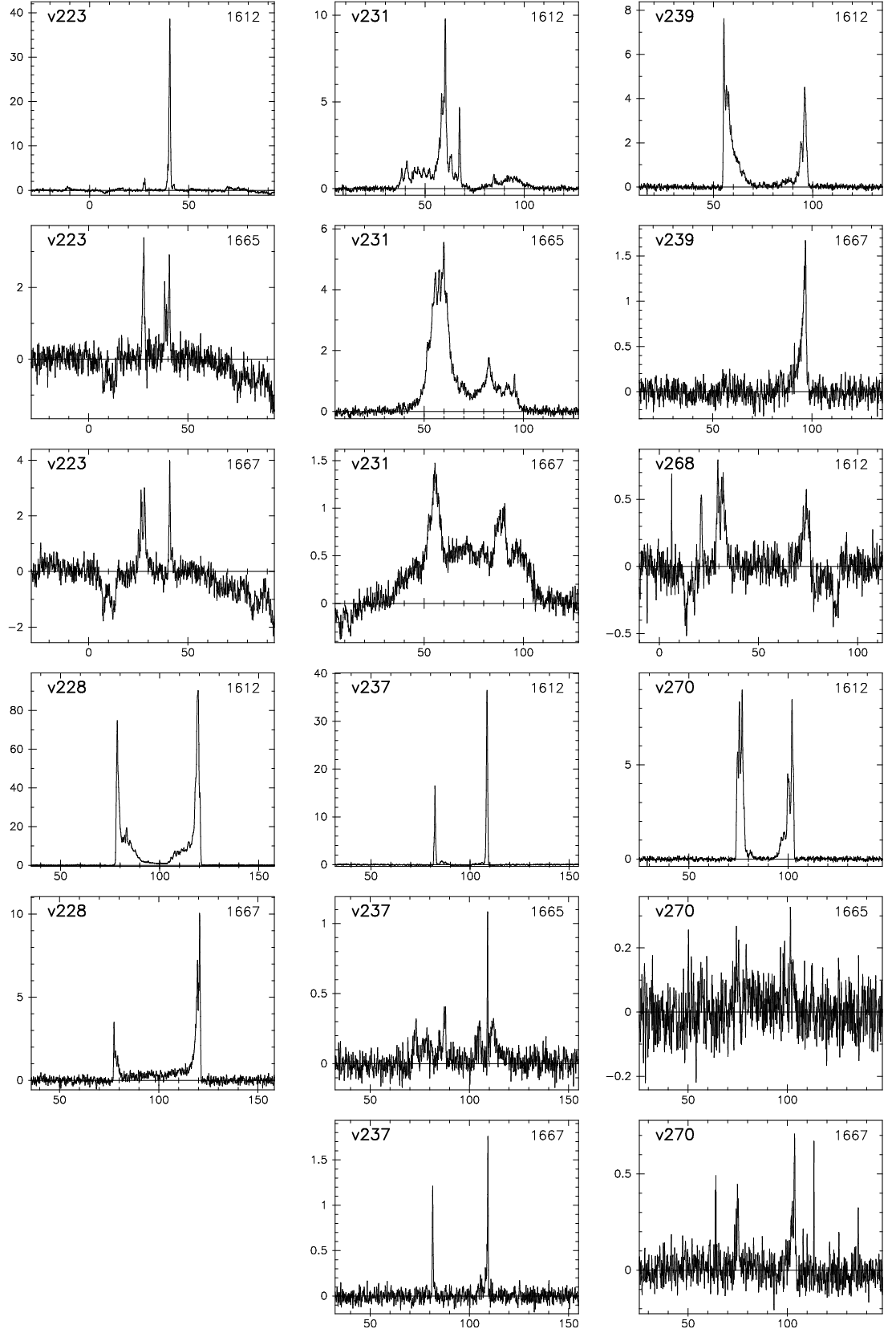}}}
	\caption{1612, 1665 and 1667 MHz spectra
	continued.\label{Fi:thumbnails:j}}
\end{figure}
\addtocounter{figure}{-1} \addtocounter{subfigure}{10}
\begin{figure}
	\centering \subfigure[]{
	\scalebox{0.8}{\includegraphics{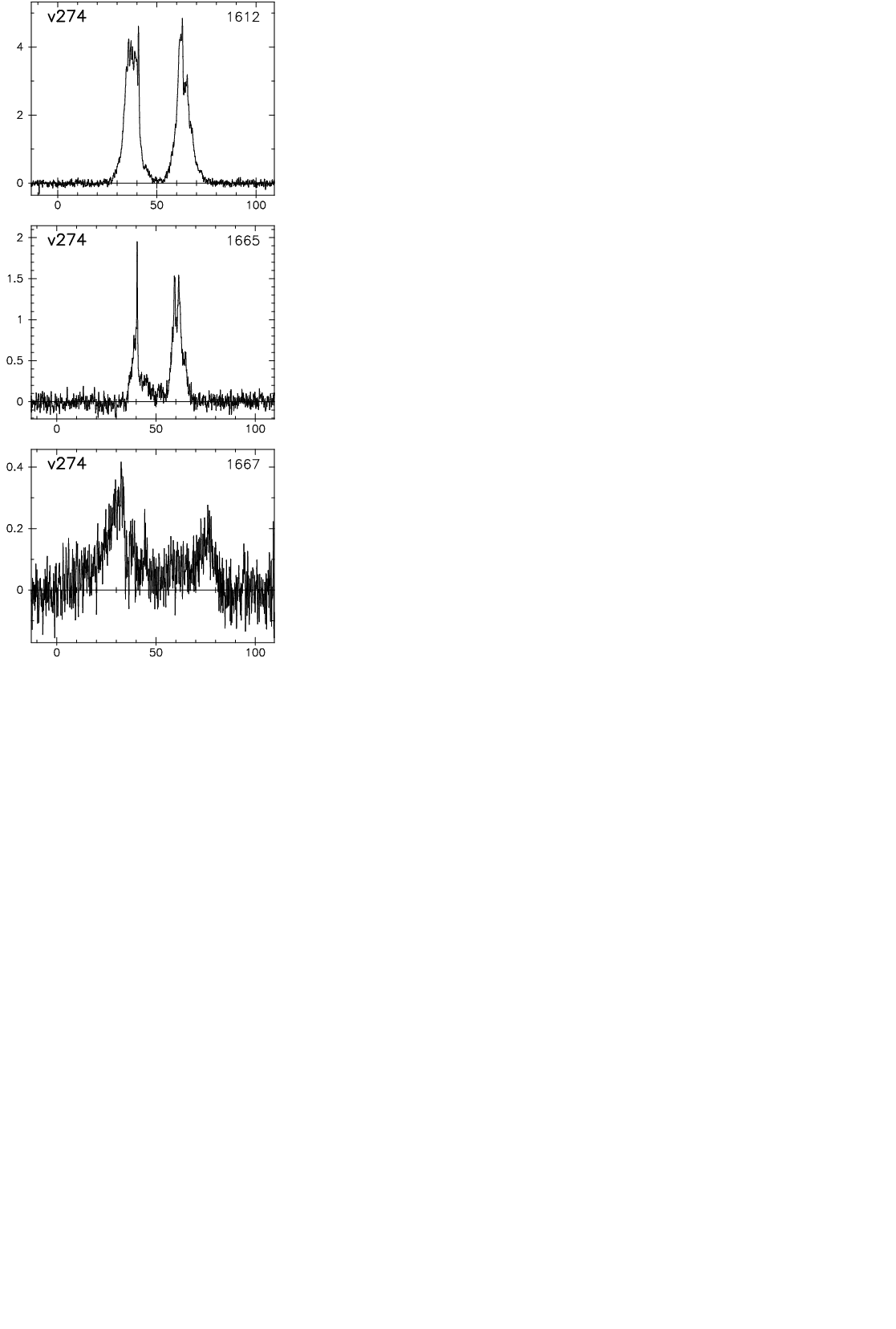}}}
	\caption{1612, 1665 and 1667 MHz spectra
	continued.\label{Fi:thumbnails:k}}
\end{figure}


\newpage

\begin{thebibliography}{}

\bibitem[Bains et al.(2003)]{Bai03} Bains, I.,
Gledhill, T. M., Yates, J. A., \& Richards, A. M. S. 2003, MNRAS, 338,
287

\bibitem[Bowers \& Johnston(1990)]{Bow90} Bowers, P. F., \& Johnston, K. J. 
1990, ApJ, 354, 676

\bibitem[Bowers(1984)]{Bow84} Bowers, P. F. 1984, ApJ, 279, 350

\bibitem[Bujarrabal et al.(1980)]{Buj80} Bujarrabal, V., Guibert, J., 
Nguyen-Q-Rieu, \& Omont, A. 1980, A\&A, 84, 311

\bibitem[Caswell(2004)]{Cas04} Caswell, J. L. 2004, MNRAS, 349, 99

\bibitem[Chan, Henning \& Schreyer(1996)]{Cha96} Chan, S. J., Henning,
T., \& Schreyer, K. 1996, A\&AS, 115, 285

\bibitem[Chapman(1988)]{Cha88} Chapman, J. M. 1988, MNRAS, 230, 415

\bibitem[Chapman \& Cohen(1985)]{Cha85} Chapman, J. M., \& Cohen,
R. J. 1985, MNRAS, 212, 375

\bibitem[Chapman \& Cohen(1986)]{Cha86} Chapman, J. M., \& Cohen, R. J. 
1986, MNRAS, 220, 513


\bibitem[Chapman, Cohen \& Saikia(1991)]{Cha91} Chapman, J. M., Cohen, R. J., 
\& Saikia, D. J. 1991, MNRAS, 249, 227

\bibitem[Chapman, Habing \& Killeen(1995)]{Cha95} Chapman, J. M.,
Habing, H. J., \& Killeen, N. E. B. 1995, in ASP Conf. Ser. 83,
Astrophysical Applications of Stellar Pulsations, eds. R. S. Stobie \&
P. A. Whitelock (San Francisco: ASP), 113

\bibitem[Chapman et. al.(1994)]{Cha94} Chapman, J. M., Sivagnanam, P., 
Cohen, R. J., \& Le Squeren, A. M. 1994, MNRAS, 268, 475

\bibitem[Cohen(1989)]{Coh89} Cohen, R. J. 1989, RPPh, 52, 881

\bibitem[Diamond, Norris \& Booth(1984)]{Dia84} Diamond, P. J., Norris, 
R. P., \& Booth, R. S. 1984, MNRAS, 207, 611 

\bibitem[Elitzur(1976)]{Eli76} Elitzur, M. 1976, ApJ, 203, 124

\bibitem[Frail, Goss \& Slysh(1994)]{Frail94} Frail, D. A., Goss, W. M., 
\& Slysh, V. I. 1994, ApJ, 424, L111

\bibitem[Frank \& Mellema(1994)]{Fra94} Frank, A., \& Mellema, G.
1994, ApJ, 430, 800

\bibitem[Frater, Brooks \& Whiteoak(1992)]{Fra92} Frater, R. H., Brooks, 
J. W., \& Whiteoak, J. B. 1992, JEEEA, 12, 103

\bibitem[Garc\'{\i}a-Segura et al.(1999)]{Gar99}
Garc\'{\i}a-Segura, G., Langer, N., R\'{o}zyczka, M., \& Franco, J.
1999, ApJ, 517, 767

\bibitem[Gaume \& Mutel(1987)]{Gau87} Gaume, R. A., \& Mutel, R. M. 1987, ApJSS, 65, 193

\bibitem[Gledhill, Yates \& Richards(2001)]{Gle01} Gledhill, T. M., Yates, 
J. A., \& Richards, A. M. S. 2001, MNRAS, 328,301

\bibitem[Haynes \& Caswell(1977)]{Cas77} Haynes, R. F., \& Caswell,
J. L. 1977, MNRAS, 178, 219

\bibitem[Herman \& Habing(1985)]{Her85} Herman, J., \& Habing, H. J.
1985, A\&AS, 59, 523

\bibitem[Kwok, Purton \& Fitzgerald(1978)]{Kwo78} Kwok, S., Purton,
C. R., \& Fitzgerald, P. M. 1978, ApJ, 219, L125

 \bibitem[Lockett, Gauthier \& Elitzur(1999)]{Loc99} Lockett, P.,
Gauthier, E., \& Elitzur, M. 1999, ApJ, 511, 235

\bibitem[Manchado et al.(2000)]{Man00} Manchado, A., Villaver,
E., Stanghellini, L., \& Guerrero, M. A. 2000, in ASP Conf. Ser. 199,
Asymmetrical Planetary Nebulae II: From Origins to Microstructures,
eds. J. H. Kastner, N. Soker, \& S. Rappaport (San Francisco: ASP), 17

\bibitem[Reid \& Muhleman(1978)]{Rei78} Reid, M. J., \& Muhleman, D. O. 
1978, ApJ, 220,229

\bibitem[Sahai \& Trauger(1998)]{Sah98} Sahai, R., \& Trauger, J. T.
1998, AJ, 116, 1357

\bibitem[Sault, Teuben \& Wright(1995)]{Sau95} Sault, R. J., Teuben, P. J., \& 
Wright, M. C. H. 1995, in ASP Conf. Ser. 77, Astronomical Data Analysis 
Software and Systems IV, eds. R. A. Shaw, H. E. Payne, \& J. J. E. Hayes 
(San Francisco: ASP), 433

\bibitem[Sevenster et al.(1997a)]{Sev97I} Sevenster, M. N.,
Chapman, J. M., Habing, H. J., Killeen, N. E. B., \& Lindqvist, M.
1997a, A\&AS, 122, 79

\bibitem[Sevenster et al.(1997b)]{Sev97II} Sevenster, M. N.,
Chapman, J. M., Habing, H. J., Killeen, N. E. B., \& Lindqvist, M.
1997b, A\&AS, 124, 509

\bibitem[Sevenster \& Chapman(2001)]{Sev01b} Sevenster, M. N., \&
Chapman, J. M. 2001, ApJ, 546, L119

\bibitem[Sevenster et al.(2001)]{Sev01} Sevenster, M. N., van
Langevelde, H. J., Moody, R. A., Chapman, J. M., Habing, H. J., \&
Killeen, N. E. B. 2001, A\&A, 366, 481

\bibitem[Sevenster(2002a)]{Sev02I} Sevenster, M. N. 2002a, AJ, 123,
2772

\bibitem[Sevenster(2002b)]{Sev02II} Sevenster, M. N. 2002b, AJ, 123,
2788

\bibitem[Sivagnanam \& David(1999)]{Siv99} Sivagnanum, P., \& David, P. 
1999, MNRAS, 304, 622

\bibitem[Soker(2001)]{Sok01} Soker, N. 2001, ApJ, 558, 157

\bibitem[Staveley-Smith(1985)]{Sta85} Staveley-Smith, L. 1985, PhD
thesis, University of Manchester

\bibitem[Su, Hrivnak \& Kwok(2001)]{Su01} Su, K. Y. L., Hrivnak,
B. J., \& Kwok, S. 2001, AJ, 122, 1525

\bibitem[Ueta, Meixner \& Bobrowsky(2000)]{Uet00} Ueta, T., Meixner,
M., \& Bobrowsky, M. 2000, ApJ, 528, 861

\bibitem[van der Veen(1989)]{Vee89} van der Veen, W. E. C. J. 1989, A\&A, 210, 127

\bibitem[van der Veen \& Habing(1988)]{Vee88} van der Veen,
W. E. C. J., \& Habing, H. J. 1988, A\&A, 194, 125

\bibitem[van Hoof, Oudmaijer \& Waters(1997)]{Hoo97} van Hoof,
P. A. M., Oudmaijer, R. D., \& Waters, L. B. F. M. 1997, MNRAS, 289,
371

\bibitem[van Langevelde, van der Heiden \& van Schooneveld(1990)]{Lan90} 
van Langevelde, H. J., van der Heiden, R., \& Schooneveld, C. 1990, A\&A, 239, 193

\bibitem[Wardle(1999)]{War99} Wardle, M. 1999, ApJ, 525, L101

\bibitem[Welty, Fix \& Mutel(1987)]{Wel87} Welty, A. D., Fix, J. D.,
\& Mutel, R. L. 1987, ApJ, 318, 852

\bibitem[Wood(2000)]{Woo00} Wood, P. R. 2000, in ASP Conf. Ser. 199,
Asymmetrical Planetary Nebulae II: From Origins to Microstructures,
eds. J. H. Kastner, N. Soker, \& S. Rappaport (San Francisco: ASP), 51

\bibitem[Zijlstra et al.(2001)]{Zij01} Zijlstra, A. A., Chapman,
J. M., te Lintel Hekkert, P., Likkel, L., Comeron, F., Norris, R. P.,
Molster, F. J., \&  Cohen, R. J. 2001, MNRAS, 322, 280

\end{thebibliography}
\end{document}